\documentclass[11pt]{article}
\catcode`@11
\def\seceqaa{\@addtoreset{equation}{section}
\def\theequation{A\arabic{equation}}}
\usepackage{amssymb}
\catcode`@11
\begin{document}
\begin{titlepage}
\begin{center}
{\Large\bf Towards MQGP}
\vskip 0.1in Mansi Dhuria \footnote{email: mansidph@iitr.ernet.in}
and
 Aalok Misra\footnote{e-mail: aalokfph@iitr.ernet.in
}\\
Department of Physics, Indian Institute of Technology,
Roorkee - 247 667, Uttarakhand, India\\
 \vskip 0.5 true in
\date{\today}
\end{center}

\thispagestyle{empty}
\begin{abstract}

Using the type IIB background of \cite{metrics} - a gravity dual to a large-$N$ thermal QCD-like theory in the presence of $N_f$ thermal quarks -
we calculate the chemical potential $\mu_C$  as a  function of temperature  due to a $U(1)$ (c.f. $U(N_f)\sim SU(N_f)\times U(1)$) gauge field living on the world volume of $N_f$ space-time-filling $D7$-branes  wrapped around a four-cycle in a resolved warped
deformed conifold along with $(M)N$ (fractional) $D3$-branes,  and show that for the Ouyang embedding,
$\frac{\partial\mu_C}{\partial T}\biggl.\biggr|_{N_f}<0$ (implying what one expects: $\frac{\partial S}{\partial N_f}\biggl.\biggr|_{T}>0$) up to linear order in the embedding parameter (which we take to be real and
slightly less than unity). By explicitly verifying that $\frac{\partial\mu_C}{\partial N_f}\biggl.\biggr|_{T}>0$ up to linear order
in the embedding parameter, we demonstrate the possible thermodynamical stability of the type IIB background to that order. Analogous to
\cite{SYZ 3 Ts}, we then obtain a local (as the resolved warped deformed conifold does not possess a `third' global killing isometry along
 the `original' angular variable $\psi\in[0,4\pi]$ for implementing SYZ mirror symmetry) $M$-theory  uplift of type $IIB$ background  by
first obtaining the local type $IIA$ mirror using SYZ mirror symmetry near $(\theta_{1,2},\psi)=(\langle\theta_{1,2}\rangle,\left\{0,2\pi,4\pi
\right\})$ and then oxidizing the so-obtained type $IIA$ background to $M$ theory.  We then take two limits of this uplift: (i)  $g_s<<1, g_sN_f<<1,\frac{g_sM^2}{N}<<1, g_s^2M N_f<<1, g_sM>>1, g_sN>>1$ similar to
\cite{metrics} effected by $M\sim \epsilon^{-{\frac{3d}{2}}}$, $N\sim {\epsilon^{-19d}} $, $g_s \sim
{\epsilon^{d}}, d>0 $ and $ \epsilon\leq{\cal O}(10^{-2})$; (ii) the second `MQGP limit'  $\frac{g_sM^2}{N}<<1, g_sN>>1$ for finite
$g_S, M$ effected by:  $g_s\sim\epsilon, M\sim\epsilon^{-\frac{3d}{2}}, N\sim\epsilon^{-39d}, d>0, \epsilon \lesssim 1$. The second limit is more suited for the study of QGP (See \cite{finite_gYM_Natsuume}) than (i), and due to the finiteness of
the string coupling can meaningfully only be addressed within an M-theoretic framework.   {\it For both limits}, in this process we obtain a
black $M3$-brane solution whose near-horizon geometry near the $\theta_{1,2}=0,\pi$ branches, preserves $\frac{1}{8}$
supersymmetry. Interestingly, assuming  the formula for $\frac{\eta}{s}$ of \cite{KovtunSonStarinets} obtained for only
radial-coordinate-dependent metric (i.e. $\mu_C=0$)  to also be valid for the type IIA mirror and its
  local M-theory uplift having frozen the angular dependence, we obtain the value of
$\frac{\eta}{s}$ using the $M$ theory uplift to be exactly equal to $\frac{1}{4\pi}$ - there is no angular dependence in $G_{tt,rr,\mathbb{R}^3}$ (partly justifying the assumption, further supported by the fact that the aforementioned $\mu_C$ can be tuned to be very small)  - {\it for both limits}. In the same spirit, the diffusion constant for both, types IIB/IIA backgrounds, comes out to be the
reciprocal of the temperature. The $D=11$ supergravity action (Einstein-Hilbert + Gibbons-Hawking-York surface + Flux +
${\cal O}(R^4)$terms) receives the dominant contributions near $\langle\theta_{1,2}\rangle=0,\pi$, where there are poles. Introducing an appropriate angular cut-off $\epsilon_\theta$ and  using the $\langle\theta_{1,2}\rangle=\epsilon_\theta,\pi-\epsilon_\theta$-local uplift the specific heat from the
finite part of the action (which  is found to be cut-off-independent) turns out to be positive indicative of the thermodynamical stability of the  uplift. An
asymptotically-linear-dilaton-gravity-type interpretation can be given to the relevant counter-terms in the limit (i).  Further, it
is verified that the black $M3$-brane entropy  $S\sim r_h^3$ from M-theoretic thermodynamical methods as well as
the horizon area  calculated from the starting type IIB, mirrory type IIA  and the black $M3$-brane solutions.
\end{abstract}
\end{titlepage}

\section{Introduction}

During the past few years there has been a lot of progress toward constructing the string theoretic dual descriptions of large-N gauge theories. The experimental data obtained at the Relativistic Heavy Ion Collider (RHIC) has provided  valuable information that shows characteristics of non-perturbative gauge theories obtained in the context of Quark Gluon plasma etc. The lack of reliable computing methods to study non-perturbative gauge theories in various theoretical models motivates oneself to study non-compact manifolds that are sufficient to give gravity duals used to studying various aspects of strongly coupled (IR) behavior of gauge theory. The AdS/CFT correspondence has been widely used in establishing, in particular, the connection between strongly coupled ${\cal N} = 4$ supersymmetric Yang-Mills (SYM) theory in large N limit and classical ten-dimensional supergravity. However, the analysis based on that relies on AdS spaces whose dual is a conformal field theory (CFT) with no running couplings \cite{maldacena, witten}. Though AdS background suffices to give complete picture of IR behavior, yet it does not provide UV completion of gauge theories. These limitations were overcome in models of gauge-gravity duality that are not supersymmetric, and are non-conformal. The work in this direction has been done by many people by choosing different non-AdS backgrounds. One of the popular background used is to study D3-branes in a conifold, which is inspired from studies of branes at conical singularities, resulting in ${\cal N}=1$ superconformal gauge theory \cite{klebanov_witten, kacru+silverstein,lawrence et al}. The non-conformal or the confining behavior of gauge theories have been well understood in `Klebanov-Strassler' background \cite{Klebanov+Strassler} (as well in \cite{Klebanov+Tseytlin}) formed by placing $M$ fractional $D5$-branes wrapping over a vanishing  $S^2$ of $T^{1,1}$ base of conifold in addition to  $D3$-branes placed at the tip of the conifold. Though in this model, quarks transform  in the bi-fundamental representations of the two possible UV gauge groups, but eventually cascade away in the far IR which is not relevant to studying thermal aspects. Therefore, in order to allow the presence of fundamental quarks at finite temperature to make the discussion more appropriate to study of the deconfined phase of strongly coupled QCD i.e ``Quark Gluon Plasma", one has to incorporate co-incident $D7$-branes in `Klebanov-Strassler' construction of warped deformed conifold. The effect of $D7$-branes in KS background has been computed locally in \cite{metrics} just by introducing back-reaction due to presence of $D7$-branes with the assumption that axio-dilation vanishes locally and therefore, is not justified once one approaches the full global solution. The other subtle issues related to the same are discussed in \cite{metrics}. The very well known background given in the presence of non-zero axio-dilaton has been formed by introducing coincident $D7$-branes in `Klebanov-Tseytlin' background/in the large-$r$ limit of `Klebanov-Strassler' background via Ouyang embedding as explained in \cite{ouyang}. However, in the presence of non-zero temperature, Ouyang geometry also gets modified and has been given in the context of type IIB string theory in \cite{metrics} by inserting a black hole in a warped resolved conifold, named as modified OKS-BH background. In that background, backreaction due to presence of black hole as well as $D$7-branes is included in 10-D warp factor. Now, by extensively using modified Ouyang-Klebanov-Strassler(OKS)-BH background  including deformations, in this paper, we attempt to investigate the possibility of getting local 11-dimensional uplift of warped deformed conifolds relevant to study of particular phase of  ${\cal N}=1$ strongly coupled QCD known as ``Quark Gluon Plasma". Since deformed conifolds do not possess global isometries along three directions, it is not possible to achieve M-theory uplift globally. However as given in \cite{SYZ 3 Ts}, by applying suitable co-ordinate transformation on complex structure of  base of $T^3$ fiber, one can obtain an isometry along the third direction locally. Therefore, the main aim is to first satisfy the requirements of implementing mirror symmetry (obtained by applying three successive T-dualities on a supersymmetric $T^3$ fiber inside the Calabi-Yau) given by Strominger-Yau-Zaslow to transform the type IIB metric to its type IIA mirror so that one can obtain a local M-theory uplift of the latter.

In this work, we will consider the following two limits:
\begin{eqnarray}
\label{limits_Dasguptaetal-i}
& (i)   &  {\rm weak}(g_s){\rm coupling-large\ t'Hooft\ coupling\ limit}:\nonumber\\
& &  g_s<<1, g_sN_f<<1, \frac{g_sM^2}{N}<<1, g_sM>>1, g_sN>>1\nonumber\\
& & {\rm effected\ by}: g_s\sim\epsilon^{d}, M\sim\epsilon^{-\frac{3d}{2}}, N\sim\epsilon^{-19d}, \epsilon\leq{\cal O}(10^{-2}), d>0
 \end{eqnarray}
(the limit in the first line of (\ref{limits_Dasguptaetal-i}) though not its realization in the second line, considered in \cite{metrics});
\begin{eqnarray}
\label{limits_Dasguptaetal-ii}
& (i i) & {\rm \bf MQGP\ limit}: \frac{g_sM^2}{N}<<1, g_sN>>1, {\rm finite}\
 g_s, M\ \nonumber\\
& &{\rm effected\ by}:  g_s\sim\epsilon^d, M\sim\epsilon^{-\frac{3d}{2}}, N\sim\epsilon^{-39d}, \epsilon\lesssim 1, d>0.
\end{eqnarray}
We calculate various thermodynamical quantities relevant to demonstrating the thermodynamical stability of the uplift. In addition to that, it has
been shown that shear viscosity-to-entropy density ratio gets a universal value of $\frac{\hbar}{4 k_B \pi}$ for a large class of strongly coupled quantum field theories whose dual description involves black holes \cite{Kovtun}. Therefore, it is interesting to evaluate $\eta/s$ to verify whether gravity dual obtained using M-theory uplift actually possess aspects of strongly interacting gauge theories.

The rest of the paper is organized as follows. In section {\bf 2}, we outline the basics of (resolved) warped deformed conifolds and mention suitable co-ordinate transformations to generate, locally, a `third' isometry and to produce a large volume of the base of a local $T^3$ fibration to permit application of mirror symmetry a al SYZ in type IIB background. In section {\bf  3}, we  calculate the chemical potential arising from a $U(1)$ gauge field arising from the $U(N_f)\sim SU(N_f)\times U(1)$ living on the $D7$-brane world volume as a function of temperature using the type IIB background of \cite{metrics}. We also examine the behavior of same as a function of increasing number of quarks i.e $N_f$ at constant temperature to investigate thermodynamical stability of system in type IIB background. In section  {\bf 4}, using the 10-dimensional metric of (resolved) warped deformed conifolds \cite{metrics} and analytic
expressions given in \cite{SYZ 3 Ts} to mirror transform, locally, the components of type IIB metric as well as NS-NS B-field components, we first obtain the form of the metric of the type IIA mirror manifold. In the process, at  $\theta_i= 0,\pi$ and $r\sim \sqrt{3}a$, we show how to construct a large base of a local $T^3$ fibration so that mirror symmetry between resolved and deformed conifolds can be established locally; in fact if one assumes that the local uplift is valid globally, we show that one can (more readily) ensure a large-base $T^3$-fibration near $\theta_{1,2}=\frac{\pi}{2},
r\sim\sqrt{3}a$ as well. Next, by applying T-duality rules given in \cite{T-dual Hassan} to T-dualize RR odd-form field strengths and performing successive three T-dualities along three, locally toroidal
 isometry directions, we get various non-zero components of type IIA  2-form field strength using which we obtain one-form gauge field potential which eventually leads to a local M-theory metric. We then consider two limits - one which involves taking the weak coupling
limit of M-theory with a small string coupling and the other, which we call as the `MQGP' limit involves considering a finite string coupling
( $ g_s \lesssim 1$ and hence relevant to strongly coupled QGP).  We argue that the M-theory uplift for both limits
yields an $M3$-brane with $\frac{1}{4}$ near-horizon supersymmetry near $\theta_{1,2}=0,\pi$.  Section {\bf 5} has three sub-sections. In {\bf 5.1,
5.2}, assuming the formula for $\frac{\eta}{s}$ and the diffusion coefficient $D$  of \cite{KovtunSonStarinets} obtained for only
 radial-coordinate-dependent
metric (i.e. $\mu_C=0$) to also be valid for the type IIA mirror and its  local M-theory uplift having frozen the angular dependence (tunably
small $\mu_C$ of Sec. {\bf 3}),  we  calculate shear viscosity-entropy density ratio $\frac{\eta}{s}$ in both limits  and show that one gets exactly
$\frac{\eta}{s}=\frac{1}{4\pi}, D\sim\frac{1}{T}$; $G^{IIA}_{tt,rr}, G^{\cal M}_{tt,rr}$ having no angular dependence and  the angular
dependences entering via the metric determinants,  cancel out neatly. After that, again using the same limits, in {\bf 5.3}  we get simplified expressions of
11-dimensional Euclideanized space-time action which includes contribution from 11-dimensional bulk and  flux terms, 10-dimensional Gibbons-Hawking
term as well as contributions arising from ${\cal O}(R^4)$-terms given that the horizon radius turns out to be of the order of the string
scale making higher derivative contributions important; despite the fact that we end up
with a string-scale-sized horizon, the finite parts of the ${\cal O}(R^4)$-terms vanish in the aforementioned limits.
 We show
that in the limits (\ref{limits_Dasguptaetal-i}) and (\ref{limits_Dasguptaetal-ii}), the IR-divergent portion of, e.g., $\int_{r=r_\Lambda}\sqrt{h}R$
(and/or $\int_{r=r_\Lambda}\sqrt{h}$ and/or $\int_{r=r_\Lambda}\sqrt{h}|G_4|^2$),
can act as an appropriate counter-term cancelling the
IR-divergent portions ($r_\Lambda\rightarrow\infty$) of the Einstein-Hilbert + Gibbons-Hawking-York + Flux + ${\cal O}(R^4)$ terms of the action; in
limit (\ref{limits_Dasguptaetal-i}), we show that one can provide an asymptotically-linear-dilaton-type interpretation to the counter term(s). In the aforementioned calculations,
we introduce  cut-offs $\epsilon_{\theta}$  near $\theta_{1,2}=0,\pi$, which after
identification with $\epsilon^{\gamma_{(i),(ii)}}$ for appropriate $\gamma_{(i),(ii)}$
for the two limits (\ref{limits_Dasguptaetal-i}) and (\ref{limits_Dasguptaetal-ii}) with the $\epsilon$ also the same as the ones appearing in the limits
(\ref{limits_Dasguptaetal-i}) and (\ref{limits_Dasguptaetal-ii}), ensure that  the
finite part of the action - generated entirely by the  Gibbons-Hawking-York surface term - is independent of the cut-offs $\epsilon_{\theta}$.
  After evaluating the 11-D action/partition function in the aforementioned limit, we
calculate various thermodynamical quantities, e.g, entropy and specific heat in 11-dimensional M-theory background the sign of which has been
considered to be a criterion to check the thermodynamical stability of the $M$-theory uplift. There is one {\bf appendix A}: for the paper to be
self-contained, in this we first  quote from \cite{SYZ 3 Ts} the analytic expressions used to triple  T-dualize, locally,
the type IIB  metric components as well B-field. We then give the expressions for the type IIA local
mirror metric, simplified and valid for both limits (\ref{limits_Dasguptaetal-i})
 and (\ref{limits_Dasguptaetal-ii}). Next, we give the general  expressions of various components of T-dualized RR form field strengths relevant to obtaining $F^{IIA}_2$, not limited to (\ref{limits_Dasguptaetal-i}) or (\ref{limits_Dasguptaetal-ii}).

\section{Basics}

The basic motivation in using non-AdS background to explain the properties of QCD was to provide a geometric background that helps to explain not just the physics of QCD in the IR region but is also  sufficient to unravel the various key points needed to explain UV completion of the theory. The knowledge of UV completion is important to handle the issues related to finiteness of the  solution at short distances as well as to capture certain aspects of large-N thermal QCD. The different models were proposed to incorporate the effect of renormalization group in the dual background by  connecting conformal fixed points at IR as well as UV \cite{Freedman_et_al, Freedman_et_al1}. The first successful attempt to explain the RG flow (without any fixed point/surface) in the dual background was made by Klebanov and Strassler in \cite{Klebanov+Strassler} by embedding D-branes in a conifold background which was further extended to OKS background in the presence of fundamental quarks and finally, followed by modified Ouyang-Klebanov-Strassler (OKS)-BH background in the presence of black-hole.  Before going into details of (resolved) warped deformed conifold relevant to study of Quark Gluon Plasma, let us first review the modified (OKS)-BH geometry in the presence of the black hole given in \cite{metrics}.

Starting from Klebanov-Strassler model \cite{Klebanov+Strassler}, in this background, $N\ D3$-branes are placed at the tip of six-dimensional conifold whereas  $M\ D5$-branes are wrapped over the vanishing two cycle $S^2$ of conifold base. Introducing $M\ D5$-branes/fractional D3-branes thus produces $SU(N +M)\times SU(N)$ supersymmetric gauge theory such that the matter fields transforming as bi-fundamental representation of UV gauge group $SU(N + M) \times SU(N)$ and under renormalization group flow, eventually cascade down to $SU(M)$ in the far IR region. However, the behavior of gauge theories so-obtained is not quite simple under RG flow. The subtle issues are explained in \cite{Klebanov+Strassler} and elaborated  in more detail in \cite{metrics} also. Due to the presence of complicated gauge theory, there are numerous points where  beta functions $\beta_{g_1}, \beta_{g_2}$ corresponding to two gauge coupling constants $(g_1, g_2)$ are very small, the RG-fixed surfaces involve infinite number of choppy Seiberg dualities and hence RG flow is not smooth as one goes from one surface to another surface at different energy scales and therefore, is not able to provide the dual gravity background.  However, it has been discussed in \cite{metrics}, that inspite of the choppy nature of RG flow at the boundary, there is a smooth RG flow if one hovers towards center of two-dimensional surface which eventually leads to give weakly-coupled gravity description. Therefore, there is a very small regime of smooth RG flow in the gauge theory side that can be captured by weakly coupled supergravity description. This means the KS picture is able to explain the UV completion of the so-formed QCD description. However, the KS description does not include fundamental quarks. To discover the background relevant to study of certain aspects of gauge theory at finite temperature, one has to introduce fundamental quarks at high temperature. This is done by insert co-incident $D7$-branes in Klebanov-Strassler warped deformed construction. There are various subtle issues explained in \cite {metrics} as related to the validity of the same once one goes to explain the full Global scenario. Therefore, $D7$ branes are embedded in large r regime of KS geometry via Ouyang embedding \cite{ouyang} ( that takes into account the effect of the axio-dilaton field on the metric) given as:
\begin{equation}
r^{\frac{3}{2}}e^{\frac{i}{2}(\psi-\phi_1-\phi_2)}\sin\frac{\theta_1}{2} \sin\frac{\theta_2}{2}=\mu
\end{equation}
where $\mu$ is embedding parameter. This discussion so far is valid at $T=0$. The extended background in the presence of non-zero temperature has been discussed in \cite{metrics} by inserting a black-hole to the OKS background which results in both resolution as well as deformation of the two and three cycles of the conifold respectively at $r=0$. In the presence of a black-hole in a warped deformed conifold, the metric is given by \cite{metrics},\cite{finite_mu_Dasguptaetal}:
\begin{equation}
\label{metric}
ds^2 = {1\over \sqrt{h}}
\left(-g_1 dt^2+dx_1^2+dx_2^2+dx_3^2\right)+\sqrt{h}\Big[g_2^{-1}dr^2+r^2 d{\cal M}_5^2\Big].
\end{equation}
 $g_i$'s demonstrate the presence of black hole in modified OKS-BH background and given as follows:
\begin{eqnarray}
& & g_1(r,\theta_1,\theta_2)= 1-\frac{r_h^4}{r^4} + {\cal O}\left(\frac{g^2_sM}{N}\right),
~~~~ g_2(r,\theta_1,\theta_2) = 1-\frac{r_h^4}{r^4} + {\cal O}\left(\frac{g^2_sM}{N}\right).
\end{eqnarray}
where $r_h$ is the horizon, and the ($\theta_1, \theta_2$) dependence come from the
${\cal O}\left(\frac{g_sM^2}{N}\right)$ corrections (We note that the `black hole' factors $g_i$ are
stated to receive ${\cal O}(g_s^2 M N_f)$ corrections in \cite{metrics}, but as shown in \cite{finite_mu_Dasguptaetal},
the six-dimensional warp factors $h_i$ are expected to receive corrections of
${\cal O}\left(\frac{g_sM^2}{N}\right)$ - we assume the same to also be true of the `black hole
functions' $g_{1,2}$)\footnote{This will have the extremely non-trivial consequence that one can use
the same choice of $h_i$ and $g_i$ in the weak($g_s$)coupling-large t'Hooft couplings as well as the
`MQGP' limits, later in the paper.} and
\begin{eqnarray}
d{\cal M}_5^2 = && h_1 (d\psi + {\rm cos}~\theta_1~d\phi_1 + {\rm cos}~\theta_2~d\phi_2)^2 +
h_2 (d\theta_1^2 + {\rm sin}^2 \theta_1 ~d\phi_1^2) +   \nonumber\\
&& + h_4 (h_3 d\theta_2^2 + {\rm sin}^2 \theta_2 ~d\phi_2^2) + h_5~{\rm cos}~\psi \left(d\theta_1 d\theta_2 -
{\rm sin}~\theta_1 {\rm sin}~\theta_2 d\phi_1 d\phi_2\right) + \nonumber\\
&&  + h_5 ~{\rm sin}~\psi \left({\rm sin}~\theta_1~d\theta_2 d\phi_1 +
{\rm sin}~\theta_2~d\theta_1 d\phi_2\right).
\end{eqnarray}
Due to presence of Black-hole, $h_i$ appearing in internal metric as well as $M, N_f$ are not constant and upto linear order depend on $g_s, M, N_f$ as given below:
\begin{eqnarray}
& & h_1 = {1\over 9} + {\cal O}(\frac{g_sM^2}{N}), ~~~~~ h_2 = h_4 = {1\over 6} + {\cal O}(\frac{g_sM^2}{N}), ~~~~~ h_3 = 1 + {\cal O}(\frac{g_sM^2}{N}),\ L=\left(4\pi g_s N\right)^{\frac{1}{4}}\nonumber\\
&& M_{\rm eff} = M + \sum_{m\ge n} a_{mn} (g_sN_f)^m (g_sM)^n, ~~~~~
N_{f}^{\rm eff} = N_f + \sum_{m \ge n} b_{mn} (g_sN_f)^m (g_sM)^n.\nonumber
\end{eqnarray}
The warp factor that includes the back-reaction due to fluxes as well as black-hole is given as:
\begin{eqnarray}
\label{eq:h}
&& h =\frac{L^4}{r^4}\Bigg[1+\frac{3g_sM_{\rm eff}^2}{2\pi N}{\rm log}r\left\{1+\frac{3g_sN^{\rm eff}_f}{2\pi}\left({\rm
log}r+\frac{1}{2}\right)+\frac{g_sN^{\rm eff}_f}{4\pi}{\rm log}\left({\rm sin}\frac{\theta_1}{2}
{\rm sin}\frac{\theta_2}{2}\right)\right\}\Bigg].\nonumber\\
\end{eqnarray}
Though the results are calculated to order ${\cal O}(g_s N_f)$ at small r,  the conjectural solution is given at large r by embedding the model in F-theory set-up so that one is able to avoid the issue of singularities appearing in the background at large r and therefore, explain the holographic renormalisibility of theory. This is connected with the fact that at large r, dilaton, warp factor as well as fluxes should not blow up as r approaches infinity. As realized explicitly in \cite{metrics}, the analytical expression of warp factor given in equation (\ref{eq:h}) does not show up singularity at large r if parameters $N, M, g_s$ are tuned in such a way that following limits are satisfied.
\begin{equation}
\label{limits_Dasguptaetal}
g_s\rightarrow 0, g_sN_f\rightarrow 0,\frac{g_sM^2}{N}\rightarrow 0,g_sM\rightarrow\infty, g_sN\rightarrow\infty.
\end{equation}
Therefore,  to get the results using OKS-BH background, one should strictly satisfy aforementioned constraints on $(g_s, M, N)$.

We utilize the above background to study the thermodynamic stability of solution in a ``(resolved) warped deformed conifold background" obtained by considering limits
\begin{equation}
h_5 \not = 0, h_3 = 1, h_4 - h_2 = a, g_i = 1-\frac{r_h^4}{r^4} + {\cal O}\left(\frac{g_sM^2}{ N}\right).
\end{equation}
 i.e by considering deformation parameter $h_5$ to be very small but not equal to zero. In the calculations of thermodynamical quantities, we choose $a=\frac{1}{r_\Lambda}$.

The three-form fluxes by including black hole factors are given by \cite{metrics}:
\begin{eqnarray}
\label{three-form fluxes}
{\widetilde F}_3 & = & 2M { A_1} \left(1 + {3g_sN_f\over 2\pi}~{\rm log}~r\right) ~e_\psi \wedge
\frac{1}{2}\left({\rm sin}~\theta_1~ d\theta_1 \wedge d\phi_1-{ B_1}~{\rm sin}~\theta_2~ d\theta_2 \wedge
d\phi_2\right)\nonumber\\
&& -{3g_s MN_f\over 4\pi} { A_2}~{dr\over r}\wedge e_\psi \wedge \left({\rm cot}~{\theta_2 \over 2}~{\rm sin}~\theta_2 ~d\phi_2
- { B_2}~ {\rm cot}~{\theta_1 \over 2}~{\rm sin}~\theta_1 ~d\phi_1\right)\nonumber \\
&& -{3g_s MN_f\over 8\pi}{ A_3} ~{\rm sin}~\theta_1 ~{\rm sin}~\theta_2 \left({\rm cot}~{\theta_2 \over 2}~d\theta_1 +
{ B_3}~ {\rm cot}~{\theta_1 \over 2}~d\theta_2\right)\wedge d\phi_1 \wedge d\phi_2\label{brend}, \nonumber\\
H_3 &=&  {6g_s { A_4} M}\Bigr(1+\frac{9g_s N_f}{4\pi}~{\rm log}~r+\frac{g_s N_f}{2\pi}
~{\rm log}~{\rm sin}\frac{\theta_1}{2}~
{\rm sin}\frac{\theta_2}{2}\Bigr)\frac{dr}{r}\nonumber \\
&& \wedge \frac{1}{2}\Bigr({\rm sin}~\theta_1~ d\theta_1 \wedge d\phi_1
- { B_4}~{\rm sin}~\theta_2~ d\theta_2 \wedge d\phi_2\Bigr)
+ \frac{3g^2_s M N_f}{8\pi} { A_5} \Bigr(\frac{dr}{r}\wedge e_\psi -\frac{1}{2}de_\psi \Bigr)\nonumber  \\
&& \hspace*{1.5cm} \wedge \Bigr({\rm cot}~\frac{\theta_2}{2}~d\theta_2
-{ B_5}~{\rm cot}~\frac{\theta_1}{2} ~d\theta_1\Bigr),\nonumber\\
& & {\rm implying}\nonumber\\
 B_2 & = & 6 g_s { A_4} M\left( ln r + \frac{9 g_s N_f}{4\pi} (ln r)^2 + \frac{g_s N_f}{4\pi}(1 + 2 ln r) ln \left(sin\frac{\theta_1}{2}sin\frac{\theta_2}{2}\right)\right)\nonumber\\
& & \times \frac{1}{2}\Bigr({\rm sin}~\theta_1~ d\theta_1 \wedge d\phi_1
- { B_4}~{\rm sin}~\theta_2~ d\theta_2 \wedge d\phi_2\Bigr) \nonumber\\
& &  + \frac{3g_s^2MN_f}{8\pi} { A_5} ln r e_\psi\wedge \left(cot\frac{\theta_2}{2} d\theta_2 - { B_5} cot\frac{\theta_1}{2} d\theta_1\right),\nonumber\\
& & e^{-\Phi} = {1\over g_s} -\frac{N_f}{8\pi} ~{\rm log} \left(r^6 + 9a^2 r^4\right) -
\frac{N_f}{2\pi} {\rm log} \left({\rm sin}~{\theta_1\over 2} ~ {\rm sin}~{\theta_2\over 2}\right).
\end{eqnarray}
where the asymmetry factors ${ A_i}, { B_i}$ encode all the information of the black hole etc in our background. To order ${\cal O}(g_sN_f)$, the same are given by:
\begin{eqnarray}
\label{asymmetry}
&& { A_1} ~=~ 1 + {9g_s N_f \over 4\pi} \cdot {a^2\over r^2}\cdot (2 - 3~{\rm log}~r) + {\cal O}(a^2 g_s^2 N_f^2) \nonumber\\
&& { B_2} ~=~ 1 + {36 a^2~{\rm log}~r \over r^3 + 18 a^2 r ~{\rm log}~r} + {\cal O}(a^2 g_s^2 N_f^2)\nonumber\\
&& { A_2} ~= ~1 + {18 a^2 \over r^2} \cdot {\rm log}~r + {\cal O}(a^2 g_s^2 N_f^2) \nonumber\\
&& { B_1} ~=~ 1 + {81\over 2} \cdot
{g_s N_f a^2 {\rm log}~r \over 4\pi r^2 + 9 g_s N_f a^2 (2 - 3~{\rm log}~r)} + {\cal O}(a^2 g_s^2 N_f^2)\nonumber\\
&& { A_3} ~=~ 1 - {18 a^2 \over r^2}\cdot {\rm log}~r +  {\cal O}(a^2 g_s^2 N_f^2)\nonumber\\
&& { B_3} ~ = ~ 1 + {36 a^2 {\rm log}~r \over r^2 - 18 a^2 {\rm log}~r} + {\cal O}(a^2 g_s^2 N_f^2)\nonumber\\
&& { A_4} ~ = ~ 1 - {3a^2 \over r^2} + {\cal O}(a^2 g_s^2 N_f^2), ~~~~~ { B_4} ~ = ~ 1 + {3g_s a^2 \over r^2 - 3 a^2} + {\cal O}(a^2 g_s^2 N_f^2)\nonumber\\
&& { A_5} ~ = ~ 1 + {36 a^2 {\rm log}~r \over r} +  {\cal O}(a^2 g_s^2 N_f^2), ~~~~
{ B_5} ~ = ~ 1 + {72 a^2 {\rm log}~r \over r + 36 a^2 {\rm log}~r} + {\cal O}(a^2 g_s^2 N_f^2).
\end{eqnarray}
The ${\cal O}(a^2/r^2)$ corrections included in asymmetry factors correspond to modified Ouyang background  in the presence of black hole. The values for the axion $C_0$ and the five form $F_5$ are given by \cite{metrics}:
\begin{eqnarray}
\label{axfive}
&&C_0 ~ = ~ {N_f \over 4\pi} (\psi - \phi_1 - \phi_2),\nonumber\\
&& F_5 ~ = ~ {1\over g_s} \left[ d^4 x \wedge d h^{-1} + \ast(d^4 x \wedge dh^{-1})\right].
\end{eqnarray}
with the dilaton to be taken as approximately a constant near the D7 brane.

Working in a local limit around: $r\approx\langle r\rangle, \theta_{1,2}\approx\langle\theta_{1,2}\rangle,\psi\approx\langle\psi\rangle$ equivalent to replacing the ${\bf CP}^1(\theta_1,\phi_1), {\bf CP}^2(\theta_2,\phi_2)$ locally by $T^2(\theta_1,x), T^2(\theta_2,y)$. Define T-duality coordinates, $(\phi_1,\phi_2,\psi)\rightarrow(x,y,z)$ (\cite{SYZ 3 Ts}):
\begin{equation}
\label{xyz defs}
x = \sqrt{h_2}h^{\frac{1}{4}}sin\langle\theta_1\rangle\langle r\rangle \phi_1,\ y = \sqrt{h_4}h^{\frac{1}{4}}sin\langle\theta_2\rangle\langle r\rangle \phi_2,\ z=\sqrt{h_1}\langle r\rangle h^{\frac{1}{4}}\psi.
\end{equation}
Interestingly, around $\psi=\langle\psi\rangle$, under the coordinate transformation (\cite{SYZ 3 Ts}):
\begin{equation}
\label{transformation_psi}
\left(\begin{array}{c} sin\theta_2 d\phi_2 \\ d\theta_2\end{array} \right)\rightarrow \left(\begin{array}{cc} cos\langle\psi\rangle & sin\langle\psi\rangle \\
- sin\langle\psi\rangle & cos\langle\psi\rangle
\end{array}\right)\left(\begin{array}{c}
sin\theta_2 d\phi_2\\
d\theta_2
\end{array}
\right),
\end{equation}
the $h_5$ term becomes $h_5\left[d\theta_1 d\theta_2 - sin\theta_1 sin\theta_2 d\phi_1d\phi_2\right]$. Further, $e_\psi\rightarrow d(\psi + sin\langle\psi\rangle ln sin\theta_2) + cos\theta_1 d\phi_1 + cos\theta_2 d\phi_2$, which under $\psi\rightarrow\psi - sin\langle\psi\rangle ln sin\theta_2$, implies:
$e_\psi\rightarrow e_\psi$. Locally, thus one introduces an isometry along $\psi$ in addition to the isometries along $\phi_{1,2}$. This clearly is not valid globally - the deformed conifold does not possess a third global isometry. We will be taking $\langle\psi\rangle\sim0,2\pi,4\pi$ so that the
one-forms $sin\theta_2 d\phi_2$ and $ d\theta_2$ do not change appreciably.

To enable use of SYZ-mirror duality via three T dualities, one needs to ensure a large base (implying large complex structures of the aforementioned two two-tori) of the $T^3(x,y,z)$ fibration. This is effected via \cite{large base}:
\begin{eqnarray}
\label{SYZ-large base}
& & d\psi\rightarrow d\psi + f_1(\theta_1)cos\theta_1 d\theta_1 + f_2(\theta_2)d\theta_2,\nonumber\\
& & d\phi_{1,2}\rightarrow d\phi_{1,2} - f_{1,2}(\theta_{1,2})d\theta_{1,2},
\end{eqnarray}
for appropriately chosen large values of $f_{1,2}(\theta_{1,2})$. The three-form fluxes of (\ref{three-form fluxes}) remain invariant under (\ref{SYZ-large base}). The fact that one can choose such large values of $f_{1,2}(\theta_{1,2})$, is justified in {\bf 4}.

\section{Obtaining the Chemical Potential in the Type IIB Background}

The chemical potential in gravity dual is generated via D7-brane gauge fields in a certain geometric background. The temporal component of bulk $U(1)$ field on the D7-brane worldvolume is related to chemical potential  since it is the conjugate field to the electric charge. So, it is defined in a
gauge-invariant manner as follows:
\begin{equation}
\label{DBI}
\mu_C=\int_{r_h}^\infty dr F_{rt}.
\end{equation}
$F_{rt}$ can be evaluated by solving Lagrange equation of motion for DBI Action. Since the background given in \cite{metrics} has been chosen to give non-conformal gauge theories, it might be the case that warp factors $h_i$ appearing in the metric vary while RG flow from IR scale to UV scale. Therefore, as similar to \cite{nonextremel_dasgpta}, the analysis given below is based on the assumption that parameters $h_i$ remain constant at a particular energy scale.

Assuming $\mu(\neq0)\in\mathbb{R}$ in Ouyang's embedding (\cite{real mu Ouyang et al}): \\ $r^{\frac{3}{2}}e^{\frac{i}{2}(\psi-\phi_1-\phi_2)}\sin\frac{\theta_1}{2} \sin\frac{\theta_2}{2}=\mu$, implies that $r^{\frac{3}{2}}\sin\left(\frac{\psi-\phi_1-\phi_2}{2}\right)\sin\frac{\theta_1}{2} \sin\frac{\theta_2}{2}=0$ which could be satisfied for $\psi=\phi_1+\phi_2$ and $r^{\frac{3}{2}}\sin\frac{\theta_1}{2} \sin\frac{\theta_2}{2}=\mu$. Using the same, one obtains the following metric for a space-time-filling wrapped $D7$-brane embedded in the warped deformed conifold:
$$ds^2 = {1\over \sqrt{h\left(r,\theta_2,\theta_1(r,\theta_2)\right)}}
\left(-g_1(r) dt^2+dx^2+dy^2+dz^2\right)+\sqrt{h\left(r,\theta_2,\theta_1(r,\theta_2)\right)}\Big[\frac{dr^2}{g_2(r)}+r^2 d{\cal M}_3^2\Big],$$
where
{\small
\begin{eqnarray}
\label{eq:metric_D7}
& & d{\cal M}_3^2 = {h_1} \left({d\phi_2} (\cos ({\theta_2})+1)+{d\phi_1} \left(2-\frac{2 \mu ^2 \csc
   ^2\left(\frac{{\theta_2}}{2}\right)}{r^3}\right)\right)^2+ \nonumber\\
   & & {h_2} \left(\left(1-\left(1-\frac{2 \mu ^2
   \csc ^2\left(\frac{{\theta_2}}{2}\right)}{r^3}\right)^2\right) {d\phi_1}^2+\frac{\mu ^2 \left(\frac{3
   {dr}}{r}+{d\theta_2} \cot \left(\frac{{\theta_2}}{2}\right)\right)^2}{r^3 \left(\sin
   ^2\left(\frac{{\theta_2}}{2}\right)-\frac{\mu ^2}{r^3}\right)}\right)\nonumber\\
   & & +{h_5} \cos
   ({\phi_1}+{\phi_2}) \left(-\frac{{d\theta_2} \mu  \left(\frac{3 {dr}}{r}+{d\theta_2} \cot
   \left(\frac{{\theta_2}}{2}\right)\right)}{r^{3/2} \sqrt{\sin
   ^2\left(\frac{{\theta_2}}{2}\right)-\frac{\mu ^2}{r^3}}}-{d\phi_1} {d\phi_2} \sqrt{1-\left(1-\frac{2
   \mu ^2 \csc ^2\left(\frac{{\theta_2}}{2}\right)}{r^3}\right)^2} \sin ({\theta_2})\right)\nonumber\\
   & & +{h_5} \sin
   ({\phi_1}+{\phi_2}) \left(-\frac{\mu  \left(\frac{3 {dr}}{r}+{d\theta_2} \cot
   \left(\frac{{\theta_2}}{2}\right)\right) \sin ({\theta_2}) {d\phi_2}}{r^{3/2} \sqrt{\sin
   ^2\left(\frac{{\theta_2}}{2}\right)-\frac{\mu ^2}{r^3}}}+{d\phi_1}{d\phi_2} \sqrt{1-\left(1-\frac{2 \mu ^2 \csc
   ^2\left(\frac{{\theta_2}}{2}\right)}{r^3}\right)^2} \right)\nonumber\\
   & & +{h_4} \left({h_3}
   {d\theta_2}^2+{d\phi_2}^2 \sin ^2({\theta_2})\right).
\end{eqnarray}}
To start with, we will neglect $B_2$ of (\ref{three-form fluxes}) in the DBI action and include a $U(1)$ field strength $F=\partial_r A_t dr\wedge dt$ which would give the following:
\begin{eqnarray}
\label{eq:DBI-i}
& &  \sqrt{det\left(i^*g + F\right)} = \frac{\sqrt{\cal G}}{2 \sqrt{2} \pi ^{3/4}},
\end{eqnarray}
where
{\small
\begin{eqnarray}
\label{eq:DBI-i_i}
& & {\cal G}\equiv \frac{1}{{g_s} N \sqrt{\frac{{g_s} N}{r^4}}}\Biggl[r^4 \Biggl\{2 \sqrt{\pi } \sqrt{\frac{{g_s} N}{r^4}} \left(\frac{\mu ^2 \cot ^2\left(\frac{\theta_2}{2}\right)}{6 r^3 \left(\sin
   ^2\left(\frac{\theta_2}{2}\right)-\frac{\mu ^2}{r^3}\right)}-\frac{{h_5} \mu  \cos (\phi_1+\phi_2) \cot
   \left(\frac{\theta_2}{2}\right)}{r^{3/2} \sqrt{\sin ^2\left(\frac{\theta_2}{2}\right)-\frac{\mu ^2}{r^3}}}+\frac{1}{6}\right)\nonumber\\
   & & \times \Biggl\{2 \sqrt{\pi }
   \sqrt{\frac{{g_s} N}{r^4}} r^2 \left(-\frac{2 \mu ^4 \csc ^4\left(\frac{\theta_2}{2}\right)}{9 r^6}-\frac{2 \mu ^2 \csc
   ^2\left(\frac{\theta_2}{2}\right)}{9 r^3}+\frac{4}{9}\right)\nonumber\\
    & & \times\Biggl(2 \sqrt{\pi } \sqrt{\frac{{g_s} N}{r^4}} r^2
   \left({F_{rt}}^2+\frac{\left(\frac{{r_h}^4}{r^4}-1\right) \left(\frac{3 \mu ^2}{2 r^3 \left(\sin ^2\left(\frac{\theta_2}{2}\right)-\frac{\mu
   ^2}{r^3}\right)}+\frac{2 \sqrt{\pi } \sqrt{\frac{{g_s} N}{r^4}}}{1-\frac{{r_h}^4}{r^4}}\right)}{2 \sqrt{\pi } \sqrt{\frac{{g_s} N}{r^4}}}\right)\nonumber\\
& & \times   \left(\frac{\cos ^2(\theta_2)}{9}+\frac{2 \cos (\theta_2)}{9}+\frac{\sin ^2(\theta_2)}{6}+\frac{1}{9}\right)-\frac{18 {h_5}^2 \mu ^2 \sqrt{\pi
   } \sqrt{\frac{{g_s} N}{r^4}} \left(\frac{{r_h}^4}{r^4}-1\right) \sin ^2(\phi_1+\phi_2) \sin ^2(\theta_2)}{r \left(\sin
   ^2\left(\frac{\theta_2}{2}\right)-\frac{\mu ^2}{r^3}\right)}\Biggr)\nonumber\\
   & & -4 {g_s} N \pi  \left({F_{rt}}^2+\frac{\left(\frac{{r_h}^4}{r^4}-1\right)
   \left(\frac{3 \mu ^2}{2 r^3 \left(\sin ^2\left(\frac{\theta_2}{2}\right)-\frac{\mu ^2}{r^3}\right)}+\frac{2 \sqrt{\pi } \sqrt{\frac{{g_s}
   N}{r^4}}}{1-\frac{{r_h}^4}{r^4}}\right)}{2 \sqrt{\pi } \sqrt{\frac{{g_s} N}{r^4}}}\right)\nonumber\\
   & & \times \biggl[\frac{2}{9} (\cos (\theta_2)+1) \left(2-\frac{2 \mu
   ^2 \csc ^2\left(\frac{\theta_2}{2}\right)}{r^3}\right)+{h_5} \sqrt{1-\left(1-\frac{2 \mu ^2 \csc ^2\left(\frac{\theta_2}{2}\right)}{r^3}\right)^2}
   \sin (\phi_1+\phi_2)\nonumber\\
   & & -{h_5} \cos (\phi_1+\phi_2) \sqrt{1-\left(1-\frac{2 \mu ^2 \csc
   ^2\left(\frac{\theta_2}{2}\right)}{r^3}\right)^2} \sin (\theta_2)\biggr]^2\Biggr\} r^2\nonumber\\
   & & +\frac{1}{\sqrt{\sin
   ^2\left(\frac{\theta_2}{2}\right)-\frac{\mu ^2}{r^3}}}\biggl\{2 {h_5} \mu  \sqrt{\pi } \sqrt{\frac{{g_s} N}{r^4}}
   \cot \left(\frac{\theta_2}{2}\right) \sin (\phi_1+\phi_2) \sin (\theta_2) \times \nonumber\\
    & & \Biggl\{\frac{4 {g_s} {h_5}\mu N}{r^{3/2} \sqrt{\sin ^2\left(\frac{\theta_2}{2}\right)-\frac{\mu
   ^2}{r^3}}}\Biggl\{   \pi  \cot
   \left(\frac{\theta_2}{2}\right) \sin (\phi_1+\phi_2) \left({F_{rt}}^2+\frac{\left(\frac{{r_h}^4}{r^4}-1\right) \left(\frac{3 \mu ^2}{2 r^3
   \left(\sin ^2\left(\frac{\theta_2}{2}\right)-\frac{\mu ^2}{r^3}\right)}+\frac{2 \sqrt{\pi } \sqrt{\frac{{g_s}
   N}{r^4}}}{1-\frac{{r_h}^4}{r^4}}\right)}{2 \sqrt{\pi } \sqrt{\frac{{g_s} N}{r^4}}}\right)\nonumber\\
   & & \times\sin (\theta_2) \biggl(\frac{2}{9} (\cos (\theta_2)+1)
   \left(2-\frac{2 \mu ^2 \csc ^2\left(\frac{\theta_2}{2}\right)}{r^3}\right)+{h_5} \sqrt{1-\left(1-\frac{2 \mu ^2 \csc
   ^2\left(\frac{\theta_2}{2}\right)}{r^3}\right)^2} \sin (\phi_1+\phi_2)\nonumber\\
   & & -{h_5} \cos (\phi_1+\phi_2) \sqrt{1-\left(1-\frac{2 \mu ^2
   \csc ^2\left(\frac{\theta_2}{2}\right)}{r^3}\right)^2} \sin (\theta_2)\biggr)\Biggr\}\nonumber\\
   & & -\frac{1}{r^{3/2} \sqrt{\sin ^2\left(\frac{\theta_2}{2}\right)-\frac{\mu ^2}{r^3}}}\Biggl\{4 {g_s} {h_5} \mu  N \pi  \cot \left(\frac{\theta_2}{2}\right) \left(-\frac{2 \mu ^4 \csc ^4\left(\frac{\theta_2}{2}\right)}{9
   r^6}-\frac{2 \mu ^2 \csc ^2\left(\frac{\theta_2}{2}\right)}{9 r^3}+\frac{4}{9}\right)\nonumber\\
    & & \times\sin (\phi_1+\phi_2)
   \left({F_{rt}}^2+\frac{\left(\frac{{r_h}^4}{r^4}-1\right) \left(\frac{3 \mu ^2}{2 r^3 \left(\sin ^2\left(\frac{\theta_2}{2}\right)-\frac{\mu
   ^2}{r^3}\right)}+\frac{2 \sqrt{\pi } \sqrt{\frac{{g_s} N}{r^4}}}{1-\frac{{r_h}^4}{r^4}}\right)}{2 \sqrt{\pi } \sqrt{\frac{{g_s} N}{r^4}}}\right) \sin
   (\theta_2)\Biggr\}\Biggr\} \sqrt{r}\biggr\}\Biggr\}\Biggr].
\end{eqnarray}}
One sees that:
\begin{equation}
\label{mu0}
\int_0^{2\pi}\int_0^{2\pi}d\phi_1d\phi_2\int_0^\pi d\theta_2\left(\sqrt{det(i^*(g) + F)}\right)^{\mu^0}\sim r^3\sqrt{1 - F_{rt}^2},
\end{equation}
and
{\small
\begin{eqnarray}
\label{mu1}
& & \left(\sqrt{det(i^*(g) + F)}\right)^{\mu}\nonumber\\
& & = \frac{1}{3 \sqrt{2}
   \sqrt{\left(1-F_{rt}^2\right) r^6 \cos ^2\left(\frac{\theta_2}{2}\right) (3 \cos (\theta_2)+1)}}\Biggl[\left(F_{rt}^2-1\right) h_5 r^{9/2} \Biggl(\sqrt{2} \cot ^3\left(\frac{\theta_2}{2}\right) \cos (\phi_1+\phi_2)\nonumber\\
   & & \times \Biggl(2 r^{3/2} \cos
   (2 \theta_2) \sqrt{-\frac{1}{r^3 (\cos (\theta_2)-1)}}+6 r^{3/2} \sqrt{-\frac{1}{r^3 (\cos (\theta_2)-1)}}\nonumber\\
   & & +\cos (\theta_2) \left(3
   \sqrt{1-\cos (\theta_2)}-8 r^{3/2} \sqrt{-\frac{1}{r^3 (\cos (\theta_2)-1)}}\right) +\sqrt{1-\cos (\theta_2)}\Biggr)\nonumber\\
   & & -8 r^{3/2} \cos
   ^2\left(\frac{\theta_2}{2}\right) \sin (\phi_1+\phi_2) \sqrt{\frac{\csc ^2\left(\frac{\theta_2}{2}\right)}{r^3}}\Biggr)\Biggr];\nonumber\\
& &  \int_0^{2\pi}\int_0^{2\pi}d\phi_1d\phi_2 \left(\sqrt{det(i^*(g) + F)}\right)^{\mu}=0.
\end{eqnarray}}

 The finite part of the DBI action for $D7$ for the Ouyang embedding is hence:
\begin{equation}
\label{DBI_finite}
\int dr e^{-\phi(r)} r^3\sqrt{1 - F_{rt}^2},
\end{equation}
where $e^{-\phi(r)}=\frac{1}{g_s} - \frac{N_f}{8\pi}ln\left(r^6 + a^2 r^4\right) - \frac{Nf}{2\pi} ln\left(\mu r^{-\frac{3}{2}}\right)\stackrel{r>>a}{\longrightarrow}\frac{1}{g_s} - \frac{N_f}{2\pi}ln\mu$. We will be assuming that the embedding parameter is (real and) less than unity; if  $\mu\sim0$ then one assumes that $\mu\sim\epsilon^\alpha,\alpha>0$ so that using (\ref{limits_Dasguptaetal}) $g_s N_f ln\mu\rightarrow0$. The Euler-Lagrange eom corresponding to (\ref{DBI_finite}) is:
\begin{equation}
\label{DBI-i}
\partial_r\left(\frac{\left[\frac{1}{g_s} - \frac{N_f}{2\pi}ln\mu\right]r^3\partial_rA_t}{\sqrt{1- \left(\partial_rA_t\right)^2}}\right)=0.
\end{equation}
We therefore obtain:
\begin{equation}
\label{DBI-ii}
\partial_r A_t(r)=\frac{C e^{\phi(r)}}{\sqrt{C^2 e^{2\phi(r)} + r^6}},
\end{equation}
for large $r$ implying
\begin{equation}
\label{DBI-iii}
\mu_C=\int_{r_h}^\infty \frac{C e^{\phi(r)}}{\sqrt{C^2 e^{2\phi(r)} + r^6}}=\frac{C g_s\pi}{2\pi - g_s N_f ln\mu}\frac{\ _2F_1\left(\frac{1}{3},\frac{1}{2};\frac{4}{3};-\frac{C^2}{\left(\frac{1}{g_s} - \frac{N_f}{2\pi}ln\mu\right)^2r_h^6}\right)}{r_h^2},
\end{equation}
where $r_h=\pi \sqrt{4\pi g_s N} T$. From (\ref{DBI-iii}) one sees that:
\begin{equation}
\label{dmuoverdT-i}
\frac{\partial\mu_C}{\partial T}\Bigg|_{N_f; C=\left(\frac{1}{g_s} - \frac{N_f}{2\pi} ln\mu\right)\pi^3\left(4\pi g_s N\right)^{\frac{3}{2}}} \sim - \frac{1}{\sqrt{1 + T^6}}<0.
\end{equation}

In order to study the thermodynamical stability of the type IIB solution, we need to consider:
\begin{equation}
\label{eq:e_-phi_improved}
 e^{-\phi(r)}\approx \frac{1}{g_s} - \frac{N_f ln\mu}{2\pi} - \frac{N_f a^2}{8\pi r^2},
\end{equation}
along the Ouyang embedding. This implies that:
\begin{eqnarray}
\label{eq:mu_C_improved}
& & \mu_C = C \int_{r_h}^\infty dr\frac{1}{\left(\frac{1}{g_s} - \frac{N_f ln\mu}{2\pi} - \frac{N_f a^2}{8\pi r^2}\right)\sqrt{r^6 + \frac{C^2}{\left(\frac{1}{g_s} - \frac{N_f ln\mu}{2\pi} - \frac{N_f a^2}{8\pi r^2}\right)^2}}}\nonumber\\
& & \approx C\int_{r_h}^\infty dr\frac{g_s}{\sqrt{C^2g_s^2 + r^6}}
+ g_s N_f\int_{r_h}^\infty dr\frac {r^4\left(a^2 C g_s + 4 C g_s r^2 ln\mu\right)}{8\pi\left(C^2 g_s^2 + r^6\right)^{\frac{3}{2}}}\nonumber\\
& &  = C g_s\left[\frac{\, _2F_1\left(\frac{1}{3},\frac{1}{2};\frac{4}{3};-\frac{C^2 g_s^2}{{r_h}^6}\right)}{2 {r_h}^2}\right] + \nonumber\\
 && g_s N_f\left[ \frac{C g_s \left(\, _2F_1\left(\frac{2}{3},\frac{3}{2};\frac{5}{3};-\frac{C^2 g_s^2}{{r_h}^6}\right) a^2+8
   {r_h}^2 \, _2F_1\left(\frac{1}{3},\frac{3}{2};\frac{4}{3};-\frac{C^2 g_s^2}{r_h^6}\right) ln (\mu
   )\right)}{32 \pi  r_h^4}\right].\nonumber\\
   & &
\end{eqnarray}
From (\ref{eq:mu_C_improved}), one sees:
\begin{equation}
\label{dmuoverdT-ii}
\frac{\partial\mu_C}{\partial T}\Biggr|_{N_f,a = f r_h:f<<<1; C=\frac{\pi^3\left(4\pi g_s N\right)^{\frac{3}{2}}}{g_s}}\sim - \frac{1}{8\pi}\frac{T^6(f^2 + 4 ln\mu)}{\left(1 + T^6\right)^{\frac{3}{2}}}
+ f^2 g_s N_f\frac{\, _2F_1\left(\frac{2}{3},\frac{3}{2};\frac{5}{3}; - \frac{1}{T^6}\right)}{16\pi T^3}<0.
\end{equation}
So, from (\ref{dmuoverdT-i}) and (\ref{dmuoverdT-ii}) it is clear that  $\frac{\partial\mu_C}{\partial T}\Biggr|_{N_f} = - \frac{\partial S}{\partial N_f}\Biggr|_T < 0$.
Apart from $C_v>0$, thermodynamic stability requires: $\frac{\partial\mu_C}{\partial N_f}\left.\right|_{T}>0$, which using (\ref{eq:mu_C_improved}) for $C>0, \mu = \lim_{\epsilon\rightarrow0^+}1 - \epsilon$ is satisfied!

Let us now analyse the inclusion of the  NS-NS  $B_2$ of (\ref{three-form fluxes}) in the DBI action. From (\ref{three-form fluxes}), using the Ouyang embedding (implying for a real $\mu$: $d\psi=d\phi_1 + d\phi_2, d\theta_1 = - \tan\left(\frac{\theta_1}{2}\right)\left(3 \frac{dr}{r} + \cot\left(\frac{\theta_2}{2}\right)d\theta_2\right)$):
\begin{eqnarray}
\label{B_Ouyang}
& & B_2 = - \frac{3}{r}\tan\frac{\theta_1}{2}\left(B_{\theta_1\phi_1} + B_{\theta_1\psi}\right)dr\wedge d\phi_1 +
 \left[B_{\theta_2\phi_1}- \tan\frac{\theta_1}{2}\cot\frac{\theta_2}{2}\left(B_{\theta_1\phi_1} + B_{\theta_1\psi}\right)\right]d\theta_2\wedge d\phi_1
 \nonumber\\
 & & - \frac{3}{r}\tan\frac{\theta_1}{2}\left(B_{\theta_1\phi_2} + B_{\theta_1\psi}\right)dr\wedge d\phi_2
+\left[ B_{\theta_2\phi_2} - \tan\frac{\theta_1}{2}\cot\frac{\theta_2}{2}\left(B_{\theta_1\phi_2} + B_{\theta_1\psi}\right)\right]d\theta_2\wedge d\phi_2.
\end{eqnarray}
Working in the $g_sM\rightarrow \infty, g_sM\rightarrow \infty, g_s\rightarrow0, \frac{g_sM^2}{N}\rightarrow0$ limit of \cite{metrics}, one obtains:
{\small
\begin{eqnarray}
\label{DBI_B-i}
& & \left(\sqrt{det\left(i^*(g + B) + F\right)}\right)^{\mu^0}=\nonumber\\
& &\hskip-0.5in \frac{\sqrt{\frac{\left({F_{rt}}^2-1\right) r^2 \left(\sin ^2(\theta_2) \left(81(g_s M)^2 \left(r^2-3 a^2\right)^2 ln ^2(r) \left(-9 a^4 {g_s}+3 a^2
   {g_s} r^2+1\right)^2+\pi  r^4 {g_s N}\right)-2 \pi  r^4 {g_s N} \cos ^2(\theta_2)-4 \pi  r^4 {g_s N} \cos (\theta_2)-2 \pi  r^4
   {g_s N}\right)}{{g_s N}}}}{9 \sqrt{\pi }};\nonumber
\end{eqnarray}}
and
{\small
\begin{eqnarray}
\label{DBI_B-i}
\int_0^{2\pi}\int_0^{2\pi}d\phi_1d\phi_2\int_0^\pi d\theta_2\left(\sqrt{det\left(i^*(g + B) + F\right)}\right)^{\mu^0}
    \sim r^3\sqrt{1 - F_{rt}^2} + {\cal O}\left(\sqrt{\frac{g_sM^2}{N}}\right);
     \end{eqnarray}
   \begin{eqnarray}
   & &  \left(\sqrt{det\left(i^*(g + B) + F\right)}\right)^{\mu}=\nonumber\\
    & & \hskip -0.4in \frac{1}{12 \sqrt{\frac{{g_sN}}{r^4}}
   \sqrt{\frac{\left(F_{rt}^2-1\right) r^2 \left(\sin ^2(\theta_2) \left(81 {g_sM}^2 \left(r^2-3 a^2\right)^2 ln ^2(r) \left(-9 a^4 {g_s}+3 a^2
   {g_s} r^2+1\right)^2+\pi  r^4 {g_sN}\right)-2 \pi  r^4 {g_sN} \cos ^2(\theta_2)-4 \pi  r^4 {g_sN} \cos (\theta_2)-2 \pi  r^4
   {g_sN}\right)}{{g_sN}}}}\nonumber\\
   & & \times\Biggl\{\left(F_{rt}^2-1\right) {h_5} \sqrt{r} \Biggl(\frac{\sqrt{\pi } {g_sN} \sqrt{\sin ^2\left(\frac{\theta_2}{2}\right)} \left(10 \cos
   \left(\frac{\theta_2}{2}\right)+3 \left(\cos \left(\frac{3 \theta_2}{2}\right)+\cos \left(\frac{5 \theta_2}{2}\right)\right)\right) \csc
   ^3\left(\frac{\theta_2}{2}\right) \cos (\phi_1+\phi_2)}{\sqrt{\frac{{g_sN}}{r^4}}} \nonumber\\
   & &   -2 \Biggl[2 \sqrt{2} \cos \left(\frac{\theta_2}{2}\right)
   \cot \left(\frac{\theta_2}{2}\right) \sin (\phi_1+\phi_2) \Biggl(9 {g_sM} \left(3 a^2-r^2\right) ln (r) \sqrt{1-\cos (\theta_2)}\cos
   \left(\frac{3 \theta_2}{2}\right)\nonumber\\
   & & \left(9 a^4 {g_s}-3 a^2 {g_s} r^2-1\right) +27 {g_sM} \left(3 a^2-r^2\right) ln (r) \cos
   \left(\frac{\theta_2}{2}\right) \sqrt{1-\cos (\theta_2)} \left(9 a^4 {g_s}-3 a^2 {g_s} r^2-1\right) \nonumber\\
   & &   +8 \sqrt{\pi } r^{11/2}
   \sqrt{\frac{{g_sN}}{r^4}} \sin \left(\frac{\theta_2}{2}\right) \sqrt{-\frac{1}{r^3 (\cos (\theta_2)-1)}}\Biggr)-\nonumber\\
   & &  \frac{\sqrt{\pi } {g_sN} \sin (2
   (\phi_1+\phi_2)) \csc (\phi_1+\phi_2) \left(2 r^{9/2} \sin ^3(\theta_2) \left(\frac{\csc
   ^2\left(\frac{\theta_2}{2}\right)}{r^3}\right)^{3/2} +\sqrt{\sin ^2\left(\frac{\theta_2}{2}\right)} \sin (2 \theta_2) \csc
   ^4\left(\frac{\theta_2}{2}\right)\right)}{\sqrt{\frac{{g_sN}}{r^4}}}\Biggr]\Biggr)\Biggr\};\nonumber\\
   \end{eqnarray}}
   and
   \begin{eqnarray}
   & & \int_0^{2\pi}\int_0^{2\pi}d\phi_1d\phi_2\left(\sqrt{det\left(i^*(g + B) + F\right)}\right)^{\mu}=0.\nonumber\\
\end{eqnarray}
Hence,
\begin{equation}
\label{eq:DBI_B-iii}
\int_0^{\pi}d\theta_2\int_0^{2\pi}\int_0^{2\pi}d\phi_1d\phi_2\left(\sqrt{det\left(i^*(g + B) + F\right)}\right)\sim \left(r^3\sqrt{1 - F_{rt}^2}+ {\cal O}\left(\sqrt{\frac{g_sM^2}{N}}\right)\right) + {\cal O}(\mu^2).
\end{equation}
such that in the $g_sM\rightarrow\infty,g_sN\rightarrow\infty$ limit, the ${\cal O}(\mu^0)$ results
and consequences thereafter, for $B=0$ and $B\neq0$, match.

 For {\bf 5.1} and {\bf 5.2} one would need to have a very small $\mu_C$, which we see from
(\ref{DBI-iii}) and (\ref{eq:mu_C_improved}), can be arranged by tuning the $C$ to an appropriately small
value, e.g., for $C=0.001$, $\mu_C=0.05, 0.0006$ respectively for the two scalings of $g_s, M, N$
referred to in the abstract and use in and after Sec. {\bf 4}. The choice of $C$ in (\ref{dmuoverdT-i})
and (\ref{dmuoverdT-ii}) was only for convenience.

\section{The Local type IIA SYZ Dual and Local M-Theory Uplift}

The starting metric in the type IIB theory has the following components
\begin{eqnarray}
&& ds^2 =
g_{\mu \nu}dx^\mu ~dx^\nu + g_{x\mu} dx ~dx^\mu +  g_{y \mu} dy~
dx^\mu +  g_{z\mu} dz ~ dx^\mu +  g_{xy} dx ~dy
  + g_{xz} dx ~dz +  g_{zy} dz ~ dy +  g_{xx} dx^2 \nonumber\\
 &&  +  g_{yy}dy^2
 +  g_{zz}~dz^2,
  \end{eqnarray}
where $\mu, \nu \neq x, y, z$.
As shown in \cite{SYZ 3 Ts}, the form of the metric of the mirror manifold after performing three T-dualities, first along $x$, then along $y$ and finally along $z$:
\begin{eqnarray}
\label{mirror_metric}
& & ds^2 =
\left( G_{\mu\nu} - {G_{z\mu}G_{z\nu} - {\cal B}_{z\mu} {\cal
B}_{z\nu} \over G_{zz}} \right) dx^\mu~dx^\nu +2 \left( G_{x\nu} -
{G_{zx}G_{z\nu} - {\cal B}_{zx} {\cal B}_{z\nu}
 \over G_{zz}} \right) dx~dx^\mu  \nonumber\\
& &  + 2\left( G_{y\nu} - {G_{zy}G_{z\nu} - {\cal B}_{zy} {\cal B}_{z\nu}
 \over G_{zz}}\right) dy~dx^\nu +
2\left( G_{xy} - {G_{zx}G_{zy} - {\cal B}_{zx} {\cal B}_{zy} \over
G_{zz}}\right) dx~dy  \nonumber\\
& &  + {dz^2\over G_{zz}} + 2{{\cal
B}_{\mu z} \over G_{zz}} dx^\mu~dz + 2{{\cal B}_{xz} \over G_{zz}}
dx~dz + 2{{\cal B}_{yz} \over G_{zz}} dy~dz \nonumber\\
& & +  \left( G_{xx}
- {G^2_{zx} - {\cal B}^2_{zx} \over G_{zz}} \right) dx^2 + \left(
G_{yy} - {G^2_{zy} - {\cal B}^2_{zy} \over G_{zz}} \right)
dy^2.
\end{eqnarray}

To implement mirror symmetry a al SYZ prescription, one needs to ensure that the base of the local $T^3$-fibration is large. Near $\theta_1=\theta_2=0$ we will show that it is possible to obtain a large base for which $f_{1,2}(\theta_{1,2})>>a$ (small resolution factor, i.e. $a<<1$). The guiding principle is that one requires that the metric obtained after SYZ-mirror transformation applied to the resolved warped deformed conifold is like a warped resolved conifold at least locally, then $G^{IIA}_{\theta_1\theta_2}$ needs to vanish. We will {\bf first} implement $\frac{g_s^2M}{N_f}\rightarrow0$
(common to both limits (\ref{limits_Dasguptaetal-i}) and (\ref{limits_Dasguptaetal-ii})) for simplifying the type IIA metric components which means that we will set $h(r,\theta_1,\theta_2)=\frac{4\pi g_s N}{r^4}$; this yields the type IIA metric components enumerated in (\ref{metric-mirror}). Interestingly, if one assumes that the local uplift is also
valid globally, then we show that the same guiding principle is also applicable at  $\theta_1=\theta_2=\frac{\pi}{2}$, in fact more readily.
Assuming one has found appropriate $f_i(\theta_i)$, after T-dualizing along $x, y, z$, equation (\ref{mirror_metric}) yields (\ref{G_munu}) - (\ref{Gzx}), which yields the components of the type IIA mirror metric as given in (\ref{metric-mirror}).

Along $r=a\sqrt{3},\theta_1\rightarrow0,\theta_2\rightarrow m\theta;\theta\rightarrow0$ where $m\sim{\cal O}(1)$, (\ref{metric-mirror}) yields:
{\small
\begin{eqnarray}
\label{metric-mirror-iii}
& & G^{IIA}_{\theta_1\theta_1},G^{IIA}_{\theta_2\theta_2}\sim\sqrt{g_sN},\nonumber\\
& & G^{IIA}_{\theta_1\theta_2}\sim \lim_{g_sN\to\infty,\theta\to0}h_5\sqrt{g_sN}\theta^2\left[f_1(0)f_2(0)
- \frac{2 }{m\theta^2}\right],\nonumber\\
& & {\rm which\ vanishes\ for}\ f_1(0)\sim f_2(0)\sim\lim_{\theta\to0}\frac{\sqrt{2}}{\sqrt{m}\theta}>>a\ {\rm for\ small\ unconstrained}\ h_5;\nonumber\\
& & G^{IIA}_{\phi_1\theta_1}\sim\lim_{\theta\to0}\frac{g_s^2MN_f  ln a}{h_5^2(g_sN)^{\frac{1}{4}}\theta^4}\sim0,\nonumber\\
& & G^{IIA}_{\phi_1\theta_2}=0,\nonumber\\
& & G^{IIA}_{\psi\theta_1}\lim_{\theta\to0}\frac{g_s^2MN_f  ln a}{h_5^2(g_sN)^{\frac{1}{4}}\theta^3}\sim0,\nonumber\\
& & G^{IIA}_{\psi\theta_2}=0,\nonumber\\
& & G^{IIA}_{\phi_2\theta_1}\sim\lim_{\theta\to0}\frac{g_s^2MN_f  ln a}{(g_sN)^{\frac{1}{4}}\theta^2}\sim0,\nonumber\\
& & G^{IIA}_{\phi_2\theta_2}\sim(g_sN)^{\frac{1}{4}}\theta f_2(0)\sim(g_sN)^{\frac{1}{4}}>>1;\nonumber\\
& & G^{IIA}_{\phi_1\phi_1}\sim G^{IIA}_{\phi_2\phi_2}\sim\theta^2;\nonumber\\
& & G^{IIA}_{\psi\psi}\sim{\cal O}(1);\nonumber\\
& & G^{IIA}_{\phi_1\psi}\sim\frac{-243 \sqrt{6} h_5^2 m^2 \theta ^5+162 h_5^2 m^2 \theta ^4+18 \sqrt{6} m (9 h_5+m) \theta ^3-72 h_5
   m \theta ^2-12 \sqrt{6} \theta +8}{1458 h_5^2 m^2 \theta ^3},\nonumber\\
& & G^{IIA}_{\phi_2\psi}\sim{\cal O}(1),\nonumber\\
& & G^{IIA}_{\phi_1\phi_2}\sim\frac{1}{27 \left(9 h_5^2-1\right) m^2 \theta ^2 \left(3
   \theta ^2+2\right) \left(-27 h_5^2 \theta ^2 m^2+\left(3 \theta ^2+2\right) m^2+12 h_5 m+2\right)}\nonumber\\
 & &\times\left[-243 h_5^3 m^3 \left(-27 \theta ^3+3 \sqrt{6} \theta ^2-36 \theta +2 \sqrt{6}\right) \theta ^4-27 h_5^2 m^2
   \left(9 \sqrt{6} \theta ^4+54 \theta ^3-6 \sqrt{6} \theta ^2+144 \theta -8 \sqrt{6}\right)\theta ^2\right.\nonumber\\
   & & \hskip-0.4in\left. -12 \sqrt{6} \theta ^2+72
   \theta -3 h_5 m \left(243 m^2 \theta ^7+162 \left(2 m^2+1\right) \theta ^5-36 \sqrt{6} \theta ^4+108 \left(m^2+3\right)
   \theta ^3-12 \sqrt{6} \theta ^2-72 \theta +8 \sqrt{6}\right)\right.\nonumber\\
   & & \left.-8 \sqrt{6}\right].
\end{eqnarray}}
If $\theta\rightarrow0,h_5\rightarrow0, m\sim{\cal O}(1)$ such that $-72 h_5
   m \theta ^2-12 \sqrt{6} \theta +8=0$, then $\theta\rightarrow\frac{-\sqrt{6} + \sqrt{2}\sqrt{3 + 8 h_5 m}}{12 h_5 m}=\frac{1}{3}\sqrt{\frac{2}{3}} - \frac{2}{9}\sqrt{\frac{2}{3}} h_5 m + {\cal O}
   \left((h_5 m)^2\right)$. So, if $h_5m<<1$ then $\theta\sim0.3<1$. The understanding now is that $\theta\rightarrow0$ by dividing out the same by a large number. So, the numerator of $G^{IIA}_{\phi_1\psi}$ will go as ${\cal O}(\theta^3)$ implying that $G^{IIA}_{\phi_1\psi}$ will be finite as $\theta_{1,2}\rightarrow0$. The terms up to ${\cal O}(\theta=\frac{-\sqrt{6} + \sqrt{2}\sqrt{3 + 8 h_5 m}}{12 h_5 m})$ in the numerator of $G^{IIA}_{\phi_1\phi_2}$ will be given by $16\sqrt{\frac{2}{3}} h_5 m
   + {\cal O}\left((h_5 m)^2\right)$, which for $h_5 m<<1$ is negligible. So, $G^{IIA}_{\phi_1\phi_2}$ is finite as $\theta_1\rightarrow0,\theta_2\rightarrow0$.

Near $r=a \sqrt{3}, \theta_1=\theta_2=\frac{\pi}{2}$, using (\ref{metric-mirror}):
{\small
\begin{eqnarray}
\label{metric-mirror-ii}
& & G^{IIA}_{\theta_1\theta_1},G^{IIA}_{\theta_2\theta_2}\sim\sqrt{g_sN},\nonumber\\
& & G^{IIA}_{\theta_1\theta_2}\sim h_5\sqrt{g_sN}\left(f_1\left(\frac{\pi}{2}\right)f_2\left(\frac{\pi}{2}\right) - 2\right)=0\ {\rm for}\ f_1\left(\frac{\pi}{2}\right)=f_2\left(\frac{\pi}{2}\right)=\sqrt{2}>>a;\nonumber\\
& & G^{IIA}_{\phi_1\theta_1}\sim\frac{h_5^2g_s^2MN_f}{(g_sN)^{\frac{1}{4}}}\sim0,\nonumber\\
& & G^{IIA}_{\phi_1\theta_2}=0,\nonumber\\
& & G^{IIA}_{\psi\theta_1}\sim\frac{g_s^2MN_f}{(g_sN)^{\frac{1}{4}}}\sim0,\nonumber\\
& & G^{IIA}_{\psi\theta_2}=0,\nonumber\\
& & G^{IIA}_{\phi_2\theta_1}=0,\nonumber\\
& & G^{IIA}_{\phi_2\theta_2}=0;\nonumber\\
& & G^{IIA}_{\phi_1\phi_1/\phi_2\phi_2}=\frac{1}{1 - 9 h_5^2},\nonumber\\
& & G^{IIA}_{\psi\psi}=\frac{1}{9},\nonumber\\
& & G^{IIA}_{\phi_1\psi}\sim h_5^2,\nonumber\\
& & G^{IIA}_{\phi_2\psi}=0,\nonumber\\
& & G^{IIA}_{\phi_1\phi_2}\sim h_5\sim0.
\end{eqnarray}
}
We see from (\ref{metric-mirror-ii}) that the metric locally will look like a resolved conifold metric along $\theta_1=\theta_2=\frac{\pi}{2}$ and $r=a\sqrt{3}$.

We will now discuss how one obtains, locally, a one-form type IIA potential from the triple T-dual (along $x, y, z$) of the type IIB $F_{1,3,5}$. From (\ref{three-form fluxes}), one sees that the following are the non-zero components with respect to the T-duality coordinates $(x,y,z)$ of $F_3$:
\begin{eqnarray}
& & F_{z\theta_i x}, F_{z\theta_i y}, F_{x\theta_i y}, F_{rzy}, F_{rzx}, F_{rxy}, F_{\theta_i xy}.
\end{eqnarray}

Using the T-duality rules for RR field strengths of \cite{T-dual Hassan}, one sees that:
\begin{eqnarray}
\label{Ftildetildetildeythetai}
& & \tilde{\tilde{\tilde{F}}}_{y\theta_i/r}=\Biggl[- 3 F_{z\theta_i/rx} - 6\biggl( \biggl\{F_z - 2 g^{-1}_{xx}\left(g_{xz}F_z - g_{xx}F_z\right)\biggr\}b_{x\theta_i/r}\biggr)\nonumber\\
& & - 2 g^{-1}_{yy}\Biggl\{\tilde{g}_{yz}\biggl[F_{y\theta_i/rx} - 2 \biggl[F_y - 2 g^{-1}_{xx}\left(g_{xy}F_x - g_{xx}F_y\right)\biggr]b_{x\theta_i/r}\biggr]\Biggr\}\nonumber\\
& & -2\Biggl(- \tilde{\tilde{B}}_{zy}\Biggl[F_{y\theta_i/rx} + 2\left(F_y - 2 g^{-1}_{xx}\left(g_{xy} F_x - g_{xx}F_y\right)\right)b_{x\theta_i/r}\Biggr]\Biggr)\Biggr],
\end{eqnarray}
where
\begin{eqnarray}
\label{eq:Btildetildezy}
& & \tilde{\tilde{B}}_{zy}=\tilde{g}^{-1}_{yy}\tilde{g}_{yz}=\left(g_{yy} - g^{-1}_{xx}g^2_{xy}\right)^{-1}
\left(g_{yz} - g^{-1}_{xx}g_{xy}g_{xz}\right).
\end{eqnarray}
The components of (\ref{Ftildetildetildeythetai}) are explicitly given in (\ref{Ftildetildetildeytheta1}) - (\ref{Ftildetildetildezr}).

Now,
\begin{eqnarray}
\label{Ftildetildetildezthetair}
& & \tilde{\tilde{\tilde{F}}}_{z\theta_i/r} = -\Biggl[F_{y\theta_i/rx} + 2\left(F_y - 2g^{-1}_{xx}\left(g_{xy}F_x - g_{xx}F_y\right)b_{x\theta_i/r}\right)\Biggr],
\end{eqnarray}
from where we obtain (\ref{Ftildetildetildeztheta1}). Similarly,
\begin{eqnarray}
\label{eq:Ftildetildetildexthetair}
& & \tilde{\tilde{\tilde{F}}}_{x\theta_i/r}=-\left(F_{yz\theta_i/r} - 3 g^{-1}_{xx}\left\{g_{xy}F_{xz\theta_i/r} + g_{xx}F_{zy\theta_i/r} + g_{xz}F_{yx\theta_i/r}\right\}\right)\nonumber\\
& & \hskip-0.4in- 3 \Biggl[-\tilde{B}_{yx}\Biggl\{-F_{z\theta_i/rx} - 2\left(F_z - 2g^{-1}_{xx}\left(g_{xz}F_x - g_{xx}F_z\right)\right)b_{x\theta_i/r} - 2\tilde{g}_{yy}^{-1}\Biggl(\tilde{g}_{yz}\left\{F_{xy\theta_i/r} + 2 b_{x\theta_i/r}\left[F_y - g^{-1}_{xx} g_{yy}F_x\right]\right\}\nonumber\\
& & + \tilde{g}_{yy}\left\{F_{z\theta_i/rx} - 2 \left[F_z - 2 g^{-1}_{xx}\left(g_{xz}F_x - g_{xx}F_z\right)\right]b_{x\theta_i/r}\right\} \Biggr)\Biggr\}\Biggr]
\end{eqnarray}
which hence yields (\ref{Ftildetildetildextheta1}) - (\ref{Ftildetildetildexr}).

We therefore can construct the following gauge field one-form in the local limit:
\begin{eqnarray}
\label{A}
& & A^{F_3} = \Biggl[\tilde{\tilde{\tilde{F}}}_{xr}x dr + \tilde{\tilde{\tilde{F}}}_{x\theta_1}x d\theta_1 + \tilde{\tilde{\tilde{F}}}_{x\theta_2}x d\theta_2 + \tilde{\tilde{\tilde{F}}}_{y\theta_1}y d\theta_1 + \tilde{\tilde{\tilde{F}}}_{y\theta_2}y d\theta_2\nonumber\\
 & &  +  \tilde{\tilde{\tilde{F}}}_{z\theta_2}z d\theta_2 +  \tilde{\tilde{\tilde{F}}}_{z\theta_1}z d\theta_1 +   \tilde{\tilde{\tilde{F}}}_{zr}z dr
 + \tilde{\tilde{\tilde{F}}}_{yr}y dr\Biggr]\left(\theta_{1,2}\rightarrow\langle\theta_{1,2}\rangle,\phi_{1,2}\rightarrow\langle \phi_{1,2}\rangle,\psi\rightarrow\right\langle\psi\rangle,r\rightarrow\langle r\rangle).
\end{eqnarray}

The two-form field strength obtained from three T-dualities to $F_1$ can be obtained via application of T-duality rules of \cite{T-dual Hassan}:
\begin{eqnarray}
\label{eq:Fitldetildetildebeta1beta2}
& & \tilde{\tilde{\tilde{F}}}_{zx} = - 2 \Biggl[F_y - g^{-1}_{xx}\left\{g_{xy} F_x - g_{xx}F_y\right\}\Biggr];\nonumber\\
& & \tilde{\tilde{\tilde{F}}}_{yx} = - 3 \left[F_z - g^{-1}_{xx}\left(g_{xz}F_x - g_{xx}F_z\right)\right]
-2\tilde{g}^{-1}_{yy}\left[- \tilde{g}_{yz}\left\{F_y - g^{-1}_{xx}\left(g_{xy}F_x - g_{xx}F_y\right)\right\}\right]\nonumber\\
& & - 4 \tilde{\tilde{b}}_{zy}\Biggl[F_y - g^{-1}_{xx}\left(g_{xy}F_x - g_{xx}F_y\right)\Biggr];\nonumber\\
& & \tilde{\tilde{\tilde{F}}}_{yz}=0.
\end{eqnarray}
where
\begin{equation}
\label{eq:btildetildeyz}
\tilde{\tilde{b}}_{yz}= - \frac{g_{yz}}{g_{yy} - \frac{g_{xy}^2}{g_{xx}}}.
\end{equation}
This gives (\ref{Ftildetildetildeyx}) - (\ref{Ftildetildetildezx}.)

One therefore obtains the following two-form field strength in the mirror type IIA:
\begin{eqnarray}
\label{F2_IIA}
& & \tilde{\tilde{\tilde{F}}}_2 = \tilde{\tilde{\tilde{F}}}_{xr}dx\wedge dr + \tilde{\tilde{\tilde{F}}}_{x\theta_1}dx\wedge d\theta_1 + \tilde{\tilde{\tilde{F}}}_{x\theta_2}dx\wedge d\theta_2 + \tilde{\tilde{\tilde{F}}}_{y\theta_1}dy\wedge d\theta_1 + \tilde{\tilde{\tilde{F}}}_{y\theta_2}dy\wedge d\theta_2\nonumber\\
 & &  +  \tilde{\tilde{\tilde{F}}}_{z\theta_2}dz\wedge d\theta_2 +  \tilde{\tilde{\tilde{F}}}_{z\theta_1}dz\wedge d\theta_1 +   \tilde{\tilde{\tilde{F}}}_{zr}dz\wedge dr
 + \tilde{\tilde{\tilde{F}}}_{yr}dy\wedge dr.
\end{eqnarray}
  Hence,
  \begin{equation}
  A^{F_1}=\left(\tilde{\tilde{\tilde{F_1}}}_{yx} y dx + \tilde{\tilde{\tilde{F_1}}}_{zx} z dx\right)\left(\theta_{1,2}\rightarrow\langle\theta_{1,2}\rangle,\phi_{1,2}\rightarrow\langle \phi_{1,2}\rangle,\psi\rightarrow\right\langle\psi\rangle,r\rightarrow\langle r\rangle).
  \end{equation}

   The two-form field strength components obtained from three T-dualities applied to the self-dual five-form field strength are, using \cite{T-dual Hassan}, given via:
   \begin{eqnarray}
   \label{eq:HodgedualFtildetildetildebeta1beta2}
  & & \widetilde{\widetilde{\widetilde{(*F_5)}}}_{\beta_1\beta_2} = - \Biggl[ - (*F_5)_{xyz\beta_1\beta_2} - 4 \Biggl(- b_{x\beta_1}\left\{F_{yz\beta_2} - 2 g_{xx}^{-1}\left[g_{xy}F_{xz\beta_2} - g_{xx}F_{z\beta_2y} + g_{xz}F_{\beta_2yx}\right]\right\}\nonumber\\
   & & - b_{x\beta_2}\left\{F_{yz\beta_1} - 2 g_{xx}^{-1}\biggl[g_{xy}F_{xz\beta_1} - g_{xx}F_{z\beta_1y} + g_{xz}F_{\beta_1yx}\biggr]\right\}\Biggr)\nonumber\\
  & & - 3 \Biggl(\tilde{b}_{y\beta_1}\left[- 3 F_{x\beta_2z} + 6 b_{x\beta_2}\left\{F_z - g^{-1}_{xx}\left(g_{xz}F_x - g_{xx}F_z\right)\right\} - 2 \tilde{g}_{yy}^{-1}\left(\tilde{g}_{yz}g^{-1}_{xx}g_{x\beta_2}F_x\right)\right]\nonumber\\
  & & - \tilde{b}_{y\beta_2}\Biggl[- 3 F_{x\beta_1z} + 6 b_{x\beta_1}\left\{F_z - g^{-1}_{xx}(g_{xz}F_x - g_{xx}F_z)\right\} - 2 \tilde{g}^{-1}_{yy}\left(\tilde{g}_{yz}g^{-1}_{xx}g_{x\beta_2}F_x\right)\Biggr]\Biggr)\Biggr]\nonumber\\
  & & - 2 \left(- \tilde{\tilde{b}}_{z\beta_1}\left[- 2 g^{-1}_{xx}g_{x\beta_2}F_x\right] - b_{z\beta_2}\left[- 2 g^{-1}_{xx}g_{x\beta_1}F_x\right]\right),
   \end{eqnarray}
 where $*F_5=F_5, \beta_{1/2}\equiv r,\theta_{1/2}$, and
 \begin{equation}
 \label{eq:btildeybetai}
 \tilde{b}_{y\beta_i} = b_{y\beta_i} - \frac{g_{xy}b_{x\beta_i}}{g_{xx}}.
 \end{equation}
Therefore using (\ref{eq:HodgedualFtildetildetildebeta1beta2}), one obtains (\ref{HodgeFtildetildetildetheta1theta2}) - (\ref{HodgeFtildetildetildertheta1}).

So,
\begin{eqnarray}
& &  A^{F_5}=\left( \tilde{\tilde{\tilde{F_5}}}_{r\theta_1}r d\theta_1 + \tilde{\tilde{\tilde{F_5}}}_{\theta_1\theta_2}\theta_1 d\theta_2 + \tilde{\tilde{\tilde{F_5}}}_{r\theta_2}r d\theta_2\right)\left(\theta_{1,2}\rightarrow\langle\theta_{1,2}\rangle,\phi_{1,2}\rightarrow\langle \phi_{1,2}\rangle,\psi\rightarrow\langle\psi\rangle,r\rightarrow\langle r\rangle\right);\nonumber\\
& &\tilde{\tilde{\tilde{\phi}}}=\frac{1}{2}\left[2\phi - ln\left(g_{xx}\left\{g_{yy} - \frac{g_{xy}^2}{g_{xx}}\right\}\left\{g_{zz}-\frac{g_{xz}^2}{g_{xx}} - \frac{\left(g_{yz} - \frac{g_{xy}g_{xz}}{g_{xx}}\right)^2}{g_{yy} - \frac{g_{xy}^2}{g_{xx}}}\right\}\right)\right].
\end{eqnarray}
Therefore, the uplifted $M-$theory metric is given by:
\begin{eqnarray}
\label{Mtheory met}
& & \hskip-0.8in ds^2_{11} = e^{-\frac{2\tilde{\tilde{\tilde{\phi}}}}{3}} \Biggl[
\frac{1}{\sqrt{h\left(r,\theta_1,\theta_2\right)}}\Bigl(-g_1 dt^2+dx_1^2+dx_2^2+dx_3^2\Bigr)+\sqrt{h\left(r,\theta_1,\theta_2\right)}\Bigl(g_2^{-1}dr^2\Bigr)+  ds^2_{{\rm equation}\ (\ref{mirror_metric})}\Biggr]\nonumber\\
 & & + e^{\frac{4\tilde{\tilde{\tilde{\phi}}}}{3}}\Bigl(dx_{11} + A^{F_1}+A^{F_3}+A^{F_5}\Bigr)^2.
\end{eqnarray}
The horizon area, using (\ref{metric}) or the black $M3$-brane metric (\ref{Mtheory met}), assuming as in \cite{Gubser+Klebanov+Peet} that the world-volume coordinates $x_{1,2,3}$ are wrapped around a $T^3$ of a very large radius, will be proportional to:
\begin{equation}
\label{horizon_area}
{\rm Horizon\ area}\sim \left(\frac{1}{h^{\frac{1}{4}}(r_h)}\right)^3\left(h^{\frac{1}{4}}(r_h)\right)^5r_h^5\ \stackrel{h\sim\frac{4\pi g_sN}{r^4}}{\longrightarrow\ }\ r_h^3.
\end{equation}

In order to check for the supersymmetry of the $M3$-brane solution of (\ref{Mtheory met}) and to get the the explicit dependence of 11-dimensional action on parameters ${g_s},M/M_{\rm eff}, N/N_{\rm eff}$, we have first simplified metric components in the (i) weak($g_s$) coupling - large t'Hooft couplings limit: (\ref{limits_Dasguptaetal-i})
as well  as the (ii) `MQGP limit': (\ref{limits_Dasguptaetal-ii}).
The simplified expressions for all non-zero 11-dimensional  metric components  in
either limit using (\ref{metric-mirror}) (which too is valid in either limit)
$\forall\theta_{1,2}\in[\epsilon_{\theta_{1,2}},\pi-\epsilon_{\theta_{1,2}}]$, assuming that globally one can replace $x,y,z$ respectively by\\ $\sqrt{h_2}\left(4\pi g_sN\right)^{\frac{1}{4}}sin\theta_1 sin\phi_1, \sqrt{h_4}\left(4\pi g_sN\right)^{\frac{1}{4}}sin\theta_2 sin\phi_2,2\sqrt{h_1}\left(4\pi g_s N\right)^{\frac{1}{4}}sin\frac{\psi}{2}$,  are as follows\footnote{Remark: For a chosen scaling, ${r_{\Lambda}\rightarrow \epsilon^{-a}}$, $M/M_{\rm eff}\sim {\epsilon^{-\frac{3}{2}}d}$, $N/N_{\rm eff}\sim{\epsilon^{-19d}} $, $g_s \sim {\epsilon^{d}} $, we observe that metric components ${G^{\cal M}_{00},G^{\cal M}_{11},G^{\cal M}_{22},G^{\cal M}_{33}}$ asymptotically approach the flat metric on $\mathbb{R}^{1,3}$ as $r_{\Lambda}\rightarrow \infty$ for $ a={\frac{25}{6}} $.}:
{\small
\begin{eqnarray}
\label{eq:simplifiedmetriccomponents}
&(i)& G^{\cal M}_{0 0}\sim  -\frac{3^{2/3} r^2 \left(1-\frac{r_h^4}{r^4}\right)}{2 {g_s}^{7/6} \sqrt{N} \sqrt{\pi }} \nonumber\\
&(ii)&
G^{\cal M}_{1 1}\sim \frac{3^{2/3} r^2}{2 {g_s}^{7/6} \sqrt{N} \sqrt{\pi }}\nonumber\\
&(iii)&
G^{\cal M}_{2 2}\sim \frac{3^{2/3} r^2}{2 {g_s}^{7/6} \sqrt{N} \sqrt{\pi }} \nonumber\\
&(iv)&
G^{\cal M}_{3 3}\sim \frac{3^{2/3} r^2}{2 {g_s}^{7/6} \sqrt{N} \sqrt{\pi }}\nonumber\\
&(v)&
G^{\cal M}_{r r}\sim \frac{2~3^{2/3} \left(1-\frac{r_h^4}{r^4}\right)^{-1} \sqrt{N} \sqrt{\pi }}{\sqrt[6]{{g_s}} r^2}\nonumber\\
&(vi)&
G^{\cal M}_{x r}\sim  \frac{1}{\sqrt[3]{3} \pi ^2 (\cos (2 {\theta_1})-5)^2}\nonumber\\
& & \times\Bigl(4 {g_s}^{4/3} {N_f}^2 {\sin (\phi_1)} \sin (\frac{\psi}{2})  \sin ({\theta_1}) \left(9 \sin ^2({\theta_1})+6 \cos ^2({\theta_1})+4 \cos ({\theta_1})\right)
   \Bigl(9 {h_5} \sin ({\theta_1}) \nonumber\\
   && +4 \cos ^2({\theta_1}) \csc ({\theta_2})-2 \cos ({\theta_1}) \cot ({\theta_2})+6 \sin ^2({\theta_1}) \csc
   ({\theta_2})\Bigr)\Bigr) \nonumber\\
&(vii)& G^{\cal M}_{ r_{\theta_1}}\sim \frac{3\ 3^{2/3} {g_s}^{4/3} {N^{2}_f}  {\sin^2 (\phi_1)}  \sin ^2({\theta_1}) (8 \cos ({\theta_1})-3 \cos (2 {\theta_1})+15)^2}{2 \pi ^2 (\cos (2
   {\theta_1})-5)^2} \nonumber\\
&(viii)& G^{\cal M}_{ r_{\theta_2}}\sim \frac{3\ 3^{2/3} {g_s}^{4/3} {N^{2}_f} {\sin^2 (\phi_1)} \sin ^2({\theta_1}) (8 \cos ({\theta_1})-3 \cos (2 {\theta_1})+15)^2}{2 \pi ^2 (\cos (2
   {\theta_1})-5)^2} \nonumber\\
&(ix)& G^{\cal M}_{10 r}\sim -\frac{\sqrt[6]{3} {g_s}^{4/3} {N_f} {\sin (\phi_1)} \sin ({\theta_1}) (8 \cos ({\theta_1})-3 \cos (2 {\theta_1})+15)}{\sqrt{2} \pi  (\cos (2
   {\theta_1})-5)}\nonumber\\
&(x)&
G^{\cal M}_{\theta_1 \theta_1}\sim \frac{10^{-4}  \left(1-\frac{r_h^4}{r^4}\right)^{-2}  {g_s}^{23/6} {M_{\rm eff}}^4 {N_f}^2 \cot ^2\left(\frac{{\theta_2}}{2}\right) ln ^2(r)}{ \left(9 {g_s}^2 {N_f}
   ln ^2(r) {M_{\rm eff}}^2+6 {g_s} \pi  ln (r) {M_{\rm eff}}^2+4 N \pi ^2\right)^{3/2} \left(\left(\sin ^2({\theta_1})-\sin ^2({\theta_2})\right)
   {f_2}({\theta_2})^2+1\right)} \nonumber\\
   && + \frac{1}{2 \sqrt[3]{3} \sqrt[6]{g_s} \sqrt{N} \pi }\Bigr( 2 N \pi  \left({f_1}({\theta_1})^2 \sin ^2({\theta_1})+1\right)- \nonumber\\
   && \frac{324 {g_s} M^2 \left(r^2-3 a^2\right)^2 ln ^2(r) \sin ^2({\theta_1}) (2 \cos ({\theta_1}) \cos ({\theta_2})-9 {h_5} \sin ({\theta_1})
   \sin ({\theta_2}))^2}{r^4 (\cos (2 {\theta_1})-5) \left(2 \cos ^2({\theta_2}) \sin ^2({\theta_1})+2 \cos ^2({\theta_1}) \sin ^2({\theta_2})+3
   \left(\sin ^2({\theta_1}) \sin ^2({\theta_2})+{h_5} \sin (2 {\theta_1}) \sin (2 {\theta_2})\right)\right)}\Bigr)\nonumber\\
&(xi)&  G^{\cal M}_{\theta_1\theta_2}\sim \frac{{g_s}^{13/3}}{r^4} \Bigr(\frac{3^{2/3} {g_s} \left(1-\frac{r_h^4}{r^4}\right)^{-2} {M_{\rm eff}}^4 {N_f}^2 \cot \left(\frac{{\theta_1}}{2}\right) \cot \left(\frac{{\theta_2}}{2}\right) ln ^2(r)}{204800  \pi
   ^{7/2} \left(\frac{{g_s} N}{r^4}\right)^{3/2} r^2} + \nonumber\\
    && \frac{1}{2 \sqrt{\pi}{g_s}^{9/2} N^{1/2}}\Bigr(\frac{{h_5} N \pi  r^4 (4 (\cos (2 {\theta_1})-5)+{f_1}({\theta_1}) {f_2}({\theta_2}) (13 \sin ({\theta_1})+\sin (3 {\theta_1})) \sin
   ({\theta_2}))}{\cos (2 {\theta_1})-5}+ \nonumber\\
   && \frac{54 {g_s} \left(9 a^4 {g_s}-1\right) M^2 \left(r^2-3 a^2\right)^2 ln ^2(r) \sin ({\theta_1}) \sin ({\theta_2}) (9 {h_5} \sin
   ({\theta_1}) \sin ({\theta_2})-2 \cos ({\theta_1}) \cos ({\theta_2}))}{2 \cos ^2({\theta_2}) \sin ^2({\theta_1})+3 \sin ^2({\theta_2}) \sin
   ^2({\theta_1})+2 \cos ^2({\theta_1}) \sin ^2({\theta_2})+3 {h_5} \sin (2 {\theta_1}) \sin (2 {\theta_2})}\Bigr)\Bigr)\nonumber\\
 &(xii)& G^{\cal M}_{x \theta_1}\sim -\frac{  {g_s} ^{4/3} }{\pi ^2 (\cos (2 {\theta_1})-5)^2 \left(3
   \sin ^2({\theta_1}) \sin ^2({\theta_2})+2 \sin ^2({\theta_1}) \cos ^2({\theta_2})+2 \cos ^2({\theta_1}) \sin ^2({\theta_2})\right)}\times \nonumber\\
   && \hskip -0.1in \Bigl(3^{2/3} {N_f}^2 \sin{\phi_1} \sin{\phi_2} \sin ({\theta_1}) \sin ^2({\theta_2}) \left(9 \sin ^2({\theta_1})+6 \cos ^2({\theta_1})+4 \cos
   ({\theta_1})\right) \Bigl(-4 \cos ^4({\theta_1}) (4 \cot ({\theta_2})-9 \sin ({\theta_2})) \nonumber\\
   && +2 \cos ^3({\theta_1}) (9 \sin ({\theta_2})+4 \cos
   ({\theta_2}) \cot ({\theta_2}))-6 \sin ({\theta_1}) \cos ^2({\theta_1}) (8 \sin ({\theta_1}) \cot ({\theta_2})-6 \sin ({\theta_1}) \cos
   ({\theta_2}) \cot ({\theta_2}))\nonumber\\
   && +3 \sin ^2({\theta_1}) \cos ({\theta_1}) (9 \sin ({\theta_2})+10 \cos ({\theta_2}) \cot ({\theta_2}))-9
   \Bigl(3 \left(-3 \sin ^4({\theta_1})-\sin ^2(2 {\theta_1})\right) \sin ({\theta_2}) \nonumber\\
   && +4 \sin ^4({\theta_1}) \cot ({\theta_2})-6 \sin
   ^4({\theta_1}) \cos ({\theta_2}) \cot ({\theta_2})\Bigr)\Bigr)\Bigr) +\nonumber\\
   && \frac{1}{16 \pi ^{5/4}\sqrt[4]{N} r}\Bigr( {g_s}^{13/12} M  {N_f} \cot \left(\frac{{\theta_1}}{2}\right) \left(9 {h_5}+\left(3 \sqrt{6}-2 \cot ({\theta_1})\right)   \cot
   ({\theta_2})\right) \csc ({\theta_1}) \csc ({\theta_2}) ln (r)\nonumber\\
 &&  \left(108 ln (r) a^2+r\right) (2 \cos ({\theta_1}) \cos ({\theta_2})-9
   {h_5} \sin ({\theta_1}) \sin ({\theta_2}))\Bigr) \nonumber\\
   &(xiii)&   G^{\cal M}_{y \theta_1}\sim \Bigr(11 {g_s}^{1/12} M \csc ({\theta_1}) ln (r) \Bigr(-3 {g_s} {h_5} {N_f} r \cot \left(\frac{{\theta_1}}{2}\right)
   \left(108 ln (r) a^2+r\right) \cos ^3({\theta_1})- 8 \pi \Bigl(3 a^2- r^2\Bigr)\nonumber\\
   && \times     \cot ({\theta_2}) \sin ^2({\theta_1}) \cos
   ({\theta_1})-\Bigr(\frac{1}{2} {g_s} {N_f} r (4-2 \cos (2 {\theta_2})) \cot \left(\frac{{\theta_1}}{2}\right) \cot ({\theta_2}) \csc
   ^2({\theta_2}) \left(108 ln (r) a^2+r\right) \nonumber\\
   && -36 {h_5} \pi  \left(3 a^2-r^2\right)\Bigr) \sin ^3({\theta_1})\Bigr) \sin ^2({\theta_2})\Bigr)/\Bigr(4\pi
   ^{5/4}\sqrt[4]{N}  r^2 \Bigr(2 \cos ^2({\theta_2}) \sin ^2({\theta_1})+2 \cos ^2({\theta_1}) \sin ^2({\theta_2}) + \nonumber\\
   && 3 \left(\sin ^2({\theta_1}) \sin
   ^2({\theta_2})+{h_5} \sin (2 {\theta_1}) \sin (2 {\theta_2})\right)\Bigr)\Bigr) \nonumber\\
  &(xiv)& G^{\cal M}_{z \theta_1}\sim \frac{1}{8 \sqrt{2} \pi ^{5/4}\sqrt[4]{N} r}\Bigr(3^{2/3} {g_s}^{13/12} M  {N_f} \cot \left(\frac{{\theta_1}}{2}\right) \csc ^2({\theta_1}) ln (r) \left(108 ln (r) a^2+r\right) \nonumber\\
  &&  \left(2
   \cos ^2({\theta_1})+\left(2 \cot ^2({\theta_2})+3\right) \sin ^2({\theta_1})+6 {h_5} \cot ({\theta_2}) \sin (2 {\theta_1})\right)\Bigr)\nonumber\\
   &(xv)& G^{\cal M}_{10 \theta_1}\sim \frac{{g_s}^{11/6}   {N_f} r ln (r)}{652 \sqrt{N} \pi ^{7/4}} \Bigr( \frac{  \left(1-\frac{r_h^4}{r^4}\right)^{-1}{g_s}^{3/4} {M_{\rm eff}}^2 \cot \left(\frac{{\theta_2}}{2}\right)}{  \sqrt[4]{N} r \sqrt{\left(\sin ^2({\theta_1})-\sin ^2({\theta_2})\right)
   {f_2}({\theta_2})^2+1}}- \nonumber\\
   && \frac{17280 M \sqrt[4]{{g_s} N} \sqrt{\pi } \left(r^2-3 a^2\right) \cos ({\theta_2}) {f_1}({\theta_1}) \sin ^3({\theta_1}) (2 \cos ({\theta_1})
   \cos ({\theta_2})-9 {h_5} \sin ({\theta_1}) \sin ({\theta_2}))}{r^2 (\cos (2 {\theta_1})-5) \left(2 \cos ^2({\theta_2}) \sin
   ^2({\theta_1})+2 \cos ^2({\theta_1}) \sin ^2({\theta_2})+3 \left(\sin ^2({\theta_1}) \sin ^2({\theta_2})+{h_5} \sin (2 {\theta_1}) \sin (2
   {\theta_2})\right)\right)}\Bigr)\nonumber\\
   & & -\frac{\sqrt[6]{3} {g_s}^{4/3} {N_f} {\sin (\phi_1)} \sin ({\theta_1}) (8 \cos ({\theta_1})-3 \cos (2 {\theta_1})+15)}{\sqrt{2} \pi  (\cos (2
   {\theta_1})-5)}\nonumber\\
   &(xvi)& G^{\cal M}_{\theta_2 \theta_2}\sim  -\frac{3^{2/3} \left(1-\frac{r_h^4}{r^4}\right)^{-2}{g_s}^{23/6} {M_{\rm eff}}^4 {N_f}^2 \cot ^2\left(\frac{{\theta_1}}{2}\right) \left((\cos (2 {\theta_1})-\cos (2 {\theta_2}))
   {f_2}({\theta_2})^2-2\right) ln ^2(r)}{51200 g^2 \sqrt{\pi } \left(9 {g_s}^2 {N_f} ln ^2(r) {M_{\rm eff}}^2+6 {g_s} \pi  ln (r)
   {M_{\rm eff}}^2+4 N \pi ^2\right)^{3/2}} + \nonumber\\
   && \frac{1}{32 \sqrt[3]{3} \pi ^{5/2}\sqrt{g_s N} r^4 (\cos (2 {\theta_1})-5)} \Bigr(-64 \sqrt[3]{{g_s}} N \pi ^3 r^4 {f_2}({\theta_2})^2 \sin ^3({\theta_1}) (2 \cos ({\theta_1}) \cot ({\theta_2})-9 {h_5} \sin
   ({\theta_1})) -\nonumber\\
   && \frac{1296 {g_s}^{4/3} M^2 \pi ^2 \left(r^2-3 a^2\right)^2 \left(-9 {g_s} a^4+3 {g_s} r^2 a^2+1\right)^2 (\cos (2 {\theta_1})-5)^2 ln ^2(r) \sin
   ^2({\theta_2})}{2 \cos ^2({\theta_2}) \sin ^2({\theta_1})+2 \cos ^2({\theta_1}) \sin ^2({\theta_2})+3 \left(\sin ^2({\theta_1}) \sin
   ^2({\theta_2})+{h_5} \sin (2 {\theta_1}) \sin (2 {\theta_2})\right)}\Bigr)\nonumber\\
   &(xvii)& G^{\cal M}_{x \theta_2}\sim -\frac{  {g_s} ^{4/3} }{\pi ^2 (\cos (2 {\theta_1})-5)^2 \left(3
   \sin ^2({\theta_1}) \sin ^2({\theta_2})+2 \sin ^2({\theta_1}) \cos ^2({\theta_2})+2 \cos ^2({\theta_1}) \sin ^2({\theta_2})\right)}\times \nonumber\\
   && \hskip -0.1in \Bigl(3^{2/3} {N_f}^2 \sin{\phi_1} \sin{\phi_2} \sin ({\theta_1}) \sin ^2({\theta_2}) \left(9 \sin ^2({\theta_1})+6 \cos ^2({\theta_1})+4 \cos
   ({\theta_1})\right) \Bigl(-4 \cos ^4({\theta_1}) (4 \cot ({\theta_2})-9 \sin ({\theta_2})) \nonumber\\
   && +2 \cos ^3({\theta_1}) (9 \sin ({\theta_2})+4 \cos
   ({\theta_2}) \cot ({\theta_2}))-6 \sin ({\theta_1}) \cos ^2({\theta_1}) (8 \sin ({\theta_1}) \cot ({\theta_2})-6 \sin ({\theta_1}) \cos
   ({\theta_2}) \cot ({\theta_2}))\nonumber\\
   && +3 \sin ^2({\theta_1}) \cos ({\theta_1}) (9 \sin ({\theta_2})+10 \cos ({\theta_2}) \cot ({\theta_2}))-9
   \Bigl(3 \left(-3 \sin ^4({\theta_1})-\sin ^2(2 {\theta_1})\right) \sin ({\theta_2}) \nonumber\\
   && +4 \sin ^4({\theta_1}) \cot ({\theta_2})-6 \sin
   ^4({\theta_1}) \cos ({\theta_2}) \cot ({\theta_2})\Bigr)\Bigr)\Bigr) +\nonumber\\
   &&   \frac{72 \sqrt[6]{3} M  {g_s}^{1/12}  \left(3 a^2-r^2\right) \left(9 {g_s} a^4-3 {g_s} r^2 a^2-1\right) \cos
   ^2({\theta_1}) \cot ({\theta_1}) \cot ({\theta_2}) ln (r)}{r^2 \sqrt[4]{\pi } \sqrt[4]{N}(\cos (2 {\theta_1})-5) \left(2 \cot ^2({\theta_1})+2 \cot
   ^2({\theta_2})+3\right)}\nonumber\\
   &(xviii)& G^{\cal M}_{y \theta_2}\sim \frac{ \sqrt[4]{N} {g_s}^{-5/12}  \pi ^{1/4} (\cos (2 {\theta_1})-5) \cos ({\theta_2}) {f_2}({\theta_2}) \sin ^2({\theta_2})}{3 \sqrt{2}
   \sqrt[3]{3}  \left(2 \cos ^2({\theta_2}) \sin ^2({\theta_1})+2 \cos ^2({\theta_1}) \sin ^2({\theta_2})+3 \left(\sin ^2({\theta_1})
   \sin ^2({\theta_2})+{h_5} \sin (2 {\theta_1}) \sin (2 {\theta_2})\right)\right)}\nonumber\\
   &(xix)& G^{\cal M}_{z \theta_2}\sim \frac{3^{2/3} {g_s}^{13/12} M  {N_f}}{256 \sqrt{2} \pi ^{5/4} \sqrt[4]{N} r (\cos (2 {\theta_1})-5)} \times \nonumber\\
    && \Bigr((6 \cos (2 {\theta_1})+(1-12 {h_5}) \cos (2 ({\theta_1}-{\theta_2}))+6 \cos (2 {\theta_2})+12 {h_5} \cos (2 ({\theta_1}+{\theta_2}))+\cos
   (2 ({\theta_1}+{\theta_2}))-14) \nonumber\\
   && \csc ^2({\theta_1}) \csc ^3\left(\frac{{\theta_2}}{2}\right) ln (r) \left(36 ln (r) a^2+r\right) \sec
   \left(\frac{{\theta_2}}{2}\right) \left(\cos ^2({\theta_1})+4 \cos ({\theta_2}) \cos ({\theta_1})-\sin ^2({\theta_1})-5\right)\Biggr) \nonumber\\
   &(xx)& G^{\cal M}_{10 \theta_2}\sim  \frac{{g_s}^{11/6} {N_f} r ln (r)}{640 \sqrt{2} \sqrt[3]{3} \sqrt{N} \pi ^{7/4}} \times \nonumber\\
   && \Bigr(\frac{2   \left(1-\frac{r_h^4}{r^4}\right)^{-1} {g_s}^{3/4} {M_{\rm eff}}^2 \cot \left(\frac{{\theta_1}}{2}\right) \sqrt{\left(\sin ^2({\theta_1})-\sin ^2({\theta_2})\right)
   {f_2}({\theta_2})^2+1}}{\sqrt[4]{N} r}- \nonumber\\
   && \frac{51840 {h_5} M \sqrt[4]{{g_s} N} \sqrt{\pi } \left(r^2-3 a^2\right) \left(-9 {g_s} a^4+3 {g_s} r^2 a^2+1\right) \cos ({\theta_2})
   {f_2}({\theta_2}) \sin ({\theta_1}) \sin ^2({\theta_2})}{r^2 \left(2 \cos ^2({\theta_2}) \sin ^2({\theta_1})+2 \cos ^2({\theta_1}) \sin
   ^2({\theta_2})+3 \left(\sin ^2({\theta_1}) \sin ^2({\theta_2})+{h_5} \sin (2 {\theta_1}) \sin (2 {\theta_2})\right)\right)}\Bigr)\nonumber\\
   & &  -\frac{\sqrt[6]{3} {g_s}^{4/3} {N_f} {\sin (\phi_1)} \sin ({\theta_1}) (8 \cos ({\theta_1})-3 \cos (2 {\theta_1})+15)}{\sqrt{2} \pi  (\cos (2
   {\theta_1})-5)}\nonumber\\
   &(xxi)& G^{\cal M}_{x x}\sim -\frac{3^{2/3} \  (\cos (2 {\theta_2})-5) \sin ^2({\theta_1})}{2 {g_s}^{2/3} \left(2 \cos ^2({\theta_2}) \sin ^2({\theta_1})+2
   \cos ^2({\theta_1}) \sin ^2({\theta_2})\right)}\nonumber\\
   &(xxii)& G^{\cal M}_{x y}\sim -\frac{ \cos ^2({\theta_1}) (\cos (2 {\theta_1})-5) \cos ^3({\theta_2}) \sin ({\theta_2})}{3 \sqrt{2} 3^{5/6} {g_s}^{2/3}
   \left(\cos ^2({\theta_2}) \sin ^2({\theta_1})+\cos ^2({\theta_1}) \sin ^2({\theta_2})\right)^2}\nonumber\\
   &(xxiii)& G^{\cal M}_{x z}\sim \frac{2 \csc ({\theta_1}) \csc ({\theta_2})}{27 \sqrt[3]{3}{g_s}^{2/3} (\cos (2 {\theta_1})-5)} \times \nonumber\\
   && \hskip -0.2in \left[2 \left(\cos ({\theta_2}) \left(6 \sqrt{6} \cot ({\theta_2})-4 \cot ({\theta_1}) \cot ({\theta_2})\right)-9 \sqrt{6} \sin ({\theta_2})\right) \cos
   ^3({\theta_1})+12 \cos ({\theta_2}) \cot ({\theta_2}) \sin ({\theta_1}) \cos ^2({\theta_1})\right. \nonumber\\
   && \left.- 27 \sin ^2({\theta_1}) \left(4 {h_5} \cos
   ({\theta_2})-\sqrt{6} \sin ({\theta_2})\right) \cos ({\theta_1})+81 {h_5} \sin ^3({\theta_1}) \left(\sqrt{6} \cos ({\theta_2})+3 {h_5}
   \sin ({\theta_2})\right)\right] \nonumber\\
   &(xxiv)& G^{\cal M}_{yy}\sim \frac{  3^{2/3} }{{g_s}^{2/3}}\nonumber\\
   &(xxv)& G^{\cal M}_{y z}\sim  -\frac{  \sqrt{2  } \csc ({\theta_1}) (3 {h_5} \cos ({\theta_1})+\cot ({\theta_2}) \sin ({\theta_1}))}{3^{5/6} {g_s}^{2/3}} \nonumber\\
   &(xxvi)& G^{\cal M}_{zz }\sim -\frac{  \csc ^2({\theta_1}) \left(-2 \cos ^2({\theta_1})+\left(-2 \cot ^2({\theta_2})-3\right) \sin ^2({\theta_1})-6 {h_5}
   \cot ({\theta_2}) \sin (2 {\theta_1})\right)}{9 \sqrt[3]{3} {g_s}^{2/3}} \nonumber\\
   &(xxvii)& G^{\cal M}_{10\ 10}\sim \frac{{g_s}^{4/3}}{3 \sqrt[3]{3}}.
   \end{eqnarray}}
This is similar in spirit to the M-theory uplift of the warped resolved
conifold in \cite{large base}. For (\ref{limits_Dasguptaetal-ii}), we also need to keep in mind the following. For
finite $g_s, N_f$, one sees that $- ln\left[\frac{1}{g_s} - \frac{N_f}{8\pi} ln(9a^2r^2+r^6)\right.$
$\left. - \frac{N_f}{2\pi}
ln\left(\sin\frac{\theta_1}{2}\sin\frac{\theta_2}{2}\right)\right]\in\tilde{\tilde{\tilde{\phi}}}$ is expected to
receive the most dominant contribution near $\theta_{1,2}=0,\pi, r=r_\Lambda\rightarrow\infty$. We will assume that
$\theta_{1,2}\rightarrow0$ as $\epsilon^\gamma$  and $r_\Lambda\sim\epsilon^{-\beta},\epsilon\rightarrow0,\gamma,\beta>0$
such that: $\gamma\sim\frac{3}{4}\beta$ implying thereby that in both limits (\ref{limits_Dasguptaetal-i}) as well
as (\ref{limits_Dasguptaetal-ii}), the aforementioned log is $ln g_s$. One hence obtains (without worrying about numerical factors in each component):
{\small
\begin{eqnarray}
\label{ds11squared}
& & ds_{11}^2= \frac{-\left(1 - \frac{r_h^4}{r^4}\right)r^2}{g_s^{\frac{7}{6}}\sqrt{N}} dt^2 + \frac{r^2}{g_s^{\frac{2}{3}}\sqrt{g_sN}}\left(ds^2_{\mathbb{R}^3}\right) + \sqrt{\frac{N}{g_s^{\frac{1}{3}}}}\frac{dr^2}{\left(1 - \frac{r_h^4}{r^4}\right)} \nonumber\\
& & + g_s^{\frac{4}{3}}\left(g_s N\right)^{\frac{1}{4}}\sin\phi_1 sin\frac{\psi}{2}\Upsilon_1(\theta_1,\theta_2) d\phi_1 dr + g_s^{\frac{4}{3}}sin^2\phi_1\Upsilon_2(\theta_1) dr d\theta_1
 + g_s^{\frac{4}{3}}sin^2\phi_1\Upsilon_3(\theta_1) dr d\theta_2\nonumber\\
& & + \left[\frac{g_s^{\frac{23}{6}}M^4N_f^2\Gamma_1(\theta_{1,2})\left(ln r\right)^2}{N^{\frac{3}{2}}\left(1 - \frac{r_h^4}{r^4}\right)^2} + \frac{\sqrt{N}\Gamma_1^\prime(\theta_{1,2})}{g_s^{\frac{1}{6}}} + \frac{g_sM^2\left(r^2 - 3 a^2\right)^2\left(ln r\right)^2\Gamma_1^{\prime\prime}(\theta_{1,2})}{r^4g_s^{\frac{1}{6}}\sqrt{N}}\right]d\theta_1^2\nonumber\\
& & +\frac{g_s^{\frac{13}{3}}}{r^4}\left[\frac{M^4N_f^2g_sr^4\left(ln r\right)^2\Gamma_2(\theta_{1,2})}{\left(1 - \frac{r_h^4}{r^4}\right)^2\left(g_sN\right)^{\frac{3}{2}}} + \frac{1}{\sqrt{N}g_s^{\frac{9}{2}}}\left\{h_5Nr^4\Gamma_3(\theta_{1,2}) + g_sM^2\left(ln r\right)^2(r^2 - 3 a^2)\Gamma_4(\theta_{1,2})\right\}\right]d\theta_1d\theta_2\nonumber\\
& & + \left[ - \frac{g_s^{\frac{23}{6}}M^4N_f^2\left(ln r\right)^2\Gamma_5(\theta_{1,2})}{N^{\frac{3}{2}}\left(1 - \frac{r_h^4}{r^4}\right)^2} + \frac{1}{\sqrt{g_sN}r^4}
\left\{ - g_s^{\frac{1}{3}}Nr^4\Gamma_2(\theta_{1,2}) - g_s^{\frac{4}{3}}M^2(r^2 - 3 a^2)^2\Gamma_7(\theta_{1,2})\left(ln r\right)^2\right\}\right]d\theta_2^2\nonumber\\
& & +(g_sN)^{\frac{1}{4}}\left\{\frac{g_s^{\frac{13}{12}}MN_f ln r \Gamma_8(\theta_{1,2})(108 a^2 ln r + r)}{N^{\frac{1}{4}}r} + g_s^{\frac{4}{3}}N_f^2\sin\phi_1\sin\phi_2\Gamma_8^\prime(\theta_{1,2})\right\}d\phi_1d\theta_1\nonumber\\
& &  + \left\{g_s^{\frac{4}{3}}N_f^2\sin\phi_1\sin\phi_2\Gamma_8^\prime(\theta_{1,2}) + ln r\frac{g_s^{\frac{1}{12}}M (3 a^2 - r^2)(9g_sa^4 - 3 g_sa^2r^2 -1)
\Gamma_8^{\prime\prime}(\theta_{1,2})}{N^{\frac{1}{4}}r^2}\right\}d\phi_1d\theta_2\nonumber\\
& & + \frac{g_s^{\frac{1}{12}}M(g_sN)^{\frac{1}{4}}}{N^{\frac{1}{4}}r^2}\Upsilon(r,\theta_{1,2})d\phi_2d\theta_1
+ \frac{N^{\frac{1}{4}}(g_sN)^{\frac{1}{4}}\Gamma_9(\theta_{1,2}}{g_s^{\frac{5}{12}}}d\phi_2d\theta_2 \nonumber\\
& &
+ \frac{(g_sN)^{\frac{1}{4}}g_s^{\frac{13}{12}}MN_f ln r (108 a^2 ln r + r)\Gamma_{10}(\theta_{1,2})}{rN^{\frac{1}{4}}}d\psi d\theta_1+ \frac{(g_sN)^{\frac{1}{4}}g_s^{\frac{13}{12}}MN_f ln r \Gamma_{11}(\theta_{1,2})}{rN^{\frac{1}{4}}}d\psi d\theta_1 +  \nonumber\\
& & \frac{\sqrt{g_sN}}{g_s^{\frac{2}{3}}}\left(\Gamma_{12}(\theta_{1,2})d\phi_1^2 + \Gamma_{13}(\theta_{1,2})d\phi_2^2 + \Gamma_{14}(\theta_{1,2})d\psi^2\right. + \Gamma_{15}(\theta_{1,2})d\phi_1d\phi_2 +  \Gamma_{16}(\theta_{1,2})d\phi_1d\psi + \nonumber\\
  && \left. \Gamma_{17}(\theta_{1,2})d\phi_2d\psi\right) + g_s^{\frac{4}{3}}dx_{10}^2 +
 - g_s^{\frac{4}{3}}sin\phi_1\Upsilon_4(\theta_1) dr dx_{10}
  \nonumber\\
   & & + \frac{g_s^{\frac{11}{6}}N_f r ln r}{\sqrt{N}}\left\{\frac{g_s^{\frac{3}{4}}M^2\Gamma_{18}(\theta_{1,2})}{N^{\frac{1}{4}}\left(1 - \frac{r_h^4}{r^4}\right)} - \frac{(g_sN)^{\frac{1}{4}}(r^2 - 3 a^2)\Gamma_{19}(\theta_{1,2})}{r^2} - g_s^{\frac{4}{3}}sin\phi_1\Upsilon_5(\theta_1)\right\}dx_{10}d\theta_1\nonumber\\
& &  + \frac{g_s^{\frac{11}{6}}N_f r ln r}{\sqrt{N}}\left\{\frac{g_s^{\frac{3}{4}}M^2\Gamma_{20}(\theta_{1,2})}{N^{\frac{1}{4}}r\left(1 - \frac{r_h^4}{r^4}\right)} - \frac{h_5M(g_sN)^{\frac{1}{4}}(r^2 - 3 a^2)\Gamma_{21}(\theta_{1,2})}{r^2} - g_s^{\frac{4}{3}}sin\phi_1\Upsilon_5(\theta_1)\right\}dx_{10}d\theta_2,
\end{eqnarray}}
\noindent [the angular parts $\Gamma_i(\theta_{1,2}), \Upsilon_j(\theta_{1,2})$, etc. can be read off from (\ref{eq:simplifiedmetriccomponents})] which, similar to \cite{Cvetic_et_al_M3-i}, is an $M3$-brane solution.

Let us look at the near horizon limit of the $G^{\cal M}_{tt}$ and $G^{\cal M}_{rr}$ of (\ref{ds11squared}). Before doing so, these components in the limits (\ref{limits_Dasguptaetal-i})  takes the form:
\begin{equation}
\label{near_horison_i}
- \epsilon^{\frac{25 d}{3}}\left(1 - \frac{r_h^4}{r^4}\right)r^2 dt^2 + \frac{dr^2}{\epsilon^{\frac{29 d}{3}}\left(1 - \frac{r_h^4}{r^4}\right)}.
\end{equation}
In the near-horizon limit $r = r_h + \epsilon^\prime\chi$, implying $1 - \frac{r_h^4}{r^4} = \frac{4 \epsilon^\prime\chi}{r_h} + {\cal O}\left(\epsilon^\prime\ ^2\right)$,
\begin{equation}
\label{near_horizon_ii}
\frac{dr^2}{\epsilon^{\frac{29 d}{3}}\left(1 - \frac{r_h^4}{r^4}\right)} \sim \xi \frac{(d\chi)^2}{\chi},\ {\rm where}\ \xi\equiv\frac{\epsilon^\prime r_h}{\epsilon^{\frac{29 d}{3}}}.
\end{equation}
Writing $\xi \frac{(d\chi)^2}{\chi} = du^2$ or $\chi=\frac{u^2}{4\xi}$, one obtains:
\begin{equation}
\label{near_horizon_iii}
- 4\pi^2 u^2 T^2 dt^2 + du^2,
\end{equation}
where $T^2(r_h\approx1\ [{\rm See\ {\bf 5}}])\sim\frac{1}{\left(\sqrt{g_sN}\right)^2}$, in conformity with \cite{metrics} as well as (\ref{T}).

Similarly, in the MQGP limit (\ref{limits_Dasguptaetal-ii}), we obtains:
$G^{\cal M}_{00}\sim\epsilon^{\frac{55d}{3}}r^2\left(1 - \frac{r_h^4}{r^4}\right),
G^{\cal M}_{rr}\sim\frac{\epsilon^{-\frac{59d}{3}}}{r^2
\left(1 - \frac{r_h^4}{r^4}\right)}$, one can rewrite $G_{rr}dr^2
=\xi^\prime\frac{d\omega^2}{\omega}$ where $\xi^\prime\sim
\frac{r_h\epsilon^\prime}{\epsilon^{\frac{59d}{3}}}$. Once again writing
$\xi^\prime \frac{d\omega^2}{\omega}=dv^2$ or $\omega=\frac{v^2}{4\xi^\prime}$,
one sees that near $r=r_h\sim1, G^{\cal M}_{tt}dt^2\sim\epsilon^{38d}u^2dt^2$ implying
again $T^2\sim\frac{1}{\left(\sqrt{g_sN}\right)^2}$ in conformity with
\cite{metrics} and (\ref{T}).

As we will see in {\bf 5.3}, the action using (\ref{eq:simplifiedmetriccomponents}) is singular at $\theta_{1,2}=0,\pi$; we regulate these
pole-singularities by introducing a small angle cut-off $\epsilon_\theta$: $\theta_{1,2}\in[\epsilon_\theta,\pi-\epsilon_\theta]$. We then show
that the finite part of the action turns out to be independent of this $\epsilon_\theta$ if one identifies $\epsilon_\theta=\epsilon^\gamma$,
for an appropriate $\gamma$. We will henceforth follow this.

We now discuss the supersymmetry of the M-theory uplift in the two limits. 
In the limit (\ref{limits_Dasguptaetal-i}), the near-horizon (with $r_h\sim1$) limit of (\ref{ds11squared}) near $\theta_{1,2}=0,\pi$,
after appropriate rescaling of $\mathbb{R}^3$-variables and using $G^M_{\bullet r}=\epsilon^{\frac{29d}{3}} u G^M_{\bullet u},
G_{\phi_{1,2}\bullet}\sim\left(g_s N\right)^{\frac{1}{4}}\sin\theta_{1,2}G_{x/y\bullet}, G_{\psi\bullet}\sim\left(g_s N\right)^{\frac{1}{4}}G_{z\bullet}$, reduces to:
{\small
\begin{eqnarray}
\label{metric_simp}
& & ds_{11}^2 = - 4\pi^2 u^2 T^2 dt^2 + du^2 +  ds^2_{\mathbb{R}^{3}} + \epsilon^{12d}u sin^2\phi_1 du (d\theta_1 + d\theta_2)
 + \frac{d\theta_1^2}{\epsilon^{\frac{29d}{3}}} + \left(\frac{h_5}{\epsilon^{\frac{29d}{3}}} \right)d\theta_1d\theta_2 + \frac{d\theta_2^2}{\epsilon^{\frac{29d}{3}}} \nonumber\\
& &  +
\frac{d\phi_1^2}{\epsilon^{\frac{26d}{3}}} + \frac{d\phi_2^2}{\epsilon^{\frac{26}{3}}} + \frac{d\psi^2}{\epsilon^{\frac{32d}{3}}} + \frac{d\phi_1d\phi_2}{\epsilon^{\frac{61d}{6}}} + \frac{d\phi_1d\psi}{\epsilon^{\frac{43d}{6}}}
+ \frac{d\phi_2d\psi}{\epsilon^{\frac{29d}{3}}} + \epsilon^{\frac{23d}{2}} u sin\phi_1 du dx_{10} +
\epsilon^{7d} u sin\phi_1 sin\frac{\psi}{2} du d\phi_1 \nonumber\\
 & & + \epsilon^{-\frac{8d}{3}}\sin\phi_1\sin\phi_2d\phi_1(d\theta_1 + d\theta_2) + \epsilon^{\frac{4d}{3}}dx_{10}^2 + d\theta_1dx_{10} \epsilon^{\frac{11d}{6}}sin\phi_1 + \epsilon^{\frac{11d}{6}} sin\phi_1d\theta_2dx_{10}. \end{eqnarray}}
The terms relevant to $G^{\cal M}_{\phi_1\bullet}, G^{\cal M}_{u\bullet}, G^{\cal M}_{\theta_{1,2}\bullet}$ in (\ref{metric_simp}) are:
\begin{eqnarray}
\label{dphi1_du}
& & \frac{d\phi_1}{\epsilon^{\frac{61d}{6}}}\left(\epsilon^{\frac{3d}{2}}d\phi_1 + d\phi_2 + \epsilon^{\frac{9d}{3}}d\psi
+ u \epsilon^{\frac{103}{6}}\sin\phi_1\sin\frac{\psi}{2}d\psi + \epsilon^{\frac{45d}{6}}\sin\phi_1\sin\phi_2[d\theta_1 + d\theta_2]
\right)\nonumber\\
& & + du\left(du + u \epsilon^{12d}\sin^2\phi_1[d\theta_1 + d\theta_2] + u \epsilon^{\frac{23d}{2}}\sin\phi_1dx_{10}\right) + \frac{d\theta_1^2}{\epsilon^{\frac{29d}{3}}} + \left(\frac{h_5}{\epsilon^{\frac{29d}{3}}} \right)d\theta_1d\theta_2 + \frac{d\theta_2^2}{\epsilon^{\frac{29d}{3}}}\nonumber\\
& & + d\theta_1dx_{10} \epsilon^{\frac{11d}{6}}sin\phi_1 + \epsilon^{\frac{11d}{6}} sin\phi_1d\theta_2dx_{10},
\end{eqnarray}
which for $h_5<<1$ will be approximated by:
\begin{equation}
\label{dphi1_du_simp}
 \frac{d\phi_1}{\epsilon^{\frac{61d}{6}}}\left(\epsilon^{\frac{3d}{2}}d\phi_1 + d\phi_2 + \epsilon^{\frac{9d}{3}}d\psi\right)+ du^2 + \frac{d\theta_1^2}{\epsilon^{\frac{29d}{3}}} + \frac{d\theta_2^2}{\epsilon^{\frac{29d}{3}}} .
\end{equation}
The metric for $d=1$ restricted to the fiber $T^3(\phi_1,\phi_2,\psi): \frac{d\phi_1}{\epsilon^{\frac{61}{6}}}\left(\epsilon^{\frac{3}{2}} d\phi_1 + d\phi_2 + \epsilon^3 d\psi\right) + \frac{d\phi_2}{\epsilon^{\frac{26}{3}}}\left(d\phi_2 + \frac{d\psi}{\epsilon}\right) + \frac{d\psi^2}{\epsilon^{\frac{32}{3}}}$ can be diagonalised. The metric of (\ref{metric_simp}) or its limit-(\ref{limits_Dasguptaetal-ii}) equivalent, implies the following elfbeins for the $D=11$ space-time which locally is  $\mathbb{R}^4(t,x^{1,2,3})\times M_7(u,\theta_{1,2},\phi_{1,2},\psi,x_{10})$:
\begin{eqnarray}
\label{elfbeins}
& & e^0 = 2\pi T u dt,\nonumber\\
& & e^{1,2,3} = dx^{1,2,3},\nonumber\\
& & e^4 = du,\nonumber\\
& & e^{5,6}=\frac{d\theta_{1,2}}{\epsilon^{\alpha_\theta}}\nonumber\\
& & e^{7} = \frac{\epsilon^{-\alpha_{\Phi}}}{2^{\frac{1}{4}}}\left(d\phi_1-d\phi_2\right),\nonumber\\
& & e^{8}=\frac{\epsilon^{-\alpha_{\Phi}}}{2^{\frac{1}{4}}}\left(-d\phi_1-d\phi_2\right),\nonumber\\
& & e^{9}=\epsilon^{-\alpha_\psi}d\psi\nonumber\\
& & e^{10} = \epsilon^{\alpha_{10}}dx_{10},
\end{eqnarray}
for appropriate positive values of $\alpha_{\Phi,\psi,10}$.

From (\ref{B-3tduals}), the non-zero components of the four-form field strength $G_4$  in the limit (\ref{limits_Dasguptaetal-i}) or (\ref{limits_Dasguptaetal-ii}),  are:
\begin{eqnarray}
\label{non_zero_G4_components}
& & G_{\theta_1\theta_2 \phi_1\phi_2}, G_{r \theta_1\theta_2 \phi_2},  G_{r \theta_1 \theta_2 \phi_1}, G_{r \theta_2 \phi_1 \psi},G_{r \theta_1 \phi_1 \psi}, G_{r \theta_2 \phi_1 \phi_2}, G_{r \theta_1 \theta_2 \psi},G_{r \theta_1 \phi_2 \psi}, G_{r \theta_2 \phi_2 \psi}, G_{\theta_1 \theta_2 \phi_2 \psi}, \nonumber\\
&& G_{\theta_1 \phi_1 \phi_2 \psi}, G_{\theta_1 \theta_2 \phi_1\psi}\nonumber\\
& & ({\rm from}\ H_3^{IIA}\wedge A_1);\nonumber\\
 & & G_{ r \theta_1 \phi_110}, G_{r \theta_2 \phi_1 10}, G_{r \theta_1 \phi_2 10}, G_{r \theta_2 \phi_2 10}, G_{r \theta_2 \psi 10}, G_{r \theta_1 \psi 10}, G_{ r \phi_1 \psi 10}, G_{ r \phi_2 \psi 10}, G_{ r \theta_1 \theta_2}, G_{\theta_1 \phi_2 \psi 10}, \nonumber\\
&& G_{\theta_2 \psi_2 \psi 10}, G_{\theta_1 \phi_1 \psi 10}, G_{\theta_2 \phi_1 \psi 10}, G_{\theta_1 \theta_2 \psi 10}, G_{\theta_1 \theta_2 \phi_1 10}, G_{\theta_1 \theta_2 \phi_2 10}\nonumber\\
& &  ({\rm from}\ H_3^{IIA}\wedge dx_{10}).
\end{eqnarray}
The dominant contribution appear from following components in the two limits (\ref{limits_Dasguptaetal-i})/(\ref{limits_Dasguptaetal-ii}):
\begin{eqnarray}
\label{G4s}
&& G_{\theta_1\theta_2 \phi_1\phi_2}= \Bigl({N_f}\left(g_s N \right)^{\frac{3}{4}} {f_ 2}({\theta_2}) \sin ({\theta_1}) \cos
   ({\theta_1}) \sin (2 {\theta_2}) \cos ({\theta_2}) \Bigl(4 \cos
   ^3({\theta_1}) \sin ^2({\theta_2}) \cos ({\theta_2}) (6 {\phi_2}
   \cos ({\theta_1})-\psi )\nonumber\\
   && +\cos ^2({\theta_2}) \left(-12 {\phi_2} \cos
   ^3({\theta_1}) \sin ^2({\theta_2})+2 \psi  \sin ^4({\theta_1})-\psi
   \sin ^2({\theta_1}) (\cos (2 {\theta_1})-9)\right) \nonumber\\
   && -4 \psi  \sin
   ^2({\theta_1}) \cos ({\theta_1}) \cos ^3({\theta_2})-2 \psi  (\cos
   (2 {\theta_1})-5) \cos ^2({\theta_1}) \sin
   ^2({\theta_2})\Bigr)\Bigr)/\Bigl(\sqrt{6} \pi  (\cos (2 {\theta_1})-5) \nonumber\\
   && \left(\sin
   ^2({\theta_1}) \cos ^2({\theta_2})+\cos ^2({\theta_1}) \sin
   ^2({\theta_2})\right)^3\Bigr)\sim\epsilon^{-16}_{(\ref{limits_Dasguptaetal-i})}/
   \epsilon^{-36}_{(\ref{limits_Dasguptaetal-ii})}\nonumber\\
   && G_{r \theta_1\theta_2 \phi_2}= \frac{\sqrt{g_s N}}{160 \sqrt{2}  \pi ^{7/4}}\Biggl(\cos ({\theta_1}) \sin ({\theta_2}) \Bigl(\Bigl(\sqrt[4]{{g_s}}
   {M_{eff}}^2 {\theta_1} {f_ 1}({\theta_1}) \csc ^2({\theta_2})
   (3 {h_ 5} \cot ({\theta_1})+\cot ({\theta_2})) \nonumber\\
  && \hskip -0.2in \left(12 {g_s}
   {N_f} \log (r)+{g_s} {N_f} \log \left(\sin
   \left(\frac{{\theta_1}}{2}\right) \sin
   \left(\frac{{\theta_2}}{2}\right)\right)+3 {g_s} {N_f}+4 \pi
   \right)\Bigr)/\Bigl(N^{3/4} r \left(2 \cot ^2({\theta_1})+2 \cot
   ^2({\theta_2})+3\right)\Bigr)\nonumber\\
   && \hskip -0.2in +\frac{120 \sqrt{\pi } {N_f}
   {f_ 2}({\theta_2}) \sin ({\theta_1}) \sin (2 {\theta_2}) \cos
   ({\theta_2}) \sqrt[4]{\frac{{g_s} N}{r^4}} \left(27 {g_s} M
   {\phi_2} \cot \left(\frac{{\theta_1}}{2}\right) \sin ({\theta_2})+4
   \sqrt{3} \sqrt[4]{\pi } {\phi_1} r \sqrt[4]{{g_s} N} \sin
   ({\theta_1})\right)}{\sqrt{{g_s} N} \left(\sin ^2({\theta_1}) \cos
   ^2({\theta_2})+\cos ^2({\theta_1}) \sin
   ^2({\theta_2})\right)^2}\Bigr)\Biggr) \nonumber\\
   &&
 \sim  \epsilon^{-\frac{59}{6}}_{(\ref{limits_Dasguptaetal-i})}/\epsilon^{-\frac{129}{6}}_{(\ref{limits_Dasguptaetal-i})}
    \nonumber\\
   &&  \hskip -0.4in G_{r \theta_1 \theta_2 \phi_1} = \Bigl(32 \sqrt{6} {g_s} \left(g_s N\right)^{\frac{1}{4}} M {N_f} {\phi_2} {f_ 1}({\theta_1})
   \cos ^4({\theta_1}) \sin ^2({\theta_2}) \cos ^2({\theta_2}) (2 \cos
   ({\theta_1})-\cos ({\theta_2}))\Bigr)/\Bigl(\pi  r (\cos (2
   ({\theta_1}-{\theta_2}))\nonumber\\
   && +\cos (2 ({\theta_1}+{\theta_2}))+6 \cos
   (2 {\theta_1})+6 \cos (2 {\theta_2})-14) \left(\sin ^2({\theta_1})
   \cos ^2({\theta_2})+\cos ^2({\theta_1}) \sin
   ^2({\theta_2})\right)\Bigr)-\nonumber\\
   && \hskip -0.2in \frac{\sqrt[4]{{g_s}} {M^{2}_{eff}} \sqrt{g_s N}
   {\theta_1} {f_ 1}({\theta_1}) \left(12 {g_s} {N_f} \log
   (r)+{g_s} {N_f} \log \left(\sin \left(\frac{{\theta_1}}{2}\right)
   \sin \left(\frac{{\theta_2}}{2}\right)\right)+3 {g_s} {N_f}+4 \pi
   \right)}{320 \sqrt{2} \pi ^{7/4} N^{3/4} r}\sim\frac{1}{\epsilon^{\frac{35}{6}} r}_{(\ref{limits_Dasguptaetal-i})}/\frac{1}{\epsilon^{\frac{25}{2}} r}_{(\ref{limits_Dasguptaetal-ii})}\nonumber\\
   && G_{\theta_1\theta_2 \phi_2 \ 10}=\frac{\sqrt{g_s N} {f_ 2}({\theta_2}) \sin ({\theta_1}) \cos ({\theta_1}) \sin
   ({\theta_2}) \sin (2 {\theta_2}) \cos ({\theta_2})}{\left(\sin
   ^2({\theta_1}) \cos ^2({\theta_2})+\cos ^2({\theta_1}) \sin
   ^2({\theta_2})\right)^2}\sim\epsilon^{-\frac{32}{3}}_{(\ref{limits_Dasguptaetal-i})}/\epsilon^{-\frac{72}{3}}_{(\ref{limits_Dasguptaetal-ii})}
   \nonumber\\
   &&  G_{\theta_1\theta_2\phi_1 \ 10}=-\frac{\sqrt{g_s N} }{\left(\left(2 \cot
   ^2({\theta_1})+3\right) \sin ^2({\theta_2})+2 \cos
   ^2({\theta_2})\right)^2}\nonumber\\
   && \times \Bigl(2 {f_ 2}({\theta_2}) \csc ({\theta_1}) \sin ^2({\theta_2})
   \cos ({\theta_2}) \Bigl(-24 {h_ 5} \cot ^3({\theta_1}) \sin
   ({\theta_2}) \cos ({\theta_2})+2 \cot ^2({\theta_1}) \cos
   ^2({\theta_2})\nonumber\\
   && +\left(-2 \cot ^4({\theta_1})+\cot
   ^2({\theta_1})+3\right) \sin ^2({\theta_2})\Bigr)\Bigr)\sim\epsilon^{-\frac{32}{3}}_{(\ref{limits_Dasguptaetal-i})}/\epsilon^{-\frac{72}{3}}_{(\ref{limits_Dasguptaetal-ii})}
   \nonumber\\
   &&  G_{r \theta_1 \theta_2 \ 10}=\frac{8 {g_s} M {f_ 1}({\theta_1}) (9 \cos ({\theta_1})-\cos (3
   {\theta_1})) \sin ({\theta_2}) \cos ({\theta_2})}{r (\cos (2
   ({\theta_1}-{\theta_2}))+\cos (2 ({\theta_1}+{\theta_2}))+6 \cos
   (2 {\theta_1})+6 \cos (2 {\theta_2})-14)}\sim\frac{1}{r\sqrt{\epsilon}}_{(\ref{limits_Dasguptaetal-i})/(\ref{limits_Dasguptaetal-ii})}
   \end{eqnarray}
The amount of near-horizon supersymmetry will be determined by solving for the killing spinor $\epsilon$ by the vanishing superysmmetric
variation of the gravitino in $D=11$ supergravity, which is given by:
\begin{equation}
\label{gravitino_susy}
\delta\psi_M = D_M\varepsilon - \frac{1}{288}\Gamma_M G^{M_1M_2M_3M_4}\Gamma_{M_1M_2M_3M_4}\varepsilon + \frac{1}{36}G_M^{\ M_1M_2M_3}\Gamma_{M_1M_2M_3}\varepsilon = 0,
\end{equation}
where $D_M\varepsilon = \partial_M\varepsilon + \frac{1}{4}\omega_M^{ab}\Gamma_{ab}\varepsilon$ where $M, M_.$ and $a,b$ are respectively the curved space and tangent space indices.

Similarly, $G_u^{\ M_1M_2M_3} = G^M_{uu}G^{uM_1M_2M_3}(M_{1,2,3}\neq u)$ and $G_{\theta_1}^{\ M_1M_2M_3} = G^M_{\theta_1\theta_1}G^{\theta_1M_1M_2M_3}(M_{1,2,3}\neq \theta_1)$.
The elfbeins near $r=r_h$ and $\theta_{1,2}=0,\pi$ are, in the limit (\ref{limits_Dasguptaetal-i}) and (\ref{limits_Dasguptaetal-ii}), constants implying the vanishing of the spin connection.

In the limit (\ref{limits_Dasguptaetal-ii}), from (\ref{three-form fluxes}), we see that that from $H_3\wedge A, H_3\wedge dx_{10}$, we end up with two types of fluxes: $G_{u M_1 M_2 M_3}(M_{1,2,3}\equiv
\theta_{1,2},\phi_{1,2},x_{10})$ and $G_{\theta_1 m_1 m_2 m_3}(m_{1,2,3}\equiv\theta_2,\phi_{1,2},x_{10})$. The former, using $G_{u\bullet\bullet\bullet} = u \epsilon^{\frac{59d}{3}}G_{r\bullet\bullet\bullet}$, is sub-dominant as compared to the latter and therefore will be dropped in the subsequent analysis. Hence only $G_{\theta_1\theta_2\phi_1\phi_2}, G_{\theta_1\theta_2\phi_1\ 10}, G_{\theta_1\theta_2\phi_2\ 10}$ survive.

Now, $\delta\psi_{u,t,x^{1,2,3}}\sim G^{M_1M_2M_3M_4}\Gamma_{t,u,x^{1,2,3}}\Gamma_{M_1M_2M_3M_4}\varepsilon $ (because
$G_{u,t,x^{1,2,3}}^{\ \ \ \ \ M_1M_2M_3}=0$), which given that in (\ref{limits_Dasguptaetal-i}) or (\ref{limits_Dasguptaetal-ii}) and near $\theta_{1,2}=0,\pi$ is approximately diagonal in $u,\theta_{1,2},x_{10}$, is proportional to $G^{\theta_1\theta_2M_1M_2}(\epsilon)\Gamma_{t,u,x^{1,2,3}}\Gamma_{\theta_1\theta_2M_1M_2}\varepsilon
\sim G^{\theta_1\theta_2M_1M_2}(\epsilon)\Gamma_{t,u,x^{1,2,3}}\Gamma_{\theta_1\theta_2}\Gamma_{M_1M_2}\varepsilon, (M_1,M_2,M_3)=(\phi_{1,2},x_{10})$. So, these variations will vanish upon imposing:
\begin{equation}
\label{G4.Gamma4ii}
\sum_{(M_1,M_2)=(\phi_{1,2},x_{10})}G^{\theta_1\theta_2M_1M_2}(\epsilon)\Gamma_{M_1M_2}\varepsilon = 0.
\end{equation}
From (\ref{G4.Gamma4ii}), we see that that $\delta\psi_{\theta_{1,2}}=0$ is automatically satisfied.

Finally, $\delta\psi_{\phi_{1,2},\psi,x_{10}}=0$ will be considered. We will work in the basis $(u,\theta_{1,2},
\phi_{1,2}\equiv \pm\phi_1 - \phi_2,\psi,x_{10})$ with corresponding tangent space indices given by $(4;5,6;7,8;9;10)$. So, we end up with the following set of equations:
\begin{eqnarray}
\label{killing-components_a}
& & \frac{\partial\varepsilon}{\partial\Phi_1} + \beta'G_{\Phi_1}^{\ \  abc}(\epsilon)\Gamma_{abc}\varepsilon = 0,\nonumber\\
& & \frac{\partial\varepsilon}{\partial\Phi_2} + \beta'G_{\Phi_2}^{\ \ abc}(\epsilon)\Gamma_{abc}\varepsilon = 0,\nonumber\\
& & \frac{\partial\varepsilon}{\partial\psi} + \beta'G_{\psi}^{\ \ abc}(\epsilon)\Gamma_{abc}\varepsilon = 0,\nonumber\\
& & \frac{\partial\varepsilon}{\partial x_{10}} + \beta'G_{x_{10}}^{\ \ \ \ abc}(\epsilon)\Gamma_{abc}\varepsilon = 0,
\end{eqnarray}
which utilising the facts that the most dominant contributions of the $G_4$ flux components of the type $G_{\psi\bullet\bullet\bullet}$ are $G_{\psi\Phi_{1,2}\theta_1\theta_2}$ and of the type $G_{x_{10}\bullet\bullet\bullet}$ are $G_{x_{10}\theta_1\theta_2\Phi_{1,2}}$, are respectively equivalent to the following set of equations:
\begin{eqnarray}
\label{killing-components_b}
& &\hskip -0.3in \frac{\partial\varepsilon}{\partial\Phi_1} +\beta g_{\Phi_1\Phi_1}(\epsilon)E^{\Phi_1}_{\ 7}(\epsilon)G^{7568}\Gamma_{568}\varepsilon=\frac{\partial\varepsilon}{\partial\Phi_1} +\beta \epsilon^{-\alpha_\Phi}G^{7568}(\epsilon)\Gamma_{568}\varepsilon=0\nonumber\\
& & \hskip -0.3in \frac{\partial\varepsilon}{\partial\Phi_2} + \beta g_{\Phi_2\Phi_2}(\epsilon)E^{\Phi_2}_{\ 8}(\epsilon)(\epsilon)G^{8567}(\epsilon)\Gamma_{567}(\epsilon)\varepsilon=\frac{\partial\varepsilon}{\partial\Phi_2} + \beta \epsilon^{-\alpha_\Phi}G^{8567}(\epsilon)\Gamma_{567}\varepsilon=0\nonumber\\
& & \hskip -0.3in \frac{\partial\varepsilon}{\partial\psi} + \beta g_{\psi\psi_1}(\epsilon)E^{\psi}_{\ 9}(\epsilon)\left(G^{9567}\Gamma_{567}(\epsilon) + G^{9568}(\epsilon)\Gamma_{568}\right)\varepsilon=\frac{\partial\varepsilon}{\partial\Phi_1} + \beta \epsilon^{-\alpha_\psi}\left(G^{9567}(\epsilon)\Gamma_{567} + G^{9568}(\epsilon)\Gamma_{568}\right)\varepsilon=0,\nonumber\\
& & \hskip -0.3in \frac{\partial\varepsilon}{\partial x_{10}} + \beta g_{x_{10}x_{10}}(\epsilon)E^{x_{10}}_{\ \ \ \overline{10}}(\epsilon)\left(G^{\overline{10}567}(\epsilon)\Gamma_{567} + G^{\overline{10}568}(\epsilon)\Gamma_{568}\right)\varepsilon\nonumber\\
& & \hskip -0.3in =\frac{\partial\varepsilon}{\partial x_{10}} + \beta\epsilon^{\alpha_{10}}\left(G^{\overline{10}567}(\epsilon)\Gamma_{567} + G^{\overline{10}568}(\epsilon)\Gamma_{568}\right)\varepsilon=0.
\end{eqnarray}
 In addition to (\ref{G4.Gamma4ii}), one imposes:
\begin{equation}
\label{Gamma7}
\Gamma_7\varepsilon=\pm\varepsilon;\ \Gamma_8\varepsilon=\pm\varepsilon.
\end{equation}
This implies that (\ref{G4.Gamma4ii}) becomes:
\begin{equation}
\label{G4.Gamma4iii}
\left(G^{5678}(\epsilon)  \pm G^{567\overline{10}}(\epsilon)\Gamma_{\overline{10}} \pm G^{568\overline{10}}(\epsilon)\Gamma_{\overline{10}}\right)\varepsilon=0,
\end{equation}
and the following solution of the killing spinor equation (near $r=r_h, \theta_{1,2}=0,\pi$) is obtained:
\begin{eqnarray}
\label{killing-spinor-solutionii}
& &\hskip -0.3in \varepsilon(\theta_{1,2},\Phi_{1,2},\psi,x_{10})=\nonumber\\
& & \hskip -0.3in e^{\mp\beta\Phi_1\epsilon^{-\alpha_\Phi}G^{5678}(\epsilon)\Gamma_{56}}.e^{\mp\beta\Phi_2\epsilon^{-\alpha_\Phi}G^{5678}(\epsilon)
\Gamma_{56}}.e^{-\beta\psi\epsilon^{-\alpha_\psi}(\mp G^{5679}(\epsilon)\Gamma_{56} \mp G^{5689}(\epsilon)\Gamma_{56})}.e^{x_{10}\epsilon^{\alpha_{10}}(G^{567\overline{10}}(\epsilon)\Gamma_{56} \pm G^{568\overline{10}}(\epsilon)\Gamma_{56})}\varepsilon_0.
\nonumber\\
& &
\end{eqnarray}
So, we obtain, once again, a near-horizon $\frac{1}{8}$-supersymmetric $M3$-brane solution near $\theta_{1,2}=0,\pi$.

 For values of $\theta_{1,2}$ away from 0, $\pi$ and
$r>r_h$, we expect a reduced amount of supersymmetry. In other words we expect a near-horizon
enhancement of supersymmetry (See \cite{enhancement-SUSY-horizon}). We hope to get back to this issue
in a subsequent work.

\section{Hydrodynamics and  Thermodynamics of M-theory Uplift}

In this section, we calculate hydrodynamical as well as thermodynamical quantities of local M-theory uplift obtained in {\bf Section 4} in
the limits (\ref{limits_Dasguptaetal-i}) and (\ref{limits_Dasguptaetal-ii}) of  M theory. From \cite{KovtunSonStarinets}, for black-brane solutions
of the type:
\begin{equation}
\label{KSS-a}
ds^2 = G_{00}(r)dt^2 + G_{rr}dr^2 + G_{xx}(r)\sum_{i=1}^p \left(dx^i\right)^2 + Z(r)K_{mn}(y)dy^i dy^j,
\end{equation}
where in the vicinity of the horizon $r=r_0$, $G_{00}$ vanishes, $G_{rr}$ diverges and $G_{xx}(r)$ and $Z(r)$ remain finite. Then demanding the absence of a conical singularity at the origin, the Hawking temperature associated with (\ref{KSS-a}) is given by (\cite{schmude}):
\begin{equation}
\label{KSS-b}
T=\frac{\partial_r G_{00}}{4\pi \sqrt{G_{00}G_{rr}}}.
\end{equation}
Now, in both limits (\ref{limits_Dasguptaetal-i}) and (\ref{limits_Dasguptaetal-ii}), $G_{00}^{\cal M}, G_{rr}^{\cal M}$ have no angular dependence and hence
the temperature (\ref{KSS-b})  of the black $M3$-brane (\ref{Mtheory met}) then turns out to be given by:
\begin{eqnarray}
\label{T}
& &\hskip-0.2in T =\frac{\sqrt{2}}{{r_h} \sqrt{\pi} \sqrt{\frac{{g_s} \left(18 {g_s}^2 {N_f} ln ^2({r_h}) {M_{\rm eff}}^2+3 {g_s} (4 \pi -{g_s} {N_f}
   (-3+ln (2))) ln ({r_h}) {M_{\rm eff}}^2+8 N \pi ^2\right)}{{r_h}^4}}}\stackrel{(\ref{limits_Dasguptaetal})}{\longrightarrow}\frac{r_h}{\pi L^2}. \nonumber\\
   \end{eqnarray}
To get a numerical estimate for $r_h$,  we see that equating $T$ to $\frac{r_h}{\pi L^2}$, (\ref{T}) is solved, in
both limits (\ref{limits_Dasguptaetal-i}) and the MQGP limit  (\ref{limits_Dasguptaetal-ii}) by $r_h=1+\epsilon$, where $0<\epsilon<1$.

\subsection{ Shear-viscosity-to-entropy-density ratio}

Now, the shear-viscosity-to-entropy-density ratio in the hydrodynamical gravity dual of \cite{KovtunSonStarinets}  is given by:
\begin{equation}
\label{etaovers-i}
\frac{\eta}{s}= T \frac{\sqrt{|G |}}{\sqrt{|G _{00}G _{rr}|}}\Biggr|_{r=r_h}\int_{r_h}^\infty dr \frac{|G_{00}G_{rr}|}{G_{\mathbb{R}^p}\sqrt{|G |}}.
\end{equation}
We first check the estimate evaluate $\eta/s$ for type IIA mirror metric of (\ref{metric-mirror}).
The simplified expressions of relevant metric components and its determinant in the limits (\ref{limits_Dasguptaetal-i}) as well as (\ref{limits_Dasguptaetal-ii}), freezing the angular dependence
on $\theta_{1,2}$ (there being no dependence on $\phi_{1,2},\psi,x_{10}$ in
 (\ref{limits_Dasguptaetal-i}) or (\ref{limits_Dasguptaetal-ii})), are:
  \begin{eqnarray}
  \label{simGIIA}
&&G^{IIA}_{00}\sim \frac{r^2 \left(\frac{{r_h}^4}{r^4}-1\right)}{2 \sqrt{\pi } \sqrt{{g_s} N}} ,\nonumber\\
&& G^{IIA}_{\mathbb{R}^3}\sim \frac{r^2}{2 \sqrt{\pi } \sqrt{{g_s} N}} ,\nonumber\\
&& G^{IIA}_{rr} \sim \frac{2 \sqrt{\pi } \sqrt{{g_s} N}}{r^2 \left(1-\frac{{r_h}^4}{r^4}\right)} ,\nonumber\\
& & \sqrt{|G^{IIA}|}\sim \nonumber\\
&& \left\{\frac{r^3 {f_2}({\theta_2}) (\cos (2 {\theta_1})-5)^3 \left(3 \sqrt{6}-2 \cot ({\theta_1})\right) \cot ({\theta_1}) \csc ^2({\theta_1})
   \sin ^4({\theta_2}) \cos ({\theta_2}) \sqrt{{f_1}({\theta_1})^2 \sin ^2({\theta_1})+1}}{11664 \sqrt{3} \sqrt[4]{\pi } \sqrt[4]{{g_s}}
   \sqrt[4]{N} \left(3 \sin ^2({\theta_1}) \sin ^2({\theta_2})+2 \sin ^2({\theta_1}) \cos ^2({\theta_2})+2 \cos ^2({\theta_1}) \sin
   ^2({\theta_2})\right)^2}\right\} .\nonumber\\
\end{eqnarray}
Again we note that in (\ref{limits_Dasguptaetal-i}) or (\ref{limits_Dasguptaetal-ii}),
$G_{00,rr,\mathbb{R}^3}^{IIA}$ are independent of the angular coordinates;
additionally it was possible to tune the chemical potential $\mu_C$ to a small value - as shown in Sec. {\bf 3}. This permits use of
(\ref{etaovers-i}). Utilizing above,
\begin{eqnarray}
\label{int}
&& \int_{r_h}^\infty dr\frac{|G^{IIA}_{00}G^{IIA}_{rr}|}{G^{IIA}_{\mathbb{R}^3}\sqrt{|G^{IIA}|}} \sim \nonumber\\
&&  \frac{23328 \sqrt{3} \pi ^{3/4} {g_s}^{3/4} N^{3/4}}{4\ {r_h}^4} \times \nonumber\\
&& \left\{\frac{ \sin ^2({\theta_1}) \tan ({\theta_1}) \csc ^4({\theta_2}) \sec ({\theta_2}) \left(3 \sin
   ^2({\theta_1}) \sin ^2({\theta_2})+2 \sin ^2({\theta_1}) \cos ^2({\theta_2})+2 \cos ^2({\theta_1}) \sin ^2({\theta_2})\right)^2}{
   {f_2}({\theta_2}) (\cos (2 {\theta_1})-5)^3 \left(3 \sqrt{6}-2 \cot ({\theta_1})\right) \sqrt{{f_1}({\theta_1})^2 \sin
   ^2({\theta_1})+1}}\right\}. \nonumber\\
&& {\rm and} \nonumber\\
&& \lim_{r\rightarrow r_h+\epsilon}\frac{\sqrt{|G^{IIA}|}}{\sqrt{|G^{IIA}_{tt}G^{IIA}_{rr}|}}\sim \frac{{r_h}^3}{11664 \sqrt{3} \sqrt[4]{\pi } \sqrt[4]{{g_s}}
   \sqrt[4]{N} } \times \nonumber\\
   && \left\{\frac{  {f_2}({\theta_2}) (\cos (2 {\theta_1})-5)^3 \left(3 \sqrt{6}-2 \cot ({\theta_1})\right) \cot ({\theta_1}) \csc ^2({\theta_1})
   \sin ^4({\theta_2}) \cos ({\theta_2}) \sqrt{{f_1}({\theta_1})^2 \sin ^2({\theta_1})+1}}{\left(3 \sin ^2({\theta_1}) \sin ^2({\theta_2})+2 \sin ^2({\theta_1}) \cos ^2({\theta_2})+2 \cos ^2({\theta_1}) \sin
   ^2({\theta_2})\right)^2}\right\}.\nonumber\\
\end{eqnarray}
 Multiplying above expressions as according to equation (\ref{etaovers-i}) and  putting value of $T \sim \frac{1}{\pi \sqrt{4 \pi g_s N}}$ for $r_h \sim 1$, we get
\begin{eqnarray}
\label{eq:etaoversIIA}
\frac{\eta}{s}=\frac{r_h}{\pi \sqrt{4 \pi g_s N}}\times \frac{1}{2}\sqrt{{g_s} N}\sqrt{\pi}=\frac{1}{4 \pi}.
\end{eqnarray}

The simplified expressions of relevant metric components as well as determinant of 11-dimensional metric corresponding to most dominant contribution in the limits (\ref{limits_Dasguptaetal-i})
or (\ref{limits_Dasguptaetal-ii}):
\begin{eqnarray}
&& G^{\cal M}_{00}\sim -\frac{3^{2/3} r^2 \left(1-\frac{r_h^4}{r^4}\right)}{2 \sqrt{\pi } {g_s}^{2/3} \sqrt{ {g_s} N}},\nonumber\\
&& G^{\cal M}_{\mathbb{R}^3}\sim\frac{3^{2/3}r^2}{2 \sqrt{\pi } {g_s}^{2/3} \sqrt{ {g_s} N}},\nonumber\\
&& G^{\cal M}_{rr} \sim \frac{2\ 3^{2/3} \sqrt{\pi } \sqrt{ {g_s} N}}{ {g_s}^{2/3} r^2\left(1-\frac{r_h^4}{r^4}\right)},\nonumber\\
& & \sqrt{|G^{\cal M}|}\sim \frac{4 r^3 {f_2}({\theta_2}) \cos ^2({\theta_1}) \cot ^2({\theta_1}) \cos ^3({\theta_2})}{27\ 3^{5/6} \sqrt[4]{\pi } {g_s}^{8/3}
   \sqrt[4]{{g_s} N} \left(2 \sin ^2({\theta_1}) \cos ^2({\theta_2})+2 \cos ^2({\theta_1}) \sin ^2({\theta_2})\right)}.
\end{eqnarray}
Utilizing above,
\begin{eqnarray}
\label{int}
&& \int_{r_h}^\infty dr \frac{|G^{\cal M}_{00}G^{\cal M}_{rr}|}{G^{\cal M}_{\mathbb{R}^3}\sqrt{|G^{\cal M}|}} \sim \nonumber\\
&& {\hskip -0.3in} \int_{1+\epsilon}^\infty \frac{81 \sqrt{3} \pi ^{3/4} {g_s}^2 \sqrt[4]{{g_s} N} \tan ^2({\theta_1}) \sec ^2({\theta_1}) \sec ^3({\theta_2}) \sqrt{ {g_s} N}
   \left( \sin ^2({\theta_1}) \cos ^2({\theta_2})+ \cos ^2({\theta_1}) \sin ^2({\theta_2})\right)}{2 r^5 {f_2}({\theta_2})} \,dr,\   \nonumber\\
   && \sim  \frac{81 \sqrt{3} \pi ^{3/4} {g_s}^2 \sqrt[4]{{g_s} N} \tan ^2({\theta_1}) \sec ^2({\theta_1}) \sec ^3({\theta_2}) \sqrt{ {g_s} N}
   \left(\sin ^2({\theta_1}) \cos ^2({\theta_2})+ \cos ^2({\theta_1}) \sin ^2({\theta_2})\right)}{4 {f_2}({\theta_2})}. \nonumber\\
&& {\rm and} \nonumber\\
&& \lim_{r\rightarrow r_h+\epsilon}\frac{\sqrt{|G^{\cal M}|}}{\sqrt{|G^{\cal M}_{00}G^{\cal M}_{rr}|}}\sim  \frac{4 {f_2}({\theta_2}) \cos ^2({\theta_2}) \cot ^2({\theta_2}) \cos ^3({\theta_2})}{81 \sqrt{3} \sqrt[4]{\pi } \sqrt[4]{{g_s} N} \left(2 \sin
   ^2({\theta_2}) \cos ^2({\theta_2})+2 \cos ^2({\theta_2}) \sin ^2({\theta_2})\right)}.
\end{eqnarray}
 Multiplying above expressions as according to equation (\ref{etaovers-i}) and  putting value of $T \sim \frac{1}{\sqrt{4 \pi g_s N}}$ for $r_h \sim 1$, we get
\begin{eqnarray}
\frac{\eta}{s}=\frac{r_h}{\pi \sqrt{4 \pi g_s N}}\times \frac{1}{2}\sqrt{{g_s} N}\sqrt{\pi}=\frac{1}{4 \pi}.
\end{eqnarray}
Needless to say, as mentioned earlier, given that $G_{00,rr,\mathbb{R}^3}^{IIA, {\cal M}}$ in both limits (\ref{limits_Dasguptaetal-i}) and (\ref{limits_Dasguptaetal-ii})
are independent of the angular coordinates and that one integrates only w.r.t. $r$ in (\ref{etaovers-i}), the angular portions of the metric's
determinants is bound to (and is explicitly verified above) to cancel out.

\subsection{Diffusion Coefficient }

The general expression of Diffusion coefficient in the hydrodynamical gravity dual of \cite{KovtunSonStarinets} is given by:
\begin{equation}
\label{D}
D= \frac{\sqrt{|G|}Z(r)}{G\sqrt{|G_{00}G_{rr}|}}\Biggr|_{r=r_h}\int_{r_h}^\infty dr
\frac{|G_{00}G_{rr}|}{\sqrt{|G|}Z(r)}.
\end{equation}
where $Z(r) =\sqrt{h}r^2$ and $\sqrt{h}$ is warp factor in type IIB/IIA background. Justifications
similar to the ones given in {\bf 5.1} also permit the use of (\ref{D}).

Here, we first check the estimate of diffusion coefficient in type IIB background using type IIB metric given in (\ref{metric}).
The simplified expressions of relevant metric components and determinant of  using the limits of (\ref{limits_Dasguptaetal}) are:
\begin{eqnarray}
&& g_{00}^{IIB}\sim -\frac{r^2 \left(1-\frac{{r_h}^4}{r^4}\right)}{2 \sqrt{\pi } \sqrt{{g_s} N}},\nonumber\\
&& g_{rr}^{IIB} \sim \frac{2 \sqrt{\pi } \sqrt{{g_s} N}}{r^2 \left(1-\frac{{r_h}^4}{r^4}\right)} ,\nonumber\\
& & \sqrt{|g^{IIB}|}\sim \frac{r^6 \left(1-{f_2}({\theta_2})^2 \left(\sin ^2({\theta_2})-\sin ^2({\theta_1})\right)\right)}{8 \sqrt{\pi } \sqrt{{g_s} N}}.
\end{eqnarray}
Incorporating above-mentioned results in equation (\ref{D}) and further simplifying, we get
\begin{equation}
\label{D1}
D= \frac{\sqrt{2} \pi ^{3/4} {r_h} ({g_s} N)^{3/4}}{ {\cal A}(\theta_1,\theta_2)}\int_{r_h}^\infty {\cal A}(\theta_1,\theta_2) \frac{\sqrt{2}}{\sqrt[4]{\pi } r^3 \sqrt[4]{{g_s} N}} dr.
\end{equation}
where
\begin{eqnarray}
{\cal A}(\theta_1,\theta_2)\sim \sqrt{1-{f_2}({\theta_2})^2 \left(\sin ^2({\theta_2})-\sin ^2({\theta_1})\right)}.\nonumber
\end{eqnarray}
On solving equation (\ref{D1}), one gets:
\begin{equation}
D = \frac{\sqrt{ {g_s} N} \sqrt{\pi }}{r_h}=\frac{1}{2 \pi T}.
\end{equation}
Similarly, for type IIA background formed by applying transformation rules  on (\ref{metric}), the simplified expressions of relevant metric components in the limits of (\ref{limits_Dasguptaetal}) are already given in equation (\ref{simGIIA}). Incorporating the same in equation (\ref{D}) and further simplifying, we have
\begin{equation}
\label{D2}
D= \frac{5832  \pi ^{3/4} {r_h} ({g_s} N)^{3/4}}{ {\cal A^{\prime}}(\theta_1,\theta_2)}\int_{r_h}^\infty {\cal A^{\prime}}(\theta_1,\theta_2) \frac{1}{2916 \sqrt[4]{\pi } r^3 \sqrt[4]{{g_s} N}} dr.
\end{equation}
where
\begin{eqnarray}
\hskip -0.2in {\cal A^{\prime}}(\theta_1,\theta_2)\sim \left\{\frac{\sin ^2({\theta_1}) \tan ({\theta_1}) \csc ^4({\theta_2}) \sec ({\theta_2}) \left(3 \sin ^2({\theta_1}) \sin ^2({\theta_2})+2 \sin
   ^2({\theta_1}) \cos ^2({\theta_2})+2 \cos ^2({\theta_1}) \sin ^2({\theta_2})\right)^2}{{f_2}({\theta_2}) (\cos (2 {\theta_1})-5)^3 \left(3
   \sqrt{6}-2 \cot ({\theta_1})\right) \sqrt{{f_1}({\theta_1})^2 \sin ^2({\theta_1})+1}}\right\}.\nonumber
\end{eqnarray}
After solving equation (\ref{D2}), one gets:
\begin{equation}
D = \frac{\sqrt{ {g_s} N} \sqrt{\pi }}{r_h}=\frac{1}{2 \pi T}.
\end{equation}

\subsection{Partition Function}

 To study, further, the thermodynamic properties of the solution (\ref{Mtheory met}), we need to
evaluate the partition
function ${\cal Z} = e^{- \cal{S}_E}$, where keeping in mind  $l_s \sim r_h$,
 higher order $\alpha^{\prime}$ corrections become important, the action we will consider
will include ${\cal O}(R^4)$-terms (See \cite{d_11_R_fourth}):
\begin{eqnarray}
  \label{eq:GravitationalAction}
  {\cal S_{\cal E}} & = & \frac{1}{16\pi}\int_{\cal{M}} d^{11}\!x \sqrt{G^{\cal M}} R^{\cal M}
   + \frac{1}{8\pi}\int_{\partial M} d^{10}\!x K^{\cal M} \sqrt{\hat{h}}-\frac{1}{4}\int_{M} \Bigl( |G_4|^2-{C_3\wedge G_4\wedge G_4}\bigr) \nonumber\\
& & \frac{T_2}{{2\pi}^4. 3^2.2^{13}}\int_{\cal{M}} d^{11}\!x \sqrt{G^{\cal M}}(J-\frac{1}{2}E_8)+T_2 \int C_3 \wedge X_8 - {\cal S}^{\rm ct},
\end{eqnarray}
where $T_2$ is $M2$-brane tension, while $(J,E_8,X_8)$ are quartic polynomials in 11-dimensional space and defined as:
\begin{equation}
\label{J}
J = 3.2^8 \Bigl(R^{mijn}R_{pijq} R_{m}^{\ rsp}R^{q}_{\ rsn}+\frac{1}{2}R^{mnij}R_{pqij} R_{m}^{\ rsp}R^{q}_{\ rsn}\Bigr),
\end{equation}
\begin{equation}
\label{E_8}
E_8= \epsilon^{abc m_1 n_1...m_4 n_4} \epsilon_{abc m^{\prime}_1 n^{\prime}_1...m^{\prime}_4 n^{\prime}_4}R^{m^{\prime}_1 n^{\prime}_1}_{\ \ \ \ \ \ m_1 n_1}...R^{m^{\prime}_4 n^{\prime}_4}_{\ \ \ \ \ \ m_4 n_4},
\end{equation}
   \begin{equation}
 X_8 \sim \frac{1}{192 \cdot (2 \pi^2)^4}\Bigl[tr(R^4)- (tr{R^2})^2\Bigr],
 \end{equation}
for Euclideanised space-time where $\cal{M}$ is a volume of spacetime defined by $r <r_\Lambda$, where the counter-term ${\cal S}^{\rm ct}$ is added such that the Euclidean action ${\cal S}_{\cal E}$ is finite \cite{S_ct-Perry et al},\cite{Mann+Mcnees}.  The action (\ref{eq:GravitationalAction}), apart from being IR-divergent (as $r\rightarrow\infty$) also possesses pole-singularities near $\theta_{1,2}=0,\pi$. We will regulate the second divergence by taking a
small $\theta_{1,2}$-cutoff $\epsilon_\theta$, $\theta_{1,2}\in[\epsilon_\theta,\pi-\epsilon_\theta]$, and demanding $\epsilon_\theta
\sim\epsilon^\gamma$, for an appropriate $\gamma$.  We will then explicitly check that the finite part
of (\ref{eq:GravitationalAction}) turns out to be independent of this cut-off $\epsilon/\epsilon_\theta$.

For M-theory thermodynamical calculations, we provide a slightly more detailed explanation of
the limits (\ref{limits_Dasguptaetal-i}) and (\ref{limits_Dasguptaetal-ii}):
\begin{itemize}
\item   In weak coupling  - large t'Hooft coupling(s) limit of M theory, we consider the limits:
$$g_s<<1, g_s M>>1, g_s N>>, \frac{g_s M^2}{N}<<1, {g_s}^2 M N_f<<1$$  similar to \cite{metrics}. Based on that, we assume that these parameters scale with $\epsilon$ as
  $g_s \sim {\epsilon^{d}} $, $N/N_{\rm eff}\sim  {\epsilon^{-a}} $, $M/M_{\rm eff}\sim{\epsilon^{-b}}$
 such that  ${2b-a-d}<0, {2d-b}<0, {b-d}>0,{a-d}>0$, $\forall$ $ \epsilon\rightarrow 0 $, $a,b,d > 0 $.  Further, for
(\ref{limits_Dasguptaetal-i}),  we choose $a=19 d, {b}={\frac{3}{2}}d, \gamma={\frac{1}{2}}d $. For the purpose of obtaining
the simplified type IIA mirror components (\ref{metric-mirror}) and its M-theory uplift simplified metric components of
(\ref{eq:simplifiedmetriccomponents}), we had set $\epsilon\leq0.01$.  However, it is understood
that in the identification $\epsilon_\theta=\epsilon^\gamma$, one will eventually have to take $\epsilon_\theta$ and
thus $\epsilon$ to be very small.
 \item
In MQGP limit:
$$g_s, g_s M, {g_s}^2 M N_f\equiv{\rm finite}, g_s N>>1, \frac{g_s M^2}{N}<<1.$$
 Now we assume the scaling of these parameters with $\epsilon$ as
 $g_s \sim {\alpha_1} {\epsilon^{d}} $, $N/N_{\rm eff}\sim\alpha_2 {\epsilon^{-a}} $, $M/M_{\rm eff}\sim\alpha_3{\epsilon^{-b}}$,  such that  ${2b-a-d}<0, {2d-b}<0, {b-d}>0,{a-d}>0$, $ a,b,d > 0 $. To obey these constraints,  we define $a=39 d, {b}={\frac{3}{2}}d, \gamma={\frac{3}{2}}d $. For the purpose of obtaining
the simplified type IIA mirror components (\ref{metric-mirror}) and its M-theory uplift simplified metric components of
(\ref{eq:simplifiedmetriccomponents}), we had taken $\alpha_{1,2,3}\sim{\cal O}(1), \epsilon\lesssim 1$. However, it is understood
that in the identification $\epsilon_{\theta_{1,2}}\sim\epsilon^\gamma$ given that one will eventually have to take $\epsilon_\theta$ and
thus $\epsilon$ to be very small, this would numerically imply taking large $\alpha_1$ and small $\alpha_3$ such that $g_s, M$ are
individually finite in the MQGP limit. Further, writing $\epsilon_{\theta_1}=\alpha_4\epsilon^\gamma, \epsilon_{\theta_2}=\alpha_5\epsilon^\gamma$, one will see in {\bf 5.3.1 - 5.3.4} that the Gibbons-Hawking-York surface term will be the only term containing an IR-finite contribution which would be proportional to $\frac{1}{\alpha_1^{\frac{9}{4}}\alpha_2^{\frac{1}{4}}\alpha_4^4\alpha_5}$. In the
$\epsilon_{\theta_{1,2}}\rightarrow0$-limit one will take $\alpha_{1,3,4,5}:\alpha_1^{\frac{9}{4}}\alpha_4^4\alpha_5\equiv$finite; $\alpha_2\sim{\cal O}(1)$ always and the simplified (\ref{metric-mirror})
and (\ref{eq:simplifiedmetriccomponents}) continue to be valid.

\end{itemize}

\subsubsection{Einstein-Hilbert Action}

We now evaluate the contribution of Einstein-Hilbert action in both limits
(\ref{limits_Dasguptaetal-i}) and (\ref{limits_Dasguptaetal-ii}) of M theory. In either limit,
the simplified expression  of determinant of 11 dimensional M theory uplift is as follows:
{\small
\begin{eqnarray}
\label{eq:g}
&& {G^{\cal M}}\sim \Biggl(r^6 \cos ^2({\theta_2}) \csc ^2({\theta_1}) {f_2}({\theta_2})^2 \left({f_1}({\theta_1})^2 \sin ^2({\theta_1})+1\right) \sin
   ^2({\theta_2}) \Bigl(\Bigl(4 \cos ({\theta_2}) \left(2 \cot ({\theta_1})-3 \sqrt{6}\right) \cot ({\theta_2})\nonumber\\
   && +18 \sqrt{6} \sin ({\theta_2})\Bigr)
   \cos ^3({\theta_1})+12 \cos ({\theta_2}) \cot ({\theta_2}) \sin ({\theta_1}) \cos ^2({\theta_1})+27 \sin ^2({\theta_1}) \left(\sqrt{6} \sin
   ({\theta_2})-4 {h_5} \cos ({\theta_2})\right)\nonumber\\
   && \cos ({\theta_1})  +81 {h_5} \sin ^3({\theta_1}) \left(\sqrt{6} \cos ({\theta_2})+3 {h_5}
   \sin ({\theta_2})\right)\Bigr)^2\Biggr)/\Biggl(8748 3^{2/3} {g_s}^{16/3} \sqrt{{g_s} N} \sqrt{\pi } \Bigl(2 \cos ^2({\theta_2}) \sin ^2({\theta_1}) \nonumber\\
   && +3 \sin
   ^2({\theta_2}) \sin ^2({\theta_1})+2 \cos ^2({\theta_1}) \sin ^2({\theta_2})+3 {h_5} \sin (2 {\theta_1}) \sin (2 {\theta_2})\Bigr)^2\Biggr).
\end{eqnarray}}
Ricci scalar is given as:
\begin{equation}
\label{eq:R}
R= G^{MN} G^{PQ} R_{M  P N Q}\ {\rm where}\ M, N, P, Q = 0 ,..10.
\end{equation}
For particular choice's of scaling parameters, we see that the contribution of Ricci scalar is dominated by $G^{\cal M}\ ^{\theta_1 \theta_1} G^{\cal M}\ ^{x z} R^{\cal M}_{ x \theta_1 z \theta_1}$ component. On simplifying the same using equation (\ref{eq:simplifiedmetriccomponents}), we have
{\small
\begin{eqnarray}
\label{eq:invYtheta2}
&& \hskip -0.2in G^{\cal M}\ ^{x z}\sim \Bigl(27 \sqrt[3]{3} {g_s}^{2/3} (\cos (2 {\theta_1})-5) \sin ({\theta_1}) \sin ({\theta_2})\Bigr)/\Biggl(2 \Bigl(\Bigl(4 \cos ({\theta_2}) \left(2 \cot
   ({\theta_1})-3 \sqrt{6}\right)  \cot ({\theta_2})+18 \sqrt{6} \sin ({\theta_2})\Bigr) \nonumber\\
   && \cos ^3({\theta_1})+12 \cos ({\theta_2}) \cot
   ({\theta_2}) \sin ({\theta_1}) \cos ^2({\theta_1})+27 \sin ^2({\theta_1}) \left(\sqrt{6} \sin ({\theta_2})-4 {h_5} \cos ({\theta_2})\right)
   \cos ({\theta_1})\nonumber\\
   && +81 {h_5} \sin ^3({\theta_1}) \left(\sqrt{6} \cos ({\theta_2})+3 {h_5} \sin ({\theta_2})\right)\Bigr)\Biggr).
\end{eqnarray}
Also,
\begin{eqnarray}
\label{eq:invtheta1theta1}
&& \hskip -3.8in G^{\cal M}\ ^{\theta_1 \theta_1}\sim \frac{\sqrt[3]{3} \sqrt[6]{{g_s}}}{\sqrt{N} \sqrt{\pi } \left({f_1}({\theta_1})^2 \sin ^2({\theta_1})+1\right)}.
\end{eqnarray}}
and
{\small
\begin{eqnarray}
\label{eq:R7696}
 &&  R_{x \theta_1 z \theta_1}^{\cal M} \sim -\Biggl(2 \csc ({\theta_1}) \csc ({\theta_2}) \Bigl(12 \cos ({\theta_2}) \cot ({\theta_2}) \cos ^3({\theta_1})-36 \cot ({\theta_2}) \sin
   ({\theta_1}) \left(\sqrt{6} \cos ({\theta_2})-6 {h_5} \sin ({\theta_2})\right) \nonumber\\
    && \cos ^2({\theta_1})+\left(\cos ({\theta_2}) \left(8 \cot
   ({\theta_2}) \cot ^2({\theta_1})+3 \left(8 \cot ({\theta_2})-81 \sqrt{6} {h_5}\right) \sin ^2({\theta_1})\right)-729 {h_5}^2 \sin
   ^2({\theta_1}) \sin ({\theta_2})\right) \nonumber\\
   && \cos ({\theta_1})+27 \sin ^3({\theta_1}) \left(\sqrt{6} \sin ({\theta_2})-4 {h_5} \cos
   ({\theta_2})\right)\Bigr)^2\Biggr)/\Biggl(27 \sqrt[3]{3} {g_s}^{2/3} (\cos (2 {\theta_1})-5) \Bigl(2 \Bigl(2 \cos ({\theta_2}) \nonumber\\
   && \left(3 \sqrt{6}-2 \cot
   ({\theta_1})\right) \cot ({\theta_2})-9 \sqrt{6} \sin ({\theta_2})\Bigr) \cos ^3({\theta_1})-12 \cos ({\theta_2}) \cot ({\theta_2}) \sin
   ({\theta_1}) \cos ^2({\theta_1}) \nonumber\\
   && -27 \sin ^2({\theta_1}) \left(\sqrt{6} \sin ({\theta_2})-4 {h_5} \cos ({\theta_2})\right) \cos
   ({\theta_1})-81 {h_5} \sin ^3({\theta_1}) \left(\sqrt{6} \cos ({\theta_2})+3 {h_5} \sin ({\theta_2})\right)\Bigr)\biggr).
\end{eqnarray}}
Using set of equations (\ref{eq:g})- (\ref{eq:R7696}), the most dominant contribution of 11-Dimensional Bulk term is given by following  analytical expression
{\small
\begin{eqnarray}
&&   \sqrt{G^{\cal M}} R^{\cal M}\sim \Biggl(r^3 \cos ({\theta_2}) \csc ({\theta_1}) {f_2}({\theta_2}) \sin ({\theta_2}) \Bigl(12 \cos ({\theta_2}) \cot ({\theta_2}) \cos
   ^3({\theta_1})-36 \cot ({\theta_2}) \sin ({\theta_1}) \Bigl(\sqrt{6} \cos ({\theta_2})\nonumber\\
   && -6 {h_5} \sin ({\theta_2})\Bigr) \cos
   ^2({\theta_1})+\Bigl(\cos ({\theta_2}) \left(8 \cot ({\theta_2}) \cot ^2({\theta_1})+3 \left(8 \cot ({\theta_2})-81 \sqrt{6} {h_5}\right) \sin
   ^2({\theta_1})\right) \nonumber\\
   &&-729 {h_5}^2 \sin ^2({\theta_1}) \sin ({\theta_2})\Bigr) \cos ({\theta_1})+27 \sin ^3({\theta_1}) \left(\sqrt{6} \sin
   ({\theta_2})-4 {h_5} \cos ({\theta_2})\right)\Bigr)^2\Biggr)/\Biggl(54 {g_s}^{11/4} N^{3/4} \pi ^{3/4}\nonumber\\
   &&  \sqrt{3 {f_1}({\theta_1})^2 \sin
   ^2({\theta_1})+3} \Bigl(\left(4 \cos ({\theta_2}) \left(2 \cot ({\theta_1})-3 \sqrt{6}\right) \cot ({\theta_2})+18 \sqrt{6} \sin
   ({\theta_2})\right) \cos ^3({\theta_1})+12 \cos ({\theta_2})\nonumber\\
    && \cot ({\theta_2}) \sin ({\theta_1}) \cos ^2({\theta_1})+27 \sin ^2({\theta_1})
   \left(\sqrt{6} \sin ({\theta_2})-4 {h_5} \cos ({\theta_2})\right) \cos ({\theta_1})+81 {h_5} \sin ^3({\theta_1}) \nonumber\\
   &&  \Bigl(\sqrt{6} \cos
   ({\theta_2})  +3 {h_5} \sin ({\theta_2})\Bigr)\Bigr) \Bigl(2 \cos ^2({\theta_2}) \sin ^2({\theta_1})+3 \sin ^2({\theta_2}) \sin
   ^2({\theta_1})+2 \cos ^2({\theta_1}) \sin ^2({\theta_2})\nonumber\\
   && +3 {h_5} \sin (2 {\theta_1}) \sin (2 {\theta_2})\Bigr)\Biggr).
   \end{eqnarray}}
The Einstein-Hilbert action receiving the most dominant contribution near $\theta_{1,2}=0,\pi$, we simplify the above near the same and obtain:
{\small
 \begin{eqnarray}
   && \sqrt{G^{\cal M}} R^{\cal M}\sim \frac{125  r^3 \cos ({\theta_2}) \cot ^2({\theta_2}) \csc ^4({\theta_1}) {f_2}({\theta_2})}{864 \sqrt{3} {g_s}^{11/4} N^{3/4} \pi ^{3/4}}.
 \end{eqnarray}}
We assume that result of
integration with respect to ${\theta_{1,2}}$ variables, is simply given by summing up the contribution of integrand near ${\theta_{1,2}}=\epsilon_{\theta_{1,2}}$ and  ${\theta_{1,2}}= \pi-\epsilon_{\theta_{1,2}}$.
Integrating other angular as well as radial variables, we have
{\small
\begin{eqnarray}
\label{eq:intSbulk}
&&  \frac{1}{16 \pi} \int_{x_{10}\in[0,{2\pi}], r\in[r_h,r_\Lambda],\theta_{1,2}\in[\epsilon_{\theta_{1,2}},\pi-\epsilon_{\theta_{1,2}}],\phi_{1,2}\in[0,2\pi], \psi\in[0,4\pi]}\sqrt{G^{\cal M}} R^{\cal M} \nonumber\\
&\sim&    \frac{125 {\pi}^{\frac{9}{4}} r^{4}_{\Lambda} \theta_1 \theta_2 \cos ({\theta_2}) \cot ^2({\theta_2}) \csc ^4({\theta_1}) {f_2}({\theta_2})}{1728 \sqrt{3} {g_s}^{11/4} {N}^{3/4}}\Biggr|_{\theta_{1,2}=\epsilon_{\theta_{1,2}} + \theta_{1,2}=\pi-\epsilon_{\theta_{1,2}}}.
\end{eqnarray}}
\begin{itemize}
\item  Limit (\ref{limits_Dasguptaetal-i}) :
Incorporating   $g_s \sim \ {\epsilon}$, $N/N_{\rm eff}\sim \ {\epsilon^{-19}} $, $M/M_{\rm eff}\sim {\epsilon^{-\frac{3}{2}}}$
$\theta_{1,2}\sim   {\epsilon^{\frac{5}{6}}}$, ${f_2}({{\theta_{1,2}}})\sim { \frac{1}{{\theta_{1,2}}}}$, we see that
\begin{eqnarray}
\label{eq:intS_EH1}
  && {\hskip -0.6in} { {\cal S}_{\rm EH}}\sim \frac{1}{16 \pi} \int_{x^0
\in \left[0,\frac {\pi \sqrt{4\pi g_s N}}{r_h}\right],x_{10}\in[0,{2\pi}], r\in[r_h,r_\Lambda],\theta_{1,2}\in[\epsilon_{\theta_{1,2}},\pi-\epsilon_{\theta_{1,2}}],\phi_{1,2}\in[0,2\pi], \psi\in[0,4\pi]} \sqrt{G^{\cal M}} R^{\cal M} \nonumber\\
 &&=a_{\rm EH}{\frac{r^{4}_{\Lambda}}{\epsilon^{\frac{5}{3}} r_h}}.
  \end{eqnarray}

  \item MQGP limit (\ref{limits_Dasguptaetal-ii}):
  Incorporating  $g_s={\alpha_1}\epsilon $, $N/N_{\rm eff}=\alpha_2  {\epsilon^{-39}} $, $M/M_{\rm eff}=\alpha_3{\epsilon^{-\frac{3}{2}}}$,   $\theta_{1,2}=\alpha_{4,5} {\epsilon^{\frac{15}{6}}}$, ${f_2}(\epsilon_{\theta_{1,2}})\sim  \frac{1}{\epsilon_{\theta_{1,2}}}$,
\begin{eqnarray}
\label{eq:intS_EH2}
  && {\hskip -0.6in} { {\cal S}_{\rm EH}}\sim \frac{1}{16 \pi} \int_{x^0
\in \left[0,\frac {\pi \sqrt{4\pi g_s N}}{r_h}\right],x_{10}\in[0,{2\pi}], r\in[r_h,r_\Lambda],\theta_{1,2}\in[\epsilon_{\theta_{1,2}},\pi-\epsilon_{\theta_{1,2}}],\phi_{1,2}\in[0,2\pi], \psi\in[0,4\pi]}\sqrt{G^{\cal M}} R^{\cal M}\nonumber\\
 &=& a_{EH}(\alpha_{1,3,4,5})\frac{r^{4}_{\Lambda}}{\epsilon^5 r_h}.
  \end{eqnarray}

 \end{itemize}
Therefore, the contribution corresponding to ${S_{\rm EH}}$ is divergent as $r_\Lambda$ becomes large and $\epsilon$ becomes small.

\subsubsection{Gibbons-Hawking-York Surface Action}

Similarly, using equations (\ref{Mtheory met}) and (\ref{eq:simplifiedmetriccomponents}), simplified form of the Gibbons-Hawking-York surface action will be given as under:
{\small
\begin{eqnarray}
\label{eq:Gibbons_Hawking_i}
& & \hskip -0.3in K^{\cal M} \sqrt{\hat{h}}\sim \Biggl(4 r^4 \left(1-\frac{r_h^4}{r^4}\right)\cos ({\theta_2}) \csc ({\theta_1}) {f_2}({\theta_2}) \sqrt{{f_1}({\theta_1})^2 \sin ^2({\theta_1})+1} \sin ({\theta_2})
   \Bigl(\Bigl(4 \cos ({\theta_2}) \left(2 \cot ({\theta_1})-3 \sqrt{6}\right) \cot ({\theta_2})\nonumber\\
   && \hskip -0.3in +18 \sqrt{6} \sin ({\theta_2})\Bigr)   \cos
   ^3({\theta_1})+12 \cos ({\theta_2}) \cot ({\theta_2}) \sin ({\theta_1}) \cos ^2({\theta_1})+27 \sin ^2({\theta_1}) \left(\sqrt{6} \sin
   ({\theta_2})-4 {h_5} \cos ({\theta_2})\right) \nonumber\\
   &&\hskip -0.3in \cos ({\theta_1})  +81 {h_5} \sin ^3({\theta_1}) \left(\sqrt{6} \cos ({\theta_2})+3 {h_5}
   \sin ({\theta_2})\right)\Bigr)\Biggr)/\Biggl(324 \sqrt{3} {g_s}^{11/4} N^{3/4} \pi ^{3/4} \Bigl(2 \cos ^2({\theta_2}) \sin ^2({\theta_1}) \nonumber\\
   &&\hskip -0.3in +3 \sin
   ^2({\theta_2}) \sin ^2({\theta_1})+2 \cos ^2({\theta_1}) \sin ^2({\theta_2})+3 {h_5} \sin (2 {\theta_1}) \sin (2 {\theta_2})\Bigr)\Biggr),
\end{eqnarray}}
where
   $K^{\cal M}\equiv\frac{1}{2}\sqrt{G^{\cal M}\ ^{rr}}\frac{\partial_r {\rm det} h_{ab}}{{\rm det} h_{ab}}\Biggr|_{r=r_\Lambda} $, $h_{ab}$ is the pull-back of $G^{\cal M}_{MN}$ on to $r=r_\Lambda$ where $M=(r,a)$.
   Further simplifying above near $\theta_{1,2}=0,\pi$, the analytical expression reduces to
   {\small
   \begin{eqnarray}
   \label{eq:Gibbons_Hawking ii}
 &&  K^{\cal M} \sqrt{\hat{h}} \sim \frac{ 4 \left(1-\frac{r_h^4}{r^4_{\Lambda}}\right) r^4 \cos ^2({\theta_1}) \cos ^3({\theta_2}) \cot ^2({\theta_1}) {f_2}({\theta_2})}{81 \sqrt{3} {g_s}^{11/4} N^{3/4} \pi ^{3/4} \left(
   \cos ^2({\theta_2}) \sin ^2({\theta_1})+  \cos ^2({\theta_1}) \sin ^2({\theta_2})\right)}.
      \end{eqnarray}}
Utilizing the same approach as used in equation (\ref{eq:intSbulk}) and integrating, we have,
 \begin{eqnarray}
  & &  \frac{1}{8 \pi} \int_{x_{10}\in[0,{2\pi}], \theta_{1,2}\in[0,\pi],\phi_{1,2}\in[0,2\pi], \psi\in[0,4\pi]} K^{\cal M}\sqrt{\hat h} \Biggr|_{r=r_\Lambda}  \nonumber\\
   & \sim & \frac{16 \pi ^{9/4} \left(1-\frac{r_h^4}{r^4_{\Lambda}}\right) {r^4_{\Lambda}} \ \theta_1 \theta_2  \cos ^2({\theta_1}) \cos ^3({\theta_2}) \cot ^2({\theta_1}) {f_2}({\theta_2})}{81\sqrt{3} {g_s}^{11/4} N^{3/4} \left(2
   \cos ^2({\theta_2}) \sin ^2({\theta_1})+2 \cos ^2({\theta_1}) \sin ^2({\theta_2})\right)}\Biggr|_{\theta_{1,2}=\epsilon_{\theta_{1,2}} + \theta_{1,2}=\pi-\epsilon_{\theta_{1,2}}}.
  \end{eqnarray}
  \begin{itemize}
\item  Limit (\ref{limits_Dasguptaetal-i}) :
Incorporating   $g_s \sim  {\epsilon} $, $N/N_{\rm eff}\sim  {\epsilon^{-19}} $, $M/M_{\rm eff}\sim  {\epsilon^{-\frac{3}{2}}}$ $\theta_{1,2 }\sim  {\epsilon^{\frac{5}{6}}}$,  ${f_2}({{\theta_{1,2}}})\sim { \frac{1}{{\theta_{1,2}}}}$, we see that:
\begin{eqnarray}
\label{eq:Gibbons_Hawking iii}
{\cal S}_{\rm GHY-boundary}&\sim& \frac{1}{8 \pi} \int_{x^0
\in \left[0,\frac {\pi \sqrt{4\pi g_s N}}{r_h}\right],x_{10}\in[0,{2\pi}], ,\theta_{1,2}\in[\epsilon_{\theta_{1,2}},\pi-\epsilon_{\theta_{1,2}}],\phi_{1,2}\in[0,2\pi], \psi\in[0,4\pi]}  K^{\cal M} \sqrt{\hat{h}}  \Biggr|_{r=r_\Lambda}\nonumber\\
 &\sim&(+{\rm ive})\bigl(\frac{ {r^{4}_\Lambda}-  {r_h}^4}{r_h}\bigr).
  \end{eqnarray}
So,
  ${\cal S}_{\rm GHY-boundary}^{\rm finite}\sim -{r_h}^3$ and ${\cal S}_{\rm GHY-boundary}^{\rm Infinite}=a_{{\rm GHY-boundary}}\frac{{r^{4}_\Lambda}}{r_h}$.

  \item MQGP limit (\ref{limits_Dasguptaetal-ii}):
  Incorporating  $g_s \sim \ {\epsilon} $, $N/N_{\rm eff}\sim   {\epsilon^{-39}} $, $M/M_{\rm eff}\sim  {\epsilon^{-\frac{3}{2}}}$,   $\epsilon_{\theta_{1,2}}\sim\epsilon^{\frac{15}{6}}$, ${f_2}({{\theta_{1,2}}})\sim { \frac{1}{{\theta_{1,2}}}}$,
\begin{eqnarray}
\label{eq:Gibbons_Hawking iv}
{\cal S}_{\rm GHY-boundary}&\sim & \frac{1}{8 \pi} \int_{x^0
\in \left[0,\frac {\pi \sqrt{4\pi g_s N}}{r_h}\right],x_{10}\in[0,{2\pi}], ,\theta_{1,2}\in[\epsilon_{\theta_{1,2}},\pi-\epsilon_{\theta_{1,2}}],\phi_{1,2}\in[0,2\pi], \psi\in[0,4\pi]}  K^{\cal M} \sqrt{\hat{h}}  \Biggr|_{r=r_\Lambda}\nonumber\\
&& \sim (+{\rm ive})  \left( \frac{r^{4}_\Lambda-  {r_h}^4}{r_h}\right).
  \end{eqnarray}
  Further, for $\epsilon_\theta<<1$, writing $g_s=\alpha_1\epsilon, N=\alpha_2\epsilon^{-39}, M=\alpha_3\epsilon^{-\frac{3}{2}}, \epsilon_{\theta_1}=\alpha_4\epsilon^\gamma, \epsilon_{\theta_2}=\alpha_5\epsilon^\gamma$, one can show that
   (\ref{eq:Gibbons_Hawking iv}) will be proportional to $\frac{1}{\alpha_1^{\frac{9}{4}}\alpha_2^{\frac{1}{4}}\alpha_4^4\alpha_5}$.  We will take the large $\alpha_1$ and the small $\alpha_{4,5} :\alpha_1^{\frac{9}{4}}\alpha_4^4\alpha_5\equiv$finite; $\alpha_2$ is always finite.
Therefore, ${\cal S}_{\rm GHY-boundary}^{\rm Infinite}\sim a_{{\rm GHY-boundary}}(\alpha_{1,2,4,5})\frac{{r^{4}_\Lambda}}{r_h}$,  and ${\cal S}_{\rm GHY-boundary}^{\rm IR-finite}\sim - r_h^3$, i.e., {\it independent of the cut-off $\epsilon_{\theta_{1,2}}/\epsilon$}.
\end{itemize}

\subsubsection{Flux Action}

Now,  $G_4=d C_3 + A_1 \wedge d B_2 + dx_{10}\wedge dB_2$, and $C_{\mu \nu 10}^M = B_{\mu \nu}^{IIA}, C_{\mu \nu \rho}^M = C_{\mu \nu \rho}^{IIA}$. Now, $F_4^{IIA}$ will be obtained via a triple
T-dual of type IIB $F_{1,3,5}$ where $F_1\sim F_{x/y/z}, F_3\sim F_{xy r/\theta_1/\theta_2}, F_{xz r/\theta_1/\theta_2},
 F_{yz r/\theta_1/\theta_2}$ and $F_5\sim F_{xyz \beta_1\beta_2}$ where $\beta_i=r/\theta_i$.

    Consider $T_x$ followed by $T_y$ followed by $T_z$ where $T_i$ means T-dualizing along i-th direction. As an example, $T_x F_x^{IIB}\rightarrow
{\rm non-dynamical\ 0-form\ field\ strength}^{IIA}\cite{kiritsis-book}, T_y T_x F_x^{IIB} \rightarrow F_y^{IIB}$,\\
 $T_z F_y^{IIB} \rightarrow F_{yz}^{IIA}$ implying one can never generate
$F_4^{IIA}$ from $F_1^{IIB}$.
    As also an example consider $T_x F_{xy\beta_i}^{IIB}\rightarrow F_{y \beta_i}^{IIA}, T_yF_{y \beta_i}^{IIA}\rightarrow F_{\beta_i}^{IIB}, T_z F_{\beta_i}^{IIB}
\rightarrow F_{\beta_i z}^{IIA}$ again not generating $F_4^{IIA}$;
   $ T_x F_{xyz \beta_1 \beta_2}^{IIB}\rightarrow F_{yz \beta_1 \beta_2}^{IIA}$,
    $T_y F_{yz \beta_1 \beta_2}^{IIA}\rightarrow F_{z \beta_1 \beta_2}^{IIB},
T_z F_{z \beta_1 \beta_2}^{IIB}\rightarrow F_{\beta_1 \beta_2}^{IIB}$; thus one can not generate $F_4^{IIA}$.
Thus, the four-form flux $G_4=d\left(C_{\mu\nu10}dx^\mu\wedge dx^\nu\wedge dx_{10}\right)
 + \left(A^{F_1}_1 + A^{F_3}_1 + A^{F_5}_1\right)\wedge H_3=H_3\wedge dx_{10} + A\wedge H_3$, where $C_{\mu\nu10}\equiv B_{\mu\nu}$ implying that the flux-dependent D=11 action is given by the following two terms:
\begin{eqnarray}
\label{flux_action_D=11-i}
& & \int C_3\wedge G_4\wedge G_4 = \int B\wedge dx_{10} \wedge \left(H\wedge dx_{10} + A\wedge H\right)\wedge (H\wedge dx_{10} + A\wedge H)=0,
\end{eqnarray}
and
\begin{eqnarray}
\label{flux_action_D=11-ii}
& & \int G_4\wedge *_{11}G_4 = \int \left(H_3\wedge dx_{10} + A\wedge H_3\right)\wedge *_{11}\left(H_3\wedge dx_{10} + A\wedge H_3\right).
\end{eqnarray}
Now, $H_3\wedge dx_{10}\wedge *_{11}\left(H_3\wedge A\right)=0$ as neither $H_3$ nor $A$ has support along $x_{10}$. Hence:
\begin{eqnarray}
\label{flux_action_D=11-iii}
& & H_3\wedge dx_{10}\wedge *_{11}\left(H_3\wedge dx_{10}\right)=\sqrt{G}H_{\mu\nu\rho10}G^{\mu\mu_1}G^{\nu\nu_1}G^{\rho\rho_1}G^{10\lambda_1}H_{\mu_1\nu_1\rho_1\lambda_1}
dt\wedge...dx_{10}\nonumber\\
& & = \sqrt{G} H_{\mu\nu\rho10}\left(-G^{\mu10}G^{\nu\nu_1}G^{\rho\rho_1}G^{10\lambda_1}H_{\nu_1\rho_1\lambda_1} + G^{\mu\mu_1}G^{\nu10}G^{\rho\rho_1}G^{10\lambda_1}H_{\mu_1\rho_1\lambda_1} \right.\nonumber\\
& & \left.- G^{\mu\mu_1}G^{\nu\nu_1}G^{\rho10}G^{10\lambda_1}H_{\mu_1\nu_1\lambda_1} + G^{\mu\mu_1}G^{\nu\nu_1}G^{\rho\rho_1}G^{10\ 10}H_{\mu_1\nu_1\rho_1}\right)dt\wedge...dx_{10},
\end{eqnarray}
where $H_{\mu\nu\rho10}=H_{\mu\nu\rho}$, and
\begin{eqnarray}
\label{flux_action_D=11-iv}
& & \left(H\wedge A\right)\wedge *_{11}\left(H\wedge A\right)=\sqrt{G} H_{[\mu\nu\rho}A_{\lambda]}G^{\mu\mu_1}G^{\nu\nu_1}G^{\lambda\lambda_1}H_{[\mu_1\nu_1\rho_1}A_{\lambda_1]},
\end{eqnarray}
where
$H_{[\mu_1\mu_2\mu_3}A_{\mu_4]}\equiv H_{\mu_1\mu_2\mu_3}A_{\mu_4} - \left(H_{\mu_2\mu_3\mu_4}A_{\mu_1} - H_{\mu_3\mu_4\mu_1}A_{\mu_2} + H_{\mu_4\mu_1\mu_2}A_{\mu_3}\right)$.
\vskip 0.2in

Considering the same scaling behavior as used to calculate the contribution of Einstein-Hilbert Action as well as Gibbons-Hawking-York surface action terms,  we see that for both limits (\ref{limits_Dasguptaetal-i}) and (\ref{limits_Dasguptaetal-ii}), in equation (\ref{flux_action_D=11-iii}), contribution of $ H_3\wedge dx^{11}\wedge *_{11}(H_3\wedge dx^{11})$ is always dominated by \\ $ \sqrt{G} H^{2}_{\theta_1 \theta_2 y} G^{\cal M}\ ^{\theta_1\theta_1}G^{\cal M}\ ^{\theta_2 \theta_2}G^{\cal M}\ ^{yy}G^{\cal M}\ ^{10\ 10}$ term  and in equation (\ref{flux_action_D=11-iv}), contribution of $(H\wedge A)\wedge *_{11}(H\wedge A)$ is dominated by $ \sqrt{G} H^{2}_{ \theta_1 \theta_2 y} {A^2_{y}} G^{\cal M}\ ^{\theta_1\theta_1}G^{\cal M}\ ^{\theta_1\theta_1} G^{\cal M}\ ^{\theta_2\theta_2}G^{yy}  $ term. Therefore, for simplicity in calculations, we assume that leading contribution in equations ($\ref{flux_action_D=11-iii}$) and ($\ref{flux_action_D=11-iv}$) are governed by aforementioned terms. The
 relevant inverses of the 11-dimensional metric components  of (\ref{metric-mirror}), in the limit (\ref{limits_Dasguptaetal}), simplify to the following expressions:
 {\small
\begin{eqnarray}
\label{eq:Gtheta2theta2etal}
&\bullet& G^{\cal M}\ ^{\theta_2 \theta_2}\sim \Biggl(216 \sqrt[3]{3} \sqrt[6]{{g_s}} \cos ^6({\theta_1}) \cot ^4({\theta_2}) \Bigl(2 \cos ^2({\theta_2}) \sin ^2({\theta_1}) + 3 \sin^2({\theta_2}) \sin ^2({\theta_1})+2 \cos ^2({\theta_1}) \sin ^2({\theta_2})\nonumber\\
&&  +3 {h_5} \sin (2 {\theta_1}) \sin (2
   {\theta_2})\Bigr)^2\Biggr)/\Biggl(\sqrt{N} \sqrt{\pi } {f_2}({\theta_2})^2 \left({f_1}({\theta_1})^2 \sin ^2({\theta_1})+1\right) \left(\cos
   ^2({\theta_2}) \sin ^2({\theta_1})+\cos ^2({\theta_1}) \sin ^2({\theta_2})\right)^2 \nonumber\\
&& \Bigl(\left(4 \cos ({\theta_2}) \left(2 \cot ({\theta_1})-3
   \sqrt{6}\right) \cot ({\theta_2})+18 \sqrt{6} \sin ({\theta_2})\right) \cos ^3({\theta_1})+12 \cos ({\theta_2}) \cot ({\theta_2}) \sin
   ({\theta_1}) \cos ^2({\theta_1}) \nonumber\\
   && +27 \sin ^2({\theta_1}) \left(\sqrt{6} \sin ({\theta_2})-4 {h_5} \cos ({\theta_2})\right) \cos
   ({\theta_1})+81 {h_5} \sin ^3({\theta_1}) \left(\sqrt{6} \cos ({\theta_2})+3 {h_5} \sin ({\theta_2})\right)\Bigr)^2\Biggr)\nonumber\\
  &\bullet& G^{\cal M}\ ^{yy} \sim \Bigl(36 \sqrt[3]{3} {g_s}^{2/3} \csc ^4({\theta_2}) \sec ^2({\theta_2}) \sin ^3({\theta_1}) (9 {h_5} \sin ({\theta_1})-2 \cos ({\theta_1})
   \cot ({\theta_2})) \Bigl(2 \cos ^2({\theta_2}) \sin ^2({\theta_1}) \nonumber\\
   && +3 \sin ^2({\theta_2}) \sin ^2({\theta_1})+2 \cos ^2({\theta_1}) \sin
   ^2({\theta_2})+3 {h_5} \sin (2 {\theta_1}) \sin (2 {\theta_2})\Bigr)^2\Biggr)/\Bigl((\cos (2 {\theta_1})-5)^3\Bigr)\nonumber\\
   &\bullet& G^{\cal M}\ ^{r r}\sim \frac{r^2 \sqrt[6]{{g_s}} \left(1-\frac{r_h^4}{r^4}\right) }{2~3^{2/3} \sqrt{\pi } \sqrt{N}  }\nonumber\\
 &\bullet& G^{\cal M}\ ^{10\ 10} \sim  3 \sqrt[3]{3} \left(\frac{1}{{g_s}}\right)^{4/3}\nonumber\\
 &\bullet&H_{r \theta_1 y}\sim \frac {{f_ 2} ({\theta_ 2}) \sin ({\theta_ 1}) \cos ({\theta_ 1}) \sin ({\theta_ 2}) \sin (2 {\theta_ 2}) \cos
      ({\theta_ 2})} {\left (\sin ^2 ({\theta_ 1}) \cos ^2 ({\theta_ 2}) + \cos ^2 ({\theta_ 1}) \sin ^2 ({\theta_ 2}) \right)^2}
   \nonumber\\
   &\bullet& A_{\theta_2}\sim \frac{\sqrt {\frac {3} {2}} {N_f} {\phi_ 2} \sin ^2 ({\theta_ 2}) \left (8 \cos ^3 ({\theta_ 1}) \cos ({\theta_ 2}) \cot ({\theta_ 2}) - 16
           \cos ^4 ({\theta_ 1}) \cot ({\theta_ 2}) \right)} {\pi  (\cos (2 {\theta_ 1}) -
       5) \left (2 \sin ^2 ({\theta_ 1}) \cos ^2 ({\theta_ 2}) + 2 \cos
            ^2 ({\theta_ 1}) \sin ^2 ({\theta_ 2}) \right)}-\nonumber\\
            && \frac{\sqrt {\frac {2} {3}} {N_f}  \psi  \csc ({\theta_ 2}) \left (-2 \cos ({\theta_ 1})
           \cos ({\theta_ 2}) + \sin ^2 ({\theta_ 1}) - \cos ^2 ({\theta_ 1}) + 5 \right)} {\pi  (\cos (2 {\theta_ 1}) - 5)} .
.
   \end{eqnarray}}
On simplifying equations (\ref{flux_action_D=11-iii}) and (\ref{flux_action_D=11-iv}) with the help of equations (\ref{eq:Gtheta2theta2etal}), the most dominant contribution near $\theta_{1,2}=0,\pi$ will be given by the following analytical expression:
{\small
\begin{eqnarray}
\label{flux_action_D=11-v}
& &  H_3\wedge dx^{11}\wedge *_{11}\left(H_3\wedge dx^{11}\right) \sim  \frac {1134 \sqrt {3}  r^3 {f_ 2} ({\theta_ 2}) (\sin ^5 ({\theta_ 1}) \cos^5 ({\theta_ 1}) \cos ^3 ({\theta_ 2}) \cot ^3 ({\theta_ 2})} {\pi ^{5/4} {g_s}^{13/ 4} N^{5/4} \left (\sin ^2 ({\theta_ 1}) \cos^2 ({\theta_ 2}) + \cos ^2 ({\theta_ 1}) \sin ^2 ({\theta_ 2}) \right)^3}
    \nonumber\\
   && \hskip 1.7in \sim \frac{1134 \sqrt{3}  r^3  {f_2}( {\theta_2}) \cos^4( {\theta_1}) \cot ( {\theta_1}) \csc ^3( {\theta_2})}{\pi ^{5/4} {g_s}^{13/ 4} N^{5/4}}.
\end{eqnarray}}
and
{\small
\begin{eqnarray}
\label{flux_action_D=11-vi}
& & \hskip -0.3in \left(H\wedge A\right)\wedge *_{11}\left(H\wedge A\right)\sim \frac {243 \sqrt {3} {N_f}^2 r^3 {f_ 2} ({\theta_ 2}) \sin ^7 ({\theta_ 1}) \cos ^3 ({\theta_ 1}) \sin ^4 (2 {\theta_ 2}) \csc({\theta_ 2}) (\psi -  6 {\phi_ 2} \cos ({\theta_ 1}))^2} {2 \pi ^{13/4} ({g_s} N)^{5/4} (\cos (2 {\theta_ 1}) - 5)^2 \left (\sin^2 ({\theta_ 1}) \cos ^2 ({\theta_ 2}) + \cos ^2 ({\theta_ 1}) \sin ^2 ({\theta_ 2}) \right)^5}\ \nonumber\\
   && \sim  \frac {243 \sqrt {3} {N_f}^2 r^3 {f_ 2} ({\theta_ 2}) \cot ^3 ({\theta_ 1}) \tan ^3 ({\theta_ 2}) \sec ^3 ({\theta_ 2}) (\psi - 6 {\phi_ 2} \cos ({\theta_ 1}))^2} {2 \pi ^{13/4} ({g_s} N)^{5/4}}.
\end{eqnarray}}
Integrating above:
{\small
\begin{eqnarray}
\label{eq:intG4wedG4i}
&&  \int_{x_{10}\in[0,{2\pi}], r\in[r_h,r_\Lambda],\theta_{1,2}\in[\epsilon_{\theta_{1,2}},\pi-\epsilon_{\theta_{1,2}}],\phi_{1,2}\in[0,2\pi], \psi\in[0,4\pi]} G_4\wedge *_{11}G_4   \sim  \nonumber\\
  && \hskip -0.3in  \frac{r^3  {f_2}( {\theta_2}) \theta_1 \theta_2 \cos^4( {\theta_1}) \cot ( {\theta_1}) \csc ^3( {\theta_2})}{{g_s}^{13/ 4} N^{5/4}}+ \frac {{N_f}^2 r^3 {f_ 2} ({\theta_ 2}) \theta_1 \theta_2 \cot ^3 ({\theta_ 1}) \tan ^3 ({\theta_ 2}) \sec ^3 ({\theta_ 2})  \cos({\theta_ 1})^2} {({g_s} N)^{5/4}}
   \Biggr|_{\theta_{1,2}=\epsilon_{\theta_{1,2}} + \theta_{1,2}=\pi-\epsilon_{\theta_{1,2}}}   .
\end{eqnarray}}

  \begin{itemize}
\item  Limit (\ref{limits_Dasguptaetal-i}) :
Incorporating   $g_s \sim  {\epsilon} $, $N/N_{\rm eff}\sim  {\epsilon^{-19}} $, $M/M_{\rm eff}\sim  {\epsilon^{-\frac{3}{2}}}$, $\theta_{1 }\sim  {\epsilon^{\frac{5}{6}}}$, $\theta_{2}\sim   {\epsilon^{\frac{5}{6}}}$, ${f_2}({{\theta_{1,2}}})\sim { \frac{1}{{\theta_{1,2}}}}$, we see that:
\begin{eqnarray}
\label{eq:intG4wedG4ii}
&& \int_{x^0
\in \left[0,\frac {\pi \sqrt{4\pi g_s N}}{r_h}\right],x_{10}\in[0,{2\pi}], r\in[r_h,r_\Lambda],\theta_{1,2}\in[\epsilon_{\theta_{1,2}},\pi-\epsilon_{\theta_{1,2}}],\phi_{1,2}\in[0,2\pi], \psi\in[0,4\pi]} G_4\wedge *_{11}G_4\nonumber\\
& &   \sim - {\epsilon}^{9}  \Bigl(\frac{r^{4}_{\Lambda}}{r_h} \Bigr).
\end{eqnarray}
 Therefore, ${\cal S}_{\rm flux}^{\rm Infinite}\sim   - {\epsilon}^{9}   \frac{r^{4}_{\Lambda}}{r_h}$.

  \item  MQGP limit (\ref{limits_Dasguptaetal-ii}):
  Incorporating  $g_s \sim \ {\alpha_1} \,{\epsilon} $, $N/N_{\rm eff}\sim \alpha_2  {\epsilon^{-39}} $, $M/M_{\rm eff}\sim  \alpha_3{\epsilon^{-\frac{3}{2}}}$,   $\theta_{1,2 }\sim \alpha_{4,5} {\epsilon^{\frac{15}{6}}}$, ${f_2}({{\theta_{1,2}}})\sim { \frac{1}{{\theta_{1,2}}}}$, we see that
\begin{eqnarray}
\label{eq:intG4wedG4iii}
&& \int_{x^0
\in \left[0,\frac {\pi \sqrt{4\pi g_s N}}{r_h}\right],x_{10}\in[0,{2\pi}], r\in[r_h,r_\Lambda],\theta_{1,2}\in[0,\pi],\phi_{1,2}\in[0,2\pi], \psi\in[0,4\pi]} G_4\wedge *_{11}G_4\nonumber\\
& &   \sim - a_{G_4}(\alpha_{1,2,3,4,5}){\epsilon}^{19}   \Bigl(\frac{r^{4}_{\Lambda}}{r_h}+ \frac{{r_h}^3} {ln(r_{\Lambda})}\Bigr).
\end{eqnarray}
Therefore,  ${\cal S}_{\rm flux}^{\rm Infinite}\sim {\epsilon}^{19}a_{G_4}(\alpha_{1,2,3,4,5}) \frac{r^{4}_{\Lambda}}{r_h}$.
 \end{itemize}

\subsubsection{${\cal O}(R^4)$ Action Terms}

In either limits (\ref{limits_Dasguptaetal-i}) and (\ref{limits_Dasguptaetal-ii}), the dominant contribution to J is given by
\begin{equation}
\label{eq:J}
 J \sim \Bigl(\bigl(G^{\cal M}\ ^{\theta_1 \theta_1}\bigr)^3  G^{\cal M}\ ^{\theta_1 \theta_2}  G^{\cal M}\ ^{\theta_2 \theta_2}G^{\cal M}\ ^{\theta_1 y}G^{\cal M}\ ^{\theta_2 y} G^{\cal M}\ ^{y y} \bigl(R_{\theta_2 \theta_1 y \theta_1}\bigr)^4 +\frac{1}{2}\bigl(R_{\theta_2 \theta_1 \theta_1 y }\bigr)^2 \bigl(R_{\theta_2 \theta_1 y \theta_1}\bigr)^2 \Bigr),
\end{equation}
and the dominant contribution to $E_8$ is given by:
\begin{eqnarray}
\label{eq:E8}
&& E_8 \sim  G^{\cal M}\ ^{0 0}G^{\cal M}\ ^{1 1}G^{\cal M}\ ^{2 2} G^{\cal M}\ ^{3 3}\bigl( G^{\cal M}\ ^{r r} G^{\cal M}\ ^{\theta_1 \theta_1}G^{\cal M}\ ^{\theta_2 \theta_2}G^{\cal M}\ ^{x x} G^{\cal M}\ ^{y y}G^{\cal M}\ ^{z z}G^{\cal M}\ ^{10 10}\bigr)^2    R_{3 r 3 r}R_{\theta_1 \theta_2 \theta_1 \theta_2}\nonumber\\
 && R_{ x y xy}R_{z 10 z 10}.
\end{eqnarray}

The simplified form of analytic expressions of $G^{\cal M}\ ^{\theta_1 \theta_1}$, $ G^{\cal M}\ ^{\theta_2 \theta_2}$ and $G^{\cal M}\ ^{r r}$ are given in equation no (\ref{eq:invtheta1theta1}) and (\ref{eq:Gtheta2theta2etal}). The simplified expressions of other inverse components as well as covariant 4-rank tensor relevant to get the estimate of J and $E_8$  are as follows:
{\small
\begin{eqnarray}
&\bullet& G^{\cal M}\ ^{\theta_1 \theta_2}=-\frac{1}{64 \sqrt{\pi } \sqrt{N} {f_2}({\theta_2})^2}\Bigl(81 \sqrt[3]{3} \sqrt[6]{{g_s}} {h_5} \csc ^4({\theta_2}) \sec ^2({\theta_2}) \left(2 \sin ^2({\theta_1}) \cos ^2({\theta_2})+2 \cos
   ^2({\theta_1}) \sin ^2({\theta_2})\right)^2 \nonumber\\
   && ({f_1}({\theta_1}) {f_2}({\theta_2}) (13 \sin ({\theta_1})+\sin (3 {\theta_1})) \sin
   ({\theta_2})-16)\bigr) \nonumber\\
&\bullet& G^{\cal M}\ ^{\theta_1 y}=-\frac{1}{\sqrt[4]{\pi } \sqrt[4]{N} {f_2}({\theta_2})}\Bigl({h_5} ({g_s} N)^{5/12} \csc ^2({\theta_2}) \sec ({\theta_2}) \left(2 \sin ^2({\theta_1}) \cos ^2({\theta_2})+2 \cos ^2({\theta_1})
   \sin ^2({\theta_2})\right) \nonumber\\
   && ({f_1}({\theta_1}) {f_2}({\theta_2}) (13 \sin ({\theta_1})+\sin (3 {\theta_1})) \sin
   ({\theta_2})-16)\Bigr)\nonumber\\
&\bullet&G^{\cal M}\ ^{\theta_2 y}= -\frac{{g_s}^{5/12} \csc ^2({\theta_2}) \sec ({\theta_2}) \left(2 \sin ^2({\theta_1}) \cos ^2({\theta_2})+2 \cos ^2({\theta_1}) \sin
   ^2({\theta_2})\right)}{\sqrt[4]{\pi } \sqrt[4]{N} {f_2}({\theta_2})}\nonumber\\
&\bullet& G^{\cal M}\ ^{y y}=
{g_s}^{2/3} \sin ^3({\theta_1}) \cos ({\theta_1}) \csc ^5({\theta_2}) \sec ({\theta_2}) \left(2 \sin ^2({\theta_1}) \cos ^2({\theta_2})+2 \cos
   ^2({\theta_1}) \sin ^2({\theta_2})\right)^2\nonumber\\
&\bullet&
G^{\cal M}\ ^{x x}=-\frac{81}{2} \sqrt[3]{3} {g_s}^{2/3} \tan ^2({\theta_1}) \sec ^6({\theta_1}) \tan ^4({\theta_2})\nonumber\\
&\bullet& G^{\cal M}\ ^{z z}=-\frac{243 \sqrt[3]{3} {g_s}^{2/3} \sin ^4({\theta_1}) \tan ^3({\theta_1}) \sin ({\theta_2}) \cos ({\theta_2})}{2 \left(\sin ^2({\theta_1}) \cos
   ^2({\theta_2})+\cos ^2({\theta_1}) \sin ^2({\theta_2})\right)^2}\nonumber\\
&\bullet& G^{\cal M}\ ^{0 0}\sim  -\frac{2 {g_s}^{7/6} \sqrt{N} \sqrt{\pi }}{3^{2/3} r^2 \left(1-\frac{r_h^4}{r^4}\right)}\nonumber\\
&\bullet& G^{\cal M}\ ^{1 1}\sim \frac{2 {g_s}^{7/6} \sqrt{N} \sqrt{\pi }}{3^{2/3} r^2}\nonumber\\
&\bullet&
G^{\cal M}\ ^{2 2}\sim \frac{2 {g_s}^{7/6} \sqrt{N} \sqrt{\pi }}{3^{2/3} r^2}\nonumber\\
&\bullet& G^{\cal M}\ ^{3 3}\sim \frac{2 {g_s}^{7/6} \sqrt{N} \sqrt{\pi }}{3^{2/3} r^2}\nonumber\\
&\bullet& R_{\theta_2 \theta_1 y \theta_1}\sim \frac{40 \sqrt{2} \sqrt[4]{\pi } {f_2}({\theta_2}) \sin ^2({\theta_1}) \cos ^2({\theta_1}) \sin ^2({\theta_2}) \cos ^5({\theta_2})}{3 \sqrt[3]{3}
   {g_s}^{5/12} \sqrt[4]{N} \left(2 \sin ^2({\theta_1}) \cos ^2({\theta_2})+2 \cos ^2({\theta_1}) \sin ^2({\theta_2})\right)^3}\nonumber\\
    &\bullet& R_{ 3 r 3 r}\sim \frac{{N_f} \sin (2 {\theta_1}) \cot \left(\frac{{\theta_1}}{2}\right)}{2 \sqrt[3]{3} \pi ^{3/2} \sqrt[6]{{g_s}} \sqrt{N}
   \left(1-\frac{{r_h}^4}{r^4}\right)}\nonumber\\
   &\bullet& R_{\theta_1 \theta_2 \theta_1 \theta_2}\sim \frac{2\ 3^{2/3} \sqrt{\pi } \sqrt{N}-\frac{2}{3} {f_2}({\theta_2})^2 \sin ^3({\theta_1}) \cos ^3({\theta_1}) \cos ^2({\theta_2}) \cot
   ({\theta_2})}{2 \sqrt[6]{{g_s}} \sin ^2({\theta_1}) \cos ^2({\theta_2})+2 \cos ^2({\theta_1}) \sin ^2({\theta_2})}\nonumber\\
   &\bullet& R_{x y x x y}\sim -\frac{128 \sin ^2({\theta_1}) \cos ^6({\theta_1}) \sin ^2({\theta_2}) \cos ^{10}({\theta_2})}{27 \sqrt[3]{3} \sqrt{\pi } {g_s}^{7/6} \sqrt{N}
   \left(\sin ^2({\theta_1}) \cos ^2({\theta_2})+\cos ^2({\theta_1}) \sin ^2({\theta_2})\right)^6}\nonumber\\
   &\bullet& R_{z 10 z 10}\sim \frac{4 {g_s}^{11/6} {N_f} \cot \left(\frac{{\theta_1}}{2}\right) \cot ^3({\theta_1})}{81 \sqrt[3]{3} \pi ^{3/2} \sqrt{N}}.
   \end{eqnarray}}
 Utilizing above and simplified form of $ G^{\cal M} $ as given in equation (\ref{eq:g}), we have
 {\small
 \begin{eqnarray}
&&  \hskip -0.5in \frac{T_2}{{2\pi}^4. 3^2.2^{13}}\int_{\cal{M}} d^{11}\!x \sqrt{G^{\cal M}}(J-\frac{1}{2}E_8)\sim \nonumber\\
 && \frac{(2 \pi^2)^{1/3}}{{2\pi}^4. 3^2.2^{13}} \Bigg(  \frac{10^5 {h_5}^3 r^{4}_{\Lambda} \theta_1 \theta_2 \sin ^9({\theta_1}) \cos ^{13}({\theta_1}) \cos ^4({\theta_2}) \cot ^{11}({\theta_2}) ({f_1}({\theta_1})
   {f_2}({\theta_2}) \sin ({\theta_1}) \sin ({\theta_2})-1)^3}{{g_s}^2 N^2 {f_2}({\theta_2})^2 \left(2 \sin ^2({\theta_1}) \cos
   ^2({\theta_2})+2 \cos ^2({\theta_1}) \sin ^2({\theta_2})\right)^4} \nonumber\\
   && +\frac{10^8 {g_s}^{67/12} {N_f}^2 \theta_1 \theta_2 \sin ^{20}({\theta_1}) \sin (2 {\theta_1}) \tan ^8({\theta_1}) \cot ^2\left(\frac{{\theta_1}}{2}\right)
   \cos ^5({\theta_2}) \cot ^3({\theta_2})}{N^{7/4} r^{2}_{\Lambda} {f_2}({\theta_2}) \left(\sin ^2({\theta_1}) \cos ^2({\theta_2})+\cos ^2({\theta_1})
   \sin ^2({\theta_2})\right)^8}\Bigg)\Biggr|_{\theta_{1,2}=\epsilon_{\theta_{1,2}} + \theta_{1,2}=\pi-\epsilon_{\theta_{1,2}}}.
   \end{eqnarray}}
 \begin{itemize}
\item  Limit (\ref{limits_Dasguptaetal-i}) :
Incorporating   $g_s \sim \ {\epsilon} $, $N/N_{\rm eff}\sim {\epsilon^{-19}} $, $M/M_{\rm eff}\sim  {\epsilon^{-\frac{3}{2}}}$
$\theta_{1,2}\sim  {\epsilon^{\frac{5}{6}}}$,, ${f_2}({{\theta_{1,2}}})\sim { \frac{1}{{\theta_{1,2}}}}$, we see that
\begin{eqnarray}
\label{eq:higherderv1}
&&  \hskip -0.5in \frac{T_2}{{2\pi}^4. 3^2.2^{13}} \int_{x^0
\in \left[0,\frac {\pi \sqrt{4\pi g_s N}}{r_h}\right],x_{10}\in[0,{2\pi}], r\in[r_h,r_\Lambda],\theta_{1,2}\in[\epsilon_{\theta_{1,2}},\pi-\epsilon_{\theta_{1,2}}],\phi_{1,2}\in[0,2\pi], \psi\in[0,4\pi]} \sqrt{G^{\cal M}}(J-\frac{1}{2}E_8)  \nonumber\\
 && \sim  \frac{{\cal O}(10^{-14})}{{\epsilon}^{-22} } \frac{r^{4}_{\Lambda}}{r_h} + \frac{{\cal O}(1 0)}{{\epsilon}^{-39} } \Bigl(\frac{1}{ r^{2}_{\Lambda} r_h}\Bigr).
\end{eqnarray}
Thus, $ {\cal S}_{\rm {\cal O}(R^4)}^{\rm finite}\sim   {{\cal O}(10 )}{{\epsilon}^{39} } \Bigl(\frac{1}{ r^{2}_{\Lambda} r_h}\Bigr)\rightarrow 0, r_{\Lambda}\rightarrow \infty~ {\rm and}~ {\epsilon}\rightarrow0$;  ${\cal S}_{\rm {\cal O}(R^4)}^{\rm Infinite}\sim   {{\cal O}(10^{-14})}{{\epsilon}^{22} } \frac{r^{4}_{\Lambda}}{r_h}$.

  \item  MQGP limit (\ref{limits_Dasguptaetal-ii}):
 Incorporating  $g_s \sim \ {\alpha_1} \,{\epsilon} $, $N/N_{\rm eff}\sim \alpha_2  {\epsilon^{-39}} $, $M/M_{\rm eff}\sim \alpha_3 {\epsilon^{-\frac{3}{2}}}$,   $\theta_{1,2}\sim \alpha_{4,5}  {\epsilon^{\frac{15}{6}}}$, ${f_{1,2}}({{\theta_{1,2}}})\sim { \frac{1}{{\theta_{1,2}}}}$,
\begin{eqnarray}
\label{eq:higherderv2}
&&  \hskip -0.5in \frac{T_2}{{2\pi}^4. 3^2.2^{13}} \int_{x^0
\in \left[0,\frac {\pi \sqrt{4\pi g_s N}}{r_h}\right],x_{10}\in[0,{2\pi}], r\in[r_h,r_\Lambda],\theta_{1,2}\in[\epsilon_{\theta_{1,2}},\pi-\epsilon_{\theta_{1,2}}],\phi_{1,2}\in[0,2\pi], \psi\in[0,4\pi]} \sqrt{G^{\cal M}}(J-\frac{1}{2}E_8)   \sim \nonumber\\
 && \sim  a_{R^4}(\alpha_{1,2,4,5}){{\cal O}(10^{-15})}{{\epsilon}^{42} } \frac{r^{4}_{\Lambda}}{r_h} +  b_{R^4}(\alpha_{1,2,4,5}){{\cal O}(10^6 )}{{\epsilon}^{81.3} } \Bigl(\frac{1}{ r^{2}_{\Lambda} r_h}\Bigr).
\end{eqnarray}
Therefore, $ {\cal S}_{\rm {\cal O}(R^4)}^{\rm finite}\sim  b_{R^4}(\alpha_{1,2,4,5}){\cal O}(10 ^6){{\epsilon}^{81.3} } \Bigl(\frac{1}{ r^{2}_{\Lambda} r_h}\Bigr) \rightarrow 0, r_{\Lambda}\rightarrow \infty$;  ${\cal S}_{\rm {\cal O}(R^4)}^{\rm Infinite}\sim  a_{R^4}(\alpha_{1,2,4,5}){{\epsilon}^{42} } \frac{r^{4}_{\Lambda}}{r_h}$.
 \end{itemize}

The other quartic term in D=11 Supergravity Action is
   $$T_2 \int C_3 \wedge X_8, $$ where $C_3$ is 3-form flux in D=11 Supergravity  defined as $C_3 =B_{\mu \nu} dx^{\mu}\wedge dx^{\nu} \wedge dx_{10}$. Now,
   $R_{\mu\nu}^{\ \ \ ab} =  R_{\mu\nu}^{\ \ \ \rho\lambda}e^{\ a}_{\rho} e^{\ b}_{\lambda}$ where $e^a_{\ \mu}$ are frames in components. Now:
$$\hskip-0.3intr(R^2) =
R_{\mu_1\nu_1}^{\ \ \ \ ab}R_{\mu_2\nu_2 b a} dx^{\mu_1}\wedge dx^{\nu_1}\wedge dx^{\mu_2}\wedge dx^{\nu_2}
= e^{\ a}_{\rho_1} e^{\ b}_{\lambda_1} E_b^{\ \rho_2}E_a^{\ \lambda_2} R_{\mu_1\nu_1}^{\rho_1\lambda_1} R_{\mu_2\nu_2\rho_2\lambda_2}
 dx^{\mu_1}\wedge dx^{\nu_1}\wedge dx^{\mu_2}\wedge dx^{\nu_2},$$
where $E_a^{\ \mu}$ are the inverse frames in components. Using:
$e^{\ a}_{\rho_1}E_a^{\ \lambda_2}=\delta^{\lambda_2}_{\rho_1}$, etc. the above gives:
$$tr(R^2)=R_{\mu_1\nu_1}^{\ \ \ \rho_1\lambda_1} R_{\mu_2\nu_2\lambda_1\rho_1}
 dx^{\mu_1}\wedge dx^{\nu_1}\wedge dx^{\mu_2}\wedge dx^{\nu_2}.$$  Writing $tr(R^4)$ and $(tr(R^2))^2$ in terms of purely covariant curvature tensor,  one similarly has:
 \begin{eqnarray}
 \label{eq:trR4}
 && tr(R^4)= G^{\cal M}\ ^{\lambda_1 \lambda^{\prime}_1} G^{\cal M}\ ^{\rho_1 \rho^{\prime}_1}G^{\cal M}\ ^{\rho_2 \rho^{\prime}_2}G^{\cal M}\ ^{\lambda_2 \lambda^{\prime}_2} R_{\rho^{\prime}_1 \lambda^{\prime}_1  \mu_1 \nu_1}R_{\mu_2 \nu_2 \lambda_1 \rho_2}  R_{\rho^{\prime}_2 \lambda^{\prime}_2 \mu_3 \nu_3  } R_{\mu_4 \nu_4 \lambda_2 \rho_1}dx^{\mu_1}\wedge dx^{\nu_1}\nonumber\\
 &&\wedge dx^{\mu_2}\wedge dx^{\nu_2}\wedge  dx^{\mu_3}\wedge dx^{\nu_3}\wedge dx^{\mu_4}\wedge dx^{\nu_4}
 \end{eqnarray}
 and
  \begin{eqnarray}
  \label{eq:trR2}
 && (tr(R^2))^2 = G^{\cal M}\ ^{\rho_1 \rho^{\prime}_1} G^{\cal M}\ ^{\lambda_1 \lambda^{\prime}_1}    G^{\cal M}\ ^{\rho_2 \rho^{\prime}_2} G^{\cal M}\ ^{\lambda_2 \lambda^{\prime}_2}R_{ \rho^{\prime}_1 \lambda^{\prime}_1 \mu_1 \nu_1 }\nonumber\\
 &&R_{\lambda_1\rho_1\mu_2 \nu_2 }R_{ \rho^{\prime}_2 \lambda^{\prime}_2 \mu_3 \nu_3 }R_{\lambda_2 \rho_2\mu_4 \nu_4 } dx^{\mu_1}\wedge dx^{\nu_1}\wedge dx^{\mu_2}\wedge dx^{\nu_2}\wedge  dx^{\mu_3}\wedge dx^{\nu_3}\wedge dx^{\mu_4}\wedge dx^{\nu_4}
 \end{eqnarray}
From equation (\ref{three-form fluxes}), the non-zero components of 2-form flux $B_2$ include  :
 $$B_{\theta_1 x},B_{\theta_2 y},B_{\theta_1 z}, B_{\theta_2 z}$$
and therefore, the non-zero components of $C_3$ are $C_{\theta_1 x 10}, C_{\theta_2 y 10}, C_{\theta_1 z 10}$ and $ C_{\theta_2 z 10}$.
Since non-zero three form flux components do not include space-time indices, the same must be included in $X_8$ to have non vanishing $ \int C_3 \wedge X_8$. Effectively, one needs to calculate $X_8$ for a Euclideanized eight-fold $M_8$ that is locally $S^1(x^0)\times\mathbb{R}^{3}_{\rm conf}\times M_4(r,\beta_i,x_{a_1},x_{a_2})$ where
$\beta_i\equiv\theta_1\ {\rm or}\ \theta_2$, $x_{a_{1,2}}\equiv (y,z)\ {\rm or}\ (x,z)\ {\rm or}\ (x,y)$ and $\mathbb{R}^3_{\rm conf}$ implies conformally Euclidean. Now, in Dasgupta et al's limit (\ref{limits_Dasguptaetal}), one makes the following observatoins:
 \begin{enumerate}
 \item
 Let $a$, etc. index the $S^1(x^0)\times\mathbb{R}^3_{\rm conf}(x^{1,2,3})$ coordinates, $m$ the $M_4(r,\beta_i,x_{a_1},x_{a_2})$ coordinates and let $\alpha=(a,m)$. From $R^{ab}_{\ \ \alpha\beta}=g^{bb}(\partial_{[\alpha}\Gamma^a_{b|\beta]} + \Gamma^\gamma_{b[\beta}\Gamma^a_{\gamma|\alpha]})$, one sees that only $R^{ab}_{\ \ ab} = - \frac{g^{bb}g^{rr}g^{aa}}{4}(\partial_rg_{bb})(\partial_rg_{aa})\neq0, a\neq b$.

 \item
 $R^{ab}{\ \ m c}=0$.

 \item
 Using $\Gamma^{\alpha^\prime}_{\beta^\prime a}=\frac{g^{aa}}{2}\partial_rg_{aa}\delta^{\alpha^\prime}_a\delta^r_{\beta^\prime} - \frac{g^{rr}}{2}\partial_rg_{aa}\delta^a_{\beta^\prime}\delta^{\alpha^\prime}_r$, one obtains:
 $R^{\alpha\beta}_{\ \ ma} = g^{rr}\partial_m\Gamma^a_{ra}\delta^\alpha_a\delta^\beta_r - g^{aa}\partial_m\Gamma^r_{aa}\delta^\alpha_r\delta^\beta_a + g^{rr}\Gamma^a_{ra}\Gamma^a_{am}\delta^\alpha_a\delta^\beta_r - g^{aa}\Gamma^r_{aa}\Gamma^n_{rm}\delta^\alpha_n\delta^\beta_a - g^{\beta\beta^\prime}\Gamma^r_{\beta^\prime m}\Gamma^a_{ra}\delta^\alpha_a + g^{aa}\Gamma^a_{am}\Gamma^r_{aa}\delta^\beta_a\delta^\alpha_r$. Therefore, e.g., $R^{ab}_{\ \ ab}\neq0,\ R^{ar}_{\ \ ar}\neq0$.
 \end{enumerate}
 Using these, and noting that $tr(R^4)$ will involve terms of the following three types:
 \begin{eqnarray}
 \label{trRfourth_types}
& & R^{\rho_1\lambda_1}_{\ \ \ \ a n_1}R_{\lambda_1\rho_2a_2n_2}R^{\rho_2\lambda_3}_{\ \ \ \ a_3n_3}R_{\lambda_3\rho_1a_4n_4}\prod_{i=1}^4dx^{a_i}\wedge dx^{n_i}\nonumber\\
& & R^{\rho_1\lambda_1}_{\ \ \ \ a_1a_2}R_{\lambda_1\rho_2m_2n_2}R^{\rho_2\lambda_3}_{\ \ \ \ a_3n_3}R_{\lambda_3\rho_1a_4n_4}dx^{a_1}\wedge dx^{a_2}\wedge dx^{m_2}\wedge dx^{n_2}\wedge dx^{a_3}\wedge dx^{n_3}\wedge dx^{a_4}\wedge dx^{n_4}\nonumber\\
& & R^{\rho_1\lambda_1}_{\ \ \ \ a_1a_2}R_{\lambda_1\rho_2a_3a_4}R^{\rho_2\lambda_3}_{\ \ \ \ m_3n_3}R_{\lambda_3\rho_1m_4n_4}\prod dx^{a_i}\wedge \prod_{i=1}^2dx^{m_i}\wedge dx^{n_i},
\end{eqnarray}
let us look at the three types of terms in (\ref{trRfourth_types}) individually below.
\begin{itemize}
\item
The first possibility will hence consist of the following set of terms:
\begin{eqnarray}
\label{trRfourth-typei}
& & R^{a_1r}_{\ \ \ a_1n_1}R_{r\rho_2a_2n_2}R^{\rho_2\lambda_3}_{\ \ \ \ a_3n_3}R_{\lambda_3a_1a_4n_4}\prod_{i=1}^4dx^{a_i}\wedge dx^{n_i},\nonumber\\
& & R^{a_1r}_{\ \ \ a_1n_1}R_{a_1\rho_2a_2n_2}R^{\rho_2\lambda_3}_{\ \ \ \ a_3n_3}R_{\lambda_3ra_4n_4}\prod_{i=1}^4dx^{a_i}\wedge dx^{n_i},\nonumber\\
& & R^{na_1}_{\ \ \ a_1n_1}R_{a_1\rho_2a_2n_2}R^{\rho_2\lambda_3}_{\ \ \ \ a_3n_3}R_{\lambda_3na_4n_4}\prod_{i=1}^4dx^{a_i}\wedge dx^{n_i},\nonumber\\
& & R^{a_1\lambda_1}_{\ \ \ a_1n_1}R_{\lambda_1\rho_2a_2n_2}R^{\rho_2\lambda_3}_{\ \ \ \ a_3n_3}R_{\lambda_3a_1a_4n_4}\prod_{i=1}^4dx^{a_i}\wedge dx^{n_i}.
\end{eqnarray}
Based on the three observations made above, each of the four terms in (\ref{trRfourth-typei})vanishes for reasons similar to:
\begin{eqnarray}
& & \hskip -0.8in R_{\lambda_3a_1a_4n_4}\prod_{i=1}^4dx^{a_i}\wedge dx^{n_i}\sim\delta^{a_4}_{\lambda_3}{\bf \delta^{a_1}_{n_4}}\prod_{i=1}^4dx^{a_i}\wedge dx^{n_i}=0\ {\rm or}\
\delta^r_{\lambda_3}\delta^r_{n_4}{\bf\delta_{a_1a_4}}\prod_{i=1}^4dx^{a_i}\wedge dx^{n_i}=0.
\end{eqnarray}

\item
The second possibility vanishes because:
\begin{eqnarray}
\label{trRfourth-typeiii}
& & R_{a_2\rho_2a_3a_4}\prod dx^{a_i}\wedge \prod_{i=1}^2dx^{m_i}\wedge dx^{n_i}\nonumber\\
& & \sim\left({\bf \delta^{a_2}_{a_3}}\delta^{a_4}_{\rho_2}\ {\rm or}\ {\bf\delta^{a_2}_{a_4}}\delta^{a_3}_{\rho_2}\right)\prod dx^{a_i}\wedge \prod_{i=1}^2dx^{m_i}\wedge dx^{n_i}=0.
\end{eqnarray}

\item
The third possibility vanishes because:
\begin{eqnarray}
\label{trRfourth-typeii}
& & \hskip -0.8in R_{\lambda_3a_1a_4n_4}dx^{a_1}\wedge dx^{a_2}\wedge dx^{m_2}\wedge dx^{n_2}\wedge dx^{a_3}\wedge dx^{n_3}\wedge dx^{a_4}\wedge dx^{n_4}\nonumber\\
& & \hskip -0.8in\sim\left(\delta^{a_4}_{\lambda_3}{\bf\delta^{a_1}_{n_4}}\ {\rm or}\ \delta^r_{\lambda_3}\delta^r_{n_4}{\bf \delta^{a_1a_4}}\right)dx^{a_1}\wedge dx^{a_2}\wedge dx^{m_2}\wedge dx^{n_2}\wedge dx^{a_3}\wedge dx^{n_3}\wedge dx^{a_4}\wedge dx^{n_4}=0.
\end{eqnarray}

\end{itemize}

Similarly, for $tr(R^2)^2=R^{\rho_1\lambda_1}_{\ \ \ \ \mu_1\nu_1}R_{\lambda_1\rho_1\mu_2\nu_2}R^{\rho_2\lambda_2}_{\ \ \ \ \mu_3\nu_3}R_{\lambda_2\rho_2\mu_4\nu_4}\prod_{i=1}^4dx^{\mu_i}\wedge dx^{\nu_i}$ there are the following types of terms:
\begin{itemize}
\item
\begin{eqnarray}
\label{trRsqsq-typei}
& & R^{\rho_1\lambda_1}_{\ \ \ \ a_1n_1}R_{\lambda_1\rho_1a_2n_2}R^{\rho_2\lambda_2}_{\ \ \ \ a_3n_3}R_{\lambda_2\rho_2a_4n_4}\prod_{i=1}^4dx^{a_i}\wedge dx^{n_i},
\end{eqnarray}
which vanishes because of reasons similar to:
\begin{eqnarray}
\label{type-i_poss}
& & R^{a_1r}_{\ \ \ a_1n_1}R_{ra_1a_2n_2}\prod_{i=1}^4dx^{a_i}\wedge dx^{n_i}\sim\delta^r_{n_2}{\bf\delta_{a_1a_2}}\prod_{i=1}^4dx^{a_i}\wedge dx^{n_i}=0.
\end{eqnarray}

\item
\begin{eqnarray}
\label{trsqsq-typeii}
& & R^{\rho_1\lambda_1}_{a_1a_2}R_{\ \ \ \lambda_1\rho_1m_2n_2}R^{\rho_2\lambda_2}_{\ \ \ \ a_3n_3}R_{\lambda_2\rho_2a_4n_4}dx^{a_1}\wedge dx^{a_2}\wedge dx^{m_2}\wedge dx^{n_2}\wedge dx^{a_3}\wedge dx^{n_3}\wedge dx^{a_4}\wedge dx^{n_4}\nonumber\\
& & \sim R^{a_1a_2}_{\ \ \ a_1a_2}{\bf R_{a_2a_1a_3n_3}}dx^{a_1}\wedge dx^{a_2}\wedge dx^{m_2}\wedge dx^{n_2}\wedge dx^{a_3}\wedge dx^{n_3}\wedge dx^{a_4}\wedge dx^{n_4}=0.
\end{eqnarray}

\item
\begin{eqnarray}
\label{trsqsq-typeiii}
& & R^{\rho_1\lambda_1}_{\ \ \ \ a_1a_2}R_{\lambda_1\rho_1a_3a_4}R^{\rho_2\lambda_2}_{\ \ \ \ m_3n_3}R_{\lambda_2\rho_2m_4n_4}\prod_{i=1}^4dx^{a_i}\wedge\prod_{j=3}^4dx^{m_j}\wedge dx^{n_j}\nonumber\\
& & \sim R^{a_1a_2}_{\ \ \ a_1a_2}R_{a_2a_1a_3a_4}\prod_{i=1}^4dx^{a_i}\wedge\prod_{j=3}^4dx^{m_j}\wedge dx^{n_j}\nonumber\\
& & \left({\bf\delta_{a_2a_3}\delta_{a_1a_4}}\ {\rm or}\ {\bf\delta_{a_2a_4}\delta_{a_1a_3}}\right)\prod_{i=1}^4dx^{a_i}\wedge\prod_{j=3}^4dx^{m_j}\wedge dx^{n_j}=0.
\end{eqnarray}

\end{itemize}

One sees, therefore,  the first and second Pontryagin classes of $TM_8$ satisfy: $$p_1^2(TM_8)=p_2(TM_8)=0$$ implying $$X_8(TM_8) = 4 p_2 - p_1^2=0,$$ and hence $$C_3\wedge X_8=0.$$

{{ \bf Counter-term evaluation :}} To summarize, in the limit (\ref{limits_Dasguptaetal-i}),
from (\ref{eq:intS_EH1}), (\ref{eq:Gibbons_Hawking iii}), (\ref{eq:intG4wedG4ii}), (\ref{eq:higherderv1}), one sees that the IR-divergent part of the action is given by:
\begin{equation}
\label{IR_i}
\frac{r_\Lambda^4}{r_h}\left(\frac{a_{EH}}{\epsilon } + a_{\rm GHY-boundary} + \epsilon^{9} a_{G_4} +
a_{R^4}\epsilon^{23} \right).
\end{equation}
The divergent part of action can be compensated by adding an appropriate counter-term corresponding to intrinsic boundary geometry so that the overall 11-dimensional Action renders finite contribution.

Using (\ref{limits_Dasguptaetal-i}) and equation (\ref{eq:simplifiedmetriccomponents}) and further simplifying near $\theta_{1,2}=0,\pi$, we have
{\small
\begin{eqnarray}
 && \sqrt{h^{\cal M}} \sim \left\{\frac{\sqrt{2} \sqrt{1 - \frac{r_h^4}{r_\lambda^4}} r^4 {f_2}({\theta_2}) \cot ^4({\theta_1}) \cos ({\theta_2})}{81 \sqrt[6]{3} \sqrt{\pi } {g_s}^{7/3} \sqrt{{g_s}
   N}}\right\}
\end{eqnarray}}
and using equation (\ref{eq:R}),
{\small
 \begin{eqnarray}
   && \sqrt{h^{\cal M}} R^{\cal M}\sim\left\{\frac{125 r^4 {f_2}({\theta_2})\sqrt{1-\frac{r^{4}_h}{r^4}} \cot ^2({\theta_1}) \csc ^4({\theta_1}) \cos ({\theta_2})}{864 \sqrt{2} 3^{5/6} \pi  {g_s}^{8/3} N}\right\}.
 \end{eqnarray} }
resulting in
{\small
\begin{eqnarray}
\label{eq:intSct}
&& {\hskip -0.4in} \frac{1}{8 \pi} \int_{x_{10}\in[0,{2\pi}], r\in[r_h,r_\Lambda],\theta_{1,2}\in[\epsilon_{\theta_{1,2}},\pi-\epsilon_{\theta_{1,2}}],\phi_{1,2}\in[0,2\pi], \psi\in[0,4\pi]}\sqrt{h^{\cal M}} R^{\cal M}|_{r=r_{\Lambda}}    \nonumber\\
&& {\hskip -0.4in} \sim \frac{125 r^{4}_{\Lambda} \pi^2 {f_2(\theta_2)} \sqrt{1-\frac{r^{4}_h}{ r^{4}_{\Lambda}}} \theta_1 \theta_2 \cot ^2({\theta_1}) \csc ^4({\theta_1}) \cos ({\theta_2})}{216 \sqrt{2} 3^{5/6} {g_s}^{8/3} N}  \Biggr|_{\theta_{1,2}=\epsilon_{\theta_{1,2}} +
\theta_{1,2}=\pi-\epsilon_{\theta_{1,2}}}. \nonumber\\
\end{eqnarray}
}
We hence see that
\begin{eqnarray}
\label{eq:ct1}
  && {\hskip -0.6in}  \frac{1}{8 \pi} \int_{x^0
\in \left[0,\frac {\pi \sqrt{4\pi g_s N}}{r_h}\right],x_{10}\in[0,{2\pi}], \theta_{1,2}\in[\epsilon_{\theta_{1,2}},\pi - \epsilon_{\theta_{1,2}}],\phi_{1,2}\in[0,2\pi], \psi\in[0,4\pi]} \sqrt{h^{\cal M}} R^{\cal M}|_{r=r_{\Lambda}} \nonumber\\
 &&\sim \epsilon^{\frac{19}{6}} \frac{r^4_{\Lambda}}{r_h}\sqrt{1 - \frac{r_h^4}{r_\Lambda^4}}\stackrel{r_\Lambda\rightarrow\infty}{\longrightarrow}\epsilon^{\frac{19}{6}}\frac{r_\Lambda^4}{r_h}.
  \end{eqnarray}
So, from (\ref{IR_i}) and (\ref{eq:ct1}), one sees that an appropriate counter term will be:
\begin{equation}
\label{ct-i}
-\epsilon^{-\frac{19}{6}}\left(\frac{a_{EH}}{\epsilon^{\frac{5}{3}} } + a_{\rm GHY-boundary} + \epsilon^{9} a_{G_4} +
a_{R^4}\epsilon^{22} \right)\int_{r=r_\Lambda}\sqrt{h}R^{\cal M}.
\end{equation}
One can show that:
\begin{eqnarray}
\label{sqrth+sqrthG4sq}
& & \int_{r=r_\Lambda}\sqrt{h}\sim\epsilon^{\kappa^{(i)}_{\rm cosmo}}\frac{r_\Lambda^4}{r_h},\nonumber\\
& & \int_{r=r_\Lambda}\sqrt{h}|G_4|^2\sim\epsilon^{\kappa^{(i)}_{\rm flux}}\frac{r_\Lambda^4}{r_h},
\end{eqnarray}
for appropriate $\kappa^{(i)}_{\rm cosmo/flux}$  using which one sees that one can also construct the following as the appropriate counter-terms:
\begin{eqnarray}
\label{ct-i2_3}
& & - \epsilon^{-\kappa^{(i)}_{\rm cosmo}}\left(\frac{a_{EH}}{\epsilon^{\frac{5}{3}} } + a_{\rm GHY-boundary} + \epsilon^{9} a_{G_4} +
a_{R^4}\epsilon^{22} \right)\int_{r=r_\Lambda}\sqrt{h},\nonumber\\
& & - \epsilon^{-\kappa^{(i)}_{\rm flux}}\left(\frac{a_{EH}}{\epsilon^{\frac{5}{3}} } + a_{\rm GHY-boundary} + \epsilon^{9} a_{G_4} +
a_{R^4}\epsilon^{22} \right)\int_{r=r_\Lambda}\sqrt{h}|G_4|^2.
\end{eqnarray}
Interestingly in the limit (\ref{limits_Dasguptaetal-i}), we argue below that one can give an asymptotically-linear-dilaton counter-term interpration to (\ref{ct-i}) and (\ref{ct-i2_3}). The behavior of surface counter-terms  in the asymptotic linear dilaton background,
 is discussed in  \cite{Mann+Mcnees}  by defining a set of linear dilaton(ADL) boundary conditions. According to the ADL boundary conditions given in  \cite{Mann+Mcnees}, the  metric is expanded as :
\begin{equation}
\label{eq:ALD1}
ds^2= d{r}^2 +r^2( h^{(0)}_{mn}+r^{2-d}( h^{(1)}_{mn}+r^{1-d}( h^{(0)}_{mn}+...)dy^m dy^n.
\end{equation}
The scalar field can be expanded as:
\begin{equation}
\label{eq:ALD2}
 \phi= {\bar \phi}ln{r} +\phi^{0} +r^{2-d} \phi^{1} + r^{1-d} \phi^{2}+....
 \end{equation}
where ${\bar \phi}$ and $\phi^{0}$ are constants, and remaining $ \phi^{i}$ are smooth functions. Similarly, the p-form field strength is expanded as
follows:
\begin{equation}
\label{eq:ALD3}
 F_{a_1....a_p}=  F^{(0)}_{a_1....a_p}+ r^{2-d}F^{(1)}_{a_1....a_p}+ r^{1-d}F^{(2)}_{a_1....a_p}+ ...
\end{equation}
Again as explained in \cite{Mann+Mcnees}, incorporating the boundary conditions (\ref{eq:ALD1})- (\ref{eq:ALD3}) in EOM corresponding to bulk integral
gives a condition
 \begin{equation}
 \label{eq:ALD4}
 \phi_0 + {\bar \phi}ln{r} = \frac{p-1}{\alpha}ln({\beta r}),
 \end{equation}
where $\beta= \frac{8(p-1)}{\alpha^2 Q^2}$ ($Q$ being related to p-form fluxes' components' magnitute - See \cite{Mann+Mcnees} for more details).
The counter-terms consistent with $\delta S=0$, are given to be:
\begin{equation}
\label{CT_ALD}
{\cal S}_{ct}= \int d^{d} x \,\sqrt{h}\left(c_0 e^{-\frac{\alpha}{p-1}\phi}+ c_1 e^{ \frac{\alpha}{p-1}\phi} R + c_2e^{ \frac{2 p-1}{p-1}\phi} F_p^2 \right)
\end{equation}
 where for the special case wherein $\alpha=\pm\left(\frac{p-1}{d-1}\right)$,
  $c_0 = 2 \beta -(d-1)^2 \beta^2 c_1$ and $c_2= -\frac{1}{2 p!}c_1$ for ALD boundary data and $c_1$ is arbitrary.

Now, the scalar field  corresponds to  geometric $x_{10}$ size modulus (when one takes the weak-coupling limit of $M$ theory compactifying it on a circle
of very small radius)  which is given by the applying triple T-duality on type IIB dilaton to yield the type IIA dilaton, i.e., $ \phi^{IIA}= \tilde {\tilde{\tilde{\phi}}}$. At least in the small $g_s$  and $g_s N_f$ limit (weakly coupled M theory description mentioned in (\ref{limits_Dasguptaetal-i})), one sees that for $r=r_\Lambda>>1$:
$\phi^{IIA}\sim  ln g_s - ln[1 - (3 g_s N_f/4\pi) ln r]$ and therefore, can be expressed as
\begin{equation}
\label{eq:ALD5}
\phi^{IIA}= \phi^{0} + \ \bar\phi \,ln{r_{\Lambda}} + {\cal O}((g_s N_f)^2)
\end{equation}
From (\ref{eq:ALD4}) and (\ref{eq:ALD5}), we get
$\phi^{IIA}=\frac{p-1}{\alpha}ln({\beta r_{\Lambda}})$. Sustituting the same in (\ref{CT_ALD}) will include the following set of terms ($p=4$):
\begin{equation}
\label{CT_ALD2}
c_1\left(-\frac{(d-1)^2\beta^2}{\beta r_\Lambda}\int_{r=r_\Lambda}\sqrt{h} + \beta r_\Lambda\int_{r=r_\Lambda}\sqrt{h}R - \frac{(\beta r_\Lambda)^7}{48}
\int_{r=r_\Lambda}\sqrt{h}|G_4|^2\right).
\end{equation}
The three terms in (\ref{CT_ALD2}) are IR-divergent and using an obvious notation, (\ref{CT_ALD2}) is given by:
\begin{equation}
\label{CT_ALD3}
c_1\frac{r_\Lambda^4}{r_h}\left(-\epsilon^{\kappa^{(i)}_{\rm cosmo}}a^{(i)}_{\rm cosmo}\frac{(d-1)^2\beta^2}{\beta r_\Lambda} + \epsilon^{\kappa^{(i)}_{\rm EH-boundary}}
a^{(i)}_{\rm EH-boundary}\beta r_\Lambda
 -\epsilon^{\kappa^{(i)}_{\rm flux}}a^{(i)}_{\rm flux} \frac{(\beta r_\Lambda)^7}{48} \right).
\end{equation}
Since the choice of $c_1$ is arbitrary, considering
\begin{equation}
\label{c_1}
c_1 =- \left(\frac{\frac{a_{EH}}{\epsilon^{\frac{5}{3}} } + a_{\rm GHY-boundary} + \epsilon^{9} a_{G_4} +
a_{R^4}\epsilon^{22} }{-\epsilon^{\kappa^{(i)}_{\rm cosmo}}a^{(i)}_{\rm cosmo}\frac{(d-1)^2\beta^2}{\beta r_\Lambda} + \epsilon^{\kappa^{(i)}_{\rm EH-boundary}}
a^{(i)}_{\rm EH-boundary}\beta r_\Lambda
 -\epsilon^{\kappa^{(i)}_{\rm flux}}a^{(i)}_{\rm flux} \frac{(\beta r_\Lambda)^7}{48}}\right)
\end{equation}
which cancels off the divergences coming from Einstein- Hilbert Action, Gibbons-Hawking-York term, flux term as well as ${\cal O}(R^4)$ as given in set of equations ((\ref{eq:intS_EH1}), (\ref{eq:Gibbons_Hawking iii}), (\ref{eq:intG4wedG4ii}), (\ref{eq:higherderv1})) in the weakly coupled description of M theory.

In limit (\ref{limits_Dasguptaetal-ii}), to summarize, from (\ref{eq:intS_EH2}), (\ref{eq:Gibbons_Hawking iv}), (\ref{eq:intG4wedG4iii}) and (\ref{eq:higherderv2}), we see that the IR-divergent contribution is given by:
\begin{equation}
\label{IR_ii}
\frac{r_\Lambda^4}{r_h}\left(\frac{a_{EH}(\alpha_{1,3,4,5})}{\epsilon^5} + a_{\rm GHY-boundary}(\alpha_{1,2,4,5}) + \epsilon^{19}\alpha_{G_4}(\alpha_{1,2,3,4,5}) + a_{R^4}(\alpha_{1,2,4,5})\epsilon^{42} \right).
\end{equation}
So, by arguments similar to the ones given to yield (\ref{ct-i}) as the counter term in limit (\ref{limits_Dasguptaetal-i}), the required counter term in limit (\ref{limits_Dasguptaetal-ii}) is:
\begin{equation}
\label{ct-ii}
-\epsilon^{-\frac{29}{6}}\left(\frac{a_{EH}(\alpha_{1,3,4,5})}{\epsilon^5} + a_{\rm GHY-boundary}(\alpha_{1,2,4,5}) + \epsilon^{19}\alpha_{G_4}(\alpha_{1,2,3,4,5}) + a_{R^4}(\alpha_{1,2,4,5})\epsilon^{42} \right)\int_{r=r_\Lambda}\sqrt{h}R^{\cal M}.
\end{equation}
Similar to (\ref{ct-i2_3}), one can also have the following counter terms:
{\small
\begin{eqnarray}
\label{ct_ii2_3}
& &\hskip -0.9in -\left(\epsilon^{-\kappa^{(ii)}_{\rm cosmo}}\int_{r=r_\Lambda}\sqrt{h},\epsilon^{-\kappa^{(ii)}_{\rm flux}}\int_{r=r_\Lambda}\sqrt{h}|G_4|^2\right)
\left(\frac{a_{EH}(\alpha_{1,3,4,5})}{\epsilon^5} + a_{\rm GHY-boundary}(\alpha_{1,2,4,5}) + \epsilon^{19}\alpha_{G_4}(\alpha_{1,2,3,4,5}) + a_{R^4}(\alpha_{1,2,4,5})\epsilon^{42} \right).\nonumber\\
& &
\end{eqnarray}}

To investigate the thermodynamic stability of the uplift, let us calculate the specific heat corresponding to the classical partition function/action calculated above wherein the finite part of the action is coming entirely from Gibbons-Hawking-York surface term, and ${\cal S}_{\cal E}^{\rm finite}\sim -r_h^3$ in the limits of (\ref{limits_Dasguptaetal-i}) and (\ref{limits_Dasguptaetal-ii}). Based on the argument given in \cite{S_ct-Perry et al}, the negative sign in the Action does not represent physical instability of the solution. Now, the average energy which is $\langle E\rangle =
 - T^2\frac{\partial {\cal S}_{\cal E}^{\rm finite}}{\partial T}
= - \frac{T^2(r_h)}{\frac{\partial T}{\partial r_h}}\frac{\partial {\cal S}_{\cal E}^{\rm finite}}{\partial r_h}$. In both limits, one sees from (exact or approximate form of) (\ref{T}) that $\frac{\partial T}{\partial r_h}>0$ implying $\langle E\rangle>0$. Now, the entropy $S = - \frac{1}{T(r_h)}\frac{T^2(r_h)}{\frac{\partial T}{\partial r_h}}
\frac{\partial {\cal S}_{\cal E}^{\rm finite}}{\partial r_h} - {\cal S}_{\cal E}^{\rm finite}$. Therefore the specific heat is given by: $C= \frac{T(r_h)}{\frac{\partial T}{\partial r_h}}\frac{\partial S}{\partial r_h}$. Again the aforementioned limit, one can show:
$\frac{T(r_h)}{\frac{\partial T}{\partial r_h}}\sim {r_h}$, which implies that entropy is positive and one can approximate the same
as $S\sim  r_h^3$. A quick check on this result arises from the fact that the entropy is expected to be proportional to the horizon area - from (\ref{horizon_area}) we see that as expected the entropy should scale like $r_h^3$. Using the same, therefore, $C\sim\frac{T(r_h)}{\frac{\partial T}{\partial r_h}}\frac{\partial \left(  r_h^3\right)}{\partial r_h}\sim  {r_h}^3 >0$ - implying a stable uplift!

\section{Results and Discussion}

We have  constructed a local M-theory uplift of a (resolved) warped deformed conifold using modified O(uyang) K(lebanov) S(trassler)-BH background given in \cite{metrics} in the context of type IIB string background relevant to study of thermal QCD with fundamental flavor thermal quarks. Following \cite{SYZ 3 Ts}, we define T-duality coordinates to have a local isometry along third direction $\psi$ in addtition to the global isometries along $\phi_1,\phi_2)$ and then apply suitable coordinate transformations on some of the angular coordinates to ensure the base of the local $T^3$-fibration is large  so that  mirror symmetry a la SYZ, can be applied (locally). Having done so, we first obtain type IIA metric formed by using analytic expressions given in \cite{SYZ 3 Ts} to mirror transform the type IIB metric components in the presence of a black hole in a warped deformed conifold and then obtain type IIA RR one-form gauge fields by applying T-duality rules on type IIB RR odd-form field strengths. Using T-dualized metric components, we calculate the contribution of the type IIA metric  along $r= \sqrt{3a}$ and $\theta_i\rightarrow \pi/2$ as well as $\theta_1\rightarrow 0~{\rm and}~\theta_2\rightarrow m \theta_1$ where $m \sim {\cal O}(1)$, and show that $G^{IIA}_{\theta_1\theta_2}$ vanishes if complex structure base/$f_i(\theta_i)$ along chosen values of $\theta_i$ is very large - one gets a warped resolved conifold. In other words, the argument that mirror of warped deformed conifold should be warped resolved conifold at least locally, automatically satisfies the condition of having large base required to implement mirror symmetry conditions proposed by Strominger-Yau-Zaslow. All of the above eventually leads to a local 11-dimensional M theory uplift.

The hydrodynamical as well as thermodynamical properties of strongly coupled (i.e. large t'Hooft coupling) gauge theories at finite temperature are governed by the presence of a black hole in the dual description, and therefore depends on horizon radius. We basically set up an approach to study the behavior of hydrodynamical as well as thermodynamical quantities in both weak coupling but large t'Hooft coupling regime of M theory accomplished by ($g_s, g_sN_f,
\frac{g_sM^2}{N}<<1,g_sM, g_sN>>1$ (c.f. \cite{metrics})) as well as MQGP limit accomplished by letting $g_s\stackrel{<}{\sim}1$  and $
\frac{g_sM^2}{N}<<1, g_sN>>1, {\rm finite}\ g_s, M, N_f$. The idea of discussing thermodynamical properties in this paper is two fold: the first is to check whether the solution  possesses the thermodynamical stability both in type IIB as well as local M theory uplift and the other is to explicitly verify whether gravity dual so obtained is able to show certain aspects of strongly coupled Plasma i.e QGP. The thermodynamical stability conditions are basically governed by inequalities imposed on certain thermodynamic quantities such as $\Delta S<0, \Delta E>0 ~{\rm and} ~\Delta H>0$ (deviations from
equilibrium values implied). In particular, considering the condition that $\Delta E(S,V,N)$ as well $\partial^2 E(S,V,N) >0$  and expanding $\partial^2 E(S,V,N)$ around equilibrium values of $(S_0, V_0, N_0)$ leads to satisfy three conditions  $C_v>0, \kappa >0 ~{\rm and}~ \frac{\partial\mu}{\partial N_f}\left.\right|_{T}>0$ for the system to be in stable thermodynamic equilibrium at constant value of S, V and N \cite{Bruno}. Keeping this in mind, we first calculate the chemical potential arising from $U(1)$ gauge field living on $N_f$ $D7$-branes wrapping a four-cycle of the warped deformed conifold as a function of temperature by considering Ouyang embedding in type IIB background and then study the behavior of the same as a function of $N_f$(number of bi-fundamental quark flavors) and show that $ \frac{\partial{\mu_C}}{\partial N_f}|_{T}>0$  upto linear order in embedding parameter $\mu$, thus obeying one of the conditions to achieve thermodynamic stability in type IIB background. Going ahead, by obtaining the local M theory uplift, we give a way to implement  both limits of M theory in thermodynamical calculations by choosing the scaling of relevant parameters $(g_s,M,N)$ scale $\epsilon$ as ($g_s \sim {\epsilon^{d}} $, $M\sim {\epsilon^{-\frac{3d}{2} }}$, $N \sim {\epsilon^{-19 d}}~ {\rm for}~  \epsilon \leq 0.01, d>0 $) and  ($g_s \sim  {\epsilon^{d}} $, $M\sim {\epsilon^{-\frac{3 d}{2}}}$, $N \sim {\epsilon^{-39 d}}~ {\rm for}~  \epsilon \lesssim 1, d>0$) consistent with
weak-string-coupling - large-t'Hooft-coupling  limit of M theory and the  `MQGP limit'
corresponding to finite-string-coupling - large-t'Hooft-coupling limit of M theory, respectively. Interestingly, we see that in both limits, the uplift produces
a black $M3$-brane whose near-horizon geometry near the $\theta_{1,2}=0,\pi$ branches, preserves $\frac{1}{8}$ supersymmetry. The other very important hydrodynamical quantities that one can obtain using dual description of 11-D supergravity background to verify the aspects of strongly coupled Q(uark) G(luon) P(lasma) is the shear viscosity-to-entropy ratio $(\eta/s)$ and diffusion coefficient D in type IIA and local M theory uplift. Considering the limits (\ref{limits_Dasguptaetal-i}) and (\ref{limits_Dasguptaetal-ii}), by first calculating the horizon radius and thereafter using the standard expressions
given in \cite{Kovtun} (the absence of angular dependence in $G^{IIA,{\cal M}}_{tt,rr,\mathbb{R}^3}$ and a tunably small chemical potential permitting the use of same), we show that shear viscosity-to-entropy ratio naturally comes out to be $\frac{1}{4 \pi}$ in type IIA and the limits (\ref{limits_Dasguptaetal-i}) and (\ref{limits_Dasguptaetal-ii}) of M theory and Diffusion coefficient D turns out to be $\frac{1}{T}$ in both type IIB and type IIA background. The results are consistent with the values that theorists expect for any quantum field theory description which has a gravity dual in (non)-extremal case \cite{buchel+liu} - the ${\cal O}(R^4)$-terms do not modify the value of $\frac{\eta}{s}$ because in both limits
(\ref{limits_Dasguptaetal-i}) and (\ref{limits_Dasguptaetal-ii}), (the IR-finite part of) the same vanishes . We next evaluate the $D=11$ Euclideanized supergravity action (Einstein-Hilbert + Hawking surface + Flux terms+${\cal O}(R^4)$terms) in the two limits. This action is expected to receive the maximum
contribution near $\theta=0,\pi$  as it possesses pole singularities near the
same - these can be regularized by considering a small ${\theta_{1,2}}$ cut-off i.e ${\theta_{1,2}} \in [\epsilon_{\theta},\pi-\epsilon_{\theta}]$ with ${\theta_{1,2}}\sim {\epsilon}^{\gamma_{(i),(ii)}}$ for $\gamma_{(i),(ii)}$ appropriate to the limits
(\ref{limits_Dasguptaetal-i}) and (\ref{limits_Dasguptaetal-ii}) in such a way
that the finite contribution (coming solely from the Gibbons-Hawking-York surface
term)  is independent of the cut-off  ${\epsilon_{\theta}}$. We show that the
IR divergence can be removed by adding appropriate surface counter terms:
 $\int_{r=r_{\Lambda}}\sqrt{h}\left(1,{R},|G_4|^2\right)$; these counter terms in
the weak-string-coupling - large-t'Hooft-coupling
limit (\ref{limits_Dasguptaetal}) can be understood in terms of asymptotically-
linear-dilaton-gravity type surface counter terms. The finite piece of the
 Gibbons-Hawking surface term turns out to be proportional to $- {r_h}^3$.  We also verify entropy as well as specific heat obtained using the partition function turns out to be positive, thus obeying one of the conditions for thermodynamical stability of 11-dimensional M-theory solution.  Therefore, local 11-Dimensional M-theory uplift so-obtained is able to provide some theoretical insight into the physics of  strongly coupled Quark Gluon Plasma.

To conclude, we will briefly discuss the construction of the inversion of the map effecting  an exact  isometry along ``$\psi^\prime$" for a deformed conifold as obtained in \cite{Minasian+Tsimpis}. The same can be shown to be along $\psi^\prime$ via the following change of coordinate system: $(\psi,\phi_2,\theta_2)\rightarrow(\psi^\prime,\phi_2^\prime,\theta_2^\prime)$ via:
\begin{eqnarray}
\label{psiprimeetal}
& & \psi^\prime = \psi + tan^{-1}\left(\frac{sin\phi_2+sin\theta_2}{cos\phi_2 cos\theta_2}\right),\nonumber\\
& & \phi^\prime_2 = \psi + \frac{1}{2}\left(tan^{-1}\left(\frac{sin\phi_2+sin\theta_2}{cos\phi_2 cos\theta_2}\right) + tan^{-1}\left(\frac{sin\phi_2-sin\theta_2}{cos\phi_2 cos\theta_2}\right)\right),\nonumber\\
& & \theta_2^\prime=h(sin\phi_2 sin\theta_2),
\end{eqnarray}
where $h(sin\phi_2 sin\theta_2)$ is an arbitrary function of $sin\phi_2sin\theta_2$; we take $h=sin\phi_2 sin\theta_2$. To invert (\ref{psiprimeetal}), by defining $\alpha_1\equiv tan\psi, \alpha_2\equiv tan\theta_2, \alpha_3=tan\phi_2$, one can convert (\ref{psiprimeetal}) to the following algebraic equations in $(\alpha_1,\alpha_2,\alpha_3)$:
\begin{eqnarray}
\label{algalphais}
& & \frac{tan\psi^\prime - \alpha_1}{1 + tan\psi^\prime \alpha_1} = \alpha_3\sqrt{1 + \alpha_2^2} + \alpha_2\sqrt{1 + \alpha_3^2},\nonumber\\
& & \frac{tan\left(2\phi_2^\prime\right) - \frac{2\alpha_1}{1 - \alpha_1^2}}{1 + \frac{2\alpha_1 tan\left(2\phi_2^\prime\right)}{1 - \alpha_1^2}} = \frac{\alpha_3\sqrt{1+\alpha_2^2}}{1 + \alpha_2^2 - \alpha_3^2},\nonumber\\
& & \theta_2^\prime = \frac{\alpha_3\alpha_2}{\sqrt{\left(1 + \alpha_3^2\right)\left(1 + \alpha_2^2\right)}}.
\end{eqnarray}
The three equations can be used to obtain the following equation in $\alpha_2$:
{\small
\begin{eqnarray}
\label{alpha2-i}
& & \hskip -0.7in \frac{\tan (2 {\Phi_2^\prime}) \left(\frac{\left(-\sqrt{{\alpha_2}^2 \left(-\left(\theta_2^\prime\ ^2-1\right)\right)-\theta_2^\prime\ ^2} \tan
   ({\psi^\prime})+{\alpha_2}^2 (\theta_2^\prime\ +1)+\theta_2^\prime\ \right)^2}{\left(\left({\alpha_2}^2
   (\theta_2^\prime\ +1)+\theta_2^\prime\ \right) \tan ({\psi^\prime})+\sqrt{{\alpha_2}^2
   \left(-\left(\theta_2^\prime\ ^2-1\right)\right)-\theta_2^\prime\ ^2}\right)^2}-1\right)-\frac{2 \left(-\sqrt{{\alpha_2}^2
   \left(-\left(\theta_2^\prime\ ^2-1\right)\right)-\theta_2^\prime\ ^2} \tan ({\psi^\prime})+{\alpha_2}^2
   (\theta_2^\prime\ +1)+\theta_2^\prime\ \right)}{\left({\alpha_2}^2 (\theta_2^\prime\ +1)+\theta_2^\prime\ \right) \tan
   ({\psi^\prime})+\sqrt{{\alpha_2}^2 \left(-\left(\theta_2^\prime\ ^2-1\right)\right)-\theta_2^\prime\ ^2}}}{\frac{2 \tan (2 {\Phi_2^\prime})
   \left(-\sqrt{{\alpha_2}^2 \left(-\left(\theta_2^\prime\ ^2-1\right)\right)-\theta_2^\prime\ ^2} \tan ({\psi^\prime})+{\alpha_2}^2
   (\theta_2^\prime\ +1)+\theta_2^\prime\ \right)}{\left({\alpha_2}^2 (\theta_2^\prime\ +1)+\theta_2^\prime\ \right) \tan
   ({\psi^\prime})+\sqrt{{\alpha_2}^2 \left(-\left(\theta_2^\prime\ ^2-1\right)\right)-\theta_2^\prime\ ^2}}+\frac{\left(-\sqrt{{\alpha_2}^2
   \left(-\left(\theta_2^\prime\ ^2-1\right)\right)-\theta_2^\prime\ ^2} \tan ({\psi^\prime})+{\alpha_2}^2
   (\theta_2^\prime\ +1)+\theta_2^\prime\ \right)^2}{\left(\left({\alpha_2}^2 (\theta_2^\prime\ +1)+\theta_2^\prime\ \right) \tan
   ({\psi^\prime})+\sqrt{{\alpha_2}^2 \left(-\left(\theta_2^\prime\ ^2-1\right)\right)-\theta_2^\prime\ ^2}\right)^2}-1}\nonumber\\
   & & \hskip -0.7in = {\alpha_2} \sqrt{-\frac{{\alpha_2}^2}{{\alpha_2}^2 \left(\theta_2^\prime\ ^2-1\right)+\theta_2^\prime\ ^2}}+\frac{\sqrt{{\alpha_2}^2+1}
   \sqrt{\left(-{\alpha_2}^2-1\right) \theta_2^\prime\ ^2}}{\sqrt{{\alpha_2}^2 \left(\theta_2^\prime\ ^2-1\right)+\theta_2^\prime\ ^2}},
\end{eqnarray}}
where $\Phi_2^\prime\equiv\phi_2^\prime - \psi^\prime$. Solving (\ref{alpha2-i}) exactly is intractable. We will solve the same near $\theta_2=0$ implying $\psi^\prime\approx\phi_2^\prime$ and $\theta_2^\prime\approx0$. In this limit working up ${\cal O}(\theta_2^\prime)$, (\ref{alpha2-i}) simplifies to:
\begin{eqnarray}
\label{alpha2-ii}
& &  \left(\alpha_2^2 (\theta_2^\prime-1)+\theta_2^\prime\right) \left(\left(\alpha_2^2-1\right) \cos (2 \Phi_2^\prime)+2 \alpha_2 \sin (2
   \Phi_2^\prime))\right)^2-2 \alpha_2^2 \cos (2 \Phi_2^\prime)\nonumber\\
   & & +\left(\alpha_2^2-1\right) \alpha_2 \sin (2
   \Phi_2^\prime) \left(\left(\alpha_2^2-1\right) \cos (2 \Phi_2^\prime)
   +2 \alpha_2 \sin (2
   \Phi_2^\prime)\right) + 2 \left(\alpha_2^2+1\right)^2 \theta_2^\prime = 0.
\end{eqnarray}
Via the transformation: $y= \alpha_2^2 + {\cal A} \alpha_2 + {\cal B}$ one obtains the quintic $y^5 + \sum_{n=1}^5a_n y^{5-n} = 0$ wherein the coefficients of $y^{3,4}$ vanish from where:
\begin{eqnarray}
\label{quintic-1}
& & a_3 = \frac{13}{12\sqrt{15}}\theta_2^\prime cosec^3\Phi_2^\prime + {\cal O}(\theta_2^\prime\ ^2),\nonumber\\
& & a_4 = \frac{3677}{1728\sqrt{15}}\theta_2^\prime cosec^3\Phi_2^\prime + {\cal O}(\theta_2^\prime\ ^2),\nonumber\\
& & a_5 = - \frac{101}{1440\sqrt{15}}\theta_2^\prime cosec^3\Phi_2^\prime + {\cal O}(\theta_2^\prime\ ^2).
\end{eqnarray}
Using the transformation $z = y^4 + a y^3 + b y^2 + c y + d$, the aforementioned quintic in y gets transformed into
$z^5 + \sum_{n=1}^5b_n z^{5-n} = 0$ wherein demanding the absence of $z^{2,3,4}$ terms yields:
\begin{eqnarray}
\label{quintic-2}
& & d = \frac{\left(3677 + 1404 a\right)}{2160\sqrt{15}}\theta_2^\prime cosec^3\Phi_2^\prime + {\cal O}(\theta_2^\prime\ ^2),\nonumber\\
& & b = \frac{303 - 735 a}{2808} + {\cal O}(\theta_2^\prime),\nonumber\\
& & a = -\frac{36681331117042986782161 y cosec^3({\Phi_2^\prime})}{714214131296639501160 \sqrt{15}}-\frac{303 \left(212706 cosec^3({\Phi_2^\prime})-13733035\right)}{102556979314}  + {\cal O}(\theta_2^\prime\ ^2),\nonumber\\
& & c = 0.
\end{eqnarray}
It turns out that the coefficient of the term linear in $z$ is  $\sim10^{-18}\theta_2^\prime\ ^2 cosec^{21}\Phi_2^\prime$ and the constant is $10^{-12}\theta_2^\prime\ ^2 cosec^{18}\Phi_2^\prime$ which we will disregard. Hence, up to ${\cal O}(\theta_2^\prime)$ one obtains $z=0$ implying a quartic in $y$ to be solved which can be shown to have $y=0$ to be a solution. This implies solving a quadratic in $\alpha_2:\alpha_2^2 + {\cal A} \alpha_2 + {\cal B}$=0 where
\begin{eqnarray}
\label{quintic-3}
& & {\cal A} \sim \frac{11}{18}\theta_2^\prime cosec\Phi_2^\prime,\nonumber\\
& & {\cal B} \sim - \frac{31}{24} + \frac{19\sqrt{15}}{120} \theta_2^\prime cosec\Phi_2^\prime.
\end{eqnarray}
This yields the following inversion map:
\begin{eqnarray}
\label{inversion}
& & \theta_2 = tan^{-1}\left(\frac{9}{8}\sqrt{\frac{3}{5}}\theta_2^\prime\right) + {\cal O}(\theta_2^\prime\ ^2),
\nonumber\\
& & \psi = tan^{-1}\left(tan\left[\psi^\prime - tan^{-1}\left(\frac{40}{\sqrt{385}}\right)\right] - \frac{243\sqrt{385}\theta_2^\prime}{8\left(\sqrt{385} cos\psi^\prime + 40 sin\psi^\prime\right)}+ {\cal O}(\theta_2^\prime\ ^2)\right),\nonumber\\
& & \phi_2 = tan^{-1}\left(8\sqrt{\frac{5}{77}} + {\cal O}(\theta_2^\prime\ ^2)\right).
\end{eqnarray}

Using the following from \cite{Minasian+Tsimpis}:
\begin{equation}
\label{differentials}
\left(\begin{array}{c} d\phi_2 \\ d\theta_2\\ d\psi \end{array}\right) = \left(\begin{array}{ccc} - cot\theta_2 sin\phi_2 & cot\theta_2 sin\phi_2 & \frac{cos\phi_2}{h^\prime\left(cos^2\phi_2 + cos^2\theta_2 sin^2\phi_2\right)} \\
cos\phi_2 & - cos\phi_2 & \frac{cos\theta_2 sin\phi_2}{h^\prime\left(cos^2\phi_2 + cos^2\theta_2 sin^2\phi_2\right)}\\
\frac{sin\phi_2}{sin\theta_2} & 1 - \frac{sin\phi_2}{sin\theta_2} & - \frac{cos\phi_2 cot\theta_2}{h^\prime\left(cos^2\phi_2 + cos^2\theta_2 sin^2\phi_2\right)}
 \end{array}\right)\left(\begin{array}{c} d\psi^\prime \\ d\phi_2^\prime \\ d\theta_2^\prime \end{array}\right)
\end{equation}
So, the $\psi$-dependent term of the deformed conifold:
\begin{equation}
\label{psidep_i}
cos\psi\left(d\theta_1d\theta_2 - sin\theta_1 sin\theta_2 d\phi_1d\phi_2\right) +
sin\psi\left(sin\theta_1 d\phi_1d\theta_1 + sin\theta_2d\phi_2d\theta_1\right)
\end{equation}
transforms to:
\begin{eqnarray}
& & \frac{sin\phi_2^\prime}{\sqrt{cos^2\theta_2 + cos^2\theta_2 sin^2\phi_2}} d\theta_1 d\theta_2^\prime -
\sin\theta_1 sin\phi_2^\prime \sqrt{cos^2\theta_2 + cos^2\theta_2 sin^2\phi_2} d\phi_1 d\Phi_2^\prime \nonumber\\
& & + cos\phi_2^\prime \sqrt{cos^2\theta_2 + cos^2\theta_2 sin^2\phi_2} d\theta_1 d\Phi_2^\prime
+ \frac{cos\phi_2^\prime}{\sqrt{cos^2\theta_2 + cos^2\theta_2 sin^2\phi_2}} d\phi_1 d\theta_2^\prime,
\end{eqnarray}
has components independent of $\psi^\prime$ though this occurs at the cost of losing the $\phi_2^\prime$ -
isometry.

\section*{Acknowledgements}

One of us (MD) is supported by a CSIR, Govt. of India, Senior Research Fellowship and the work of AM is partly supported by a CSIR project number CSR-656-PHY as well as an associateship at the Abdus Salam International Centre for Theoretical Physics, Trieste, Italy. MD would like to thank J.Schmude for a useful clarification and AM would like to thank K.Dasgupta and A.Das for useful clarifications, and B.K.Patra (co-investigator on the aforementioned project) for discussing the basics of QGP with him.

\appendix
\setcounter{equation}{0} \seceqaa
\section{Details of Local type IIA SYZ Mirror}

In this appendix, after applying T-duality, locally, along $x, y, z$ and in this order, we give explicitly  (i) the type IIA components obtained, and (ii) the components of type IIA two-form fluxes obtained from type IIB one-form, three-form and (self-dual) five-form field strengths.

\subsection{Type IIA Metric Components}

 The various components of the metric after three successive T-dualities along $x, y$ and $z$ respectively, can be written as (\cite{SYZ 3 Ts}):
 {\small
\begin{eqnarray}
\label{G_munu}
& & G_{\mu\nu} = {g_{\mu\nu}g_{xx} -
g_{x\mu}g_{x\nu} + b_{x\mu}b_{x\nu} \over g_{xx}} -
{(g_{y\mu}g_{xx} - g_{xy} g_{x \mu} + b_{xy} b_{x\mu})
(g_{y\nu}g_{xx}
 - g_{xy} g_{x \nu} + b_{xy} b_{x\nu}) \over
g_{xx}(g_{yy}g_{xx}- g_{xy}^2 + b_{xy}^2)} \nonumber\\
&& + {(b_{y\mu}g_{xx} - g_{xy} b_{x \mu} + b_{xy}
g_{x\mu})(b_{y\nu}g_{xx}
 - g_{xy} b_{x \nu} + b_{xy} g_{x\nu})\over
g_{xx}(g_{yy}g_{xx}- g_{xy}^2 + b_{xy}^2)},
\end{eqnarray}

\begin{eqnarray}
\label{Gmuz}
& & G_{\mu z} = {g_{\mu z}g_{xx} -
g_{x\mu}g_{xz} + b_{x \mu}b_{xz} \over g_{xx}} - {(g_{y\mu}g_{xx}
- g_{xy} g_{x \mu} + b_{xy} b_{x\mu}) (g_{yz}g_{xx} - g_{xy} g_{x
z} + b_{xy} b_{xz}) \over g_{xx}(g_{yy}g_{xx}- g_{xy}^2 +
b_{xy}^2)} \nonumber\\
& &  + {(b_{y\mu}g_{xx} - g_{xy} b_{x \mu} +
b_{xy} g_{x\mu})(b_{yz}g_{xx} - g_{xy} b_{x z} + b_{xy}
g_{xz})\over g_{xx}(g_{yy}g_{xx}- g_{xy}^2 + b_{xy}^2)},
\end{eqnarray}

\begin{eqnarray}
\label{Gzz}
& & G_{zz} =  {g_{zz}g_{xx} - j^2_{xz} +
b^2_{xz}\over g_{xx}} - {(g_{yz}g_{xx} - g_{xy} g_{xz} + b_{xy}
b_{xz})^2 \over g_{xx}(g_{yy}g_{xx}- g_{xy}^2 + b_{xy}^2)} + {(b_{yz}g_{xx} - g_{xy} b_{x z} + b_{xy} g_{xz})^2 \over
g_{xx}(g_{yy}g_{xx}- g_{xy}^2 + b_{xy}^2)},
\end{eqnarray}

\begin{eqnarray}
\label{Gymu}
& & G_{y \mu} = -{b_{y \mu} g_{xx} - b_{x \mu} g_{xy} + b_{xy}
 g_{\mu x} \over g_{yy}g_{xx}- g_{xy}^2 + b_{xy}^2},
~ G_{y z} = -{b_{y z} g_{xx} - b_{x z} g_{xy} + b_{xy} g_{z x}
\over g_{yy}g_{xx}- g_{xy}^2 + b_{xy}^2},
\end{eqnarray}

\begin{eqnarray}
\label{Gyy}
& & G_{yy} = {g_{xx} \over g_{yy}g_{xx}- g_{xy}^2 +
b_{xy}^2},~ G_{xx} = {g_{yy} \over g_{yy}g_{xx}- g_{xy}^2 +
b_{xy}^2}, ~G_{xy} = {-g_{xy} \over g_{yy}g_{xx}- g_{xy}^2 +
b_{xy}^2},
\end{eqnarray}

\begin{eqnarray}
\label{Gmux}
& & G_{\mu x} = {b_{\mu x} \over g_{xx}} + {(g_{\mu y} g_{xx} -
 g_{xy} g_{x \mu} + b_{xy} b_{x \mu}) b_{xy} \over
g_{xx}(g_{yy}g_{xx}- g_{xy}^2 + b_{xy}^2)}
+ {(b_{y \mu} g_{xx} - g_{xy} b_{x \mu} + b_{xy} g_{x \mu}) g_{xy}
 \over g_{xx}(g_{yy}g_{xx}- g_{xy}^2 + b_{xy}^2)},
 \end{eqnarray}

\begin{eqnarray}
\label{Gzx}
& & G_{z x} = {b_{z x} \over g_{xx}} + {(g_{z y} g_{xx} -
g_{xy} g_{x z} + b_{xy} b_{x z}) b_{xy} \over
g_{xx}(g_{yy}g_{xx}- g_{xy}^2 + b_{xy}^2)}
 + {(b_{y z} g_{xx} - g_{xy} b_{x z} + b_{xy} g_{xz}) g_{xy}
  \over g_{xx}(g_{yy}g_{xx}- g_{xy}^2 + b_{xy}^2)}.
\end{eqnarray}}
In the above formulae we have denoted the type IIB
$B$ fields as $b_{mn}$.  For the generic case we will switch on all the
components of the $B$ field
{\small
\begin{eqnarray}
b & = &
b_{\mu\nu} ~ dx^\mu \wedge dx^\nu + b_{x \mu} dx \wedge dx^\mu +  b_{y \mu}
~ dy~\wedge dx^\mu + b_{z \mu} ~ dz \wedge dx^\mu \nonumber\\
& & + ~ b_{xy}
~ dx \wedge dy +
 b_{xz} ~ dx  \wedge dz +  b_{zy}~ dz  \wedge dy.
 \end{eqnarray}}
\noindent After applying again the T-dualities, the type IIA NS-NS $B$ field in the mirror set-up will take the form
{\small
\begin{eqnarray}
\label{B}
B^{IIA} & = & \left( {\cal B}_{\mu\nu}
+ {2 {\cal B}_{z[\mu} G_{\nu]z} \over G_{zz}} \right) dx^\mu
\wedge dx^\nu + \left( {\cal B}_{\mu x} + {2 {\cal B}_{z[\mu}
G_{x]z} \over G_{zz}}\right)
 dx^\mu \wedge dx  \nonumber\\
& & \left( {\cal B}_{\mu y} + {2 {\cal B}_{z[\mu} G_{y]z} \over G_{zz}}
 \right) dx^\mu \wedge dy
+ \left( {\cal B}_{xy} + {2 {\cal B}_{z[x} G_{y]z} \over G_{zz}}
\right) dx \wedge dy \nonumber\\
& &  + {G_{z \mu} \over G_{zz}} dx^\mu
\wedge dz + {G_{z x} \over G_{zz}} dx \wedge dz + {G_{z y} \over
G_{zz}} dy \wedge dz.
\end{eqnarray}}
 Here the $G_{mn}$ components have been
given above, and the various ${\cal B}$  components can now be
written as
{\small
\begin{eqnarray}
\label{Bmunu}
{\cal B}_{\mu\nu} & = &
{b_{\mu\nu} g_{xx} + b_{x \mu} g_{\nu x} - b_{x \nu} g_{\mu x}
\over g_{xx}} \nonumber\\
& & +  {2 (g_{y[\mu}g_{xx} - g_{xy}g_{x[\mu} +
b_{xy} b_{x[\mu}) (b_{\nu]y}g_{xx} - b_{\nu]x}g_{xy} - b_{xy}
g_{\nu]x}) \over g_{xx}(g_{yy}g_{xx}- g_{xy}^2 + b_{xy}^2)},
\end{eqnarray}

\begin{eqnarray}
\label{Bmuz}
{\cal B}_{\mu z} & = &  {b_{\mu z} g_{xx} +
b_{x \mu} g_{z x} - b_{x z} g_{\mu x} \over g_{xx}} \nonumber\\
& & +  {2
(g_{y[\mu}g_{xx} - g_{xy}g_{x[\mu} + b_{xy} b_{x[\mu})
(b_{z]y}g_{xx} - b_{z]x}g_{xy} - b_{xy} g_{z]x}) \over
g_{xx}(g_{yy}g_{xx}- g_{xy}^2 + b_{xy}^2)},
\end{eqnarray}

\begin{eqnarray}
\label{Bmuy}
{\cal B}_{\mu y} & = & {g_{\mu y} g_{xx} - g_{xy} g_{x \mu} + b_{xy} b_{x \mu}
 \over g_{yy}g_{xx}- g_{xy}^2 + b_{xy}^2},\nonumber\\
{\cal B}_{z y} & = & {g_{z y} g_{xx} - g_{xy} g_{x z} + b_{xy} b_{x z} \over g_{yy}g_{xx}-
 g_{xy}^2 + b_{xy}^2},
 \end{eqnarray}

\begin{eqnarray}
\label{Bmux}
{\cal B}_{\mu x} & = & {g_{\mu x} \over g_{xx}} - {g_{xy} (g_{\mu y} g_{xx} -
g_{xy} g_{x \mu} + b_{xy} b_{x \mu}) \over
g_{xx}(g_{yy}g_{xx}- g_{xy}^2 + b_{xy}^2)} + {b_{xy} (b_{x\mu}g_{xy} - b_{y\mu}g_{xx} -
b_{xy}g_{xz})
 \over g_{xx}(g_{yy}g_{xx}- g_{xy}^2 + b_{xy}^2)},
 \end{eqnarray}

\begin{eqnarray}
\label{Bzx}
{\cal B}_{z x} & = & {g_{z x} \over g_{xx}} - {g_{xy} (g_{z y} g_{xx} -
g_{xy} g_{xz} + b_{xy} b_{x z}) \over
g_{xx}(g_{yy}g_{xx}- g_{xy}^2 + b_{xy}^2)} + {b_{xy} (b_{xz}g_{xy} - b_{yz}g_{xx} -
 b_{xy}g_{xz})
\over g_{xx}(g_{yy}g_{xx}- g_{xy}^2 + b_{xy}^2)},
\end{eqnarray}

\begin{eqnarray}
\label{Bxy}
{\cal B}_{xy} & = & {-b_{xy} \over g_{yy}g_{xx}- g_{xy}^2 + b_{xy}^2}.
\end{eqnarray}}
The analytic expressions of non-zero type IIA ${\cal B}$ components are:
\begin{eqnarray}
\label{B-3tduals}
& (i) & B^{IIA}_{xz}=
 \Bigl(-54 \sin ^3({\theta_1}) \cos ({\theta_1}) \left(\sqrt{6} \left(9 h_ 5^2-1\right)+4 h_ 5 \cot
   ({\theta_2})\right)+81 h_ 5 \bigl(2 \sin ^4({\theta_1}) \left(3 h_ 5+\sqrt{6} \cot
   ({\theta_2})\right) \nonumber\\
   && +h_5 \sin ^2(2 {\theta_1})\bigr)-12 \sin ({\theta_1}) \cos ^3({\theta_1}) \left(12
   h_ 5 \cot ({\theta_2})+2 \sqrt{6} \cot ^2({\theta_2})-3 \sqrt{6}\right)+12 \sin ^2({\theta_1}) \cos
   ^2({\theta_1}) \nonumber\\
   && \cot ({\theta_2}) \left(27 \sqrt{6} h_ 5+2 \cot ({\theta_2})\right)+16 \cos ^4({\theta_1})
   \cot ^2({\theta_2}) \Bigr)/\Bigl(3 (\cos (2 {\theta_1})-5) \Bigl(\sin ^2({\theta_1}) \bigl(-27 h_ 5^2+2 \cot
   ^2({\theta_2}) \nonumber\\
   &&+3\bigr) +6 h_ 5 \sin (2 {\theta_1}) \cot ({\theta_2})+2 \cos ^2({\theta_1})\Bigr)\Bigr), \nonumber\\
  & (ii) &B^{IIA}_{yz}= -\frac{3 \sqrt{6} (3 h_ 5 \cot ({\theta_1})+\cot ({\theta_2}))}{-27 h_ 5^2+12 h_ 5 \cot ({\theta_1})
   \cot ({\theta_2})+2 \cot ^2({\theta_1})+2 \cot ^2({\theta_2})+3} \nonumber\\
   & (iii) & B^{IIA}_{\theta1 x}= \Bigl({f_ 1}({\theta_1}) \sin ^2({\theta_1}) \left(\sin ({\theta_1}) \left(27 h_ 5^2-2 \cot
   ^2({\theta_2})-3\right)-6 h_ 5 \cos ({\theta_1}) \cot ({\theta_2})\right)\nonumber\\
   && \Bigl({g_s} \Bigl(18
   {g_s}^2 {M_{eff}}^2 {N_f} \log ^2(r)+3 {g_s} {M_{eff}}^2 \log (r) \left({g_s} {N_f} \log \left(\sin
   \left(\frac{{\theta_1}}{2}\right) \sin \left(\frac{{\theta_2}}{2}\right)\right)+3 {g_s} {N_f}+4 \pi \right) \nonumber\\
   && +8
   \pi ^2 N\Bigr)\Bigr)^{\frac{1}{4}}\Bigr)/\Bigl(2^{3/4} \sqrt{3} \sqrt[4]{\pi } \left(\sin ^2({\theta_1}) \left(-27 h_ 5^2+2 \cot
   ^2({\theta_2})+3\right)+6 h_ 5 \sin (2 {\theta_1}) \cot ({\theta_2})+2 \cos ^2({\theta_1})\right)\Bigr) \nonumber\\
& (iv) & B^{IIA}_{\theta2 x}= \Bigl(\sqrt[4]{\frac{2}{\pi }} r {f_ 2}({\theta_2}) \sin ({\theta_1}) \sin ({\theta_2}) \cos ({\theta_2}) (3
   h_ 5 \sin ({\theta_1}) \cos ({\theta_2})+\cos ({\theta_1}) \sin ({\theta_2}))\nonumber\\
   &&  \Bigl({g_s}
   \Bigl(18 {g_s}^2 {M_{eff}}^2 {N_f} \log ^2(r)+3 {g_s} {M_{eff}}^2 \log (r) \left({g_s} {N_f} \log
   \left(\sin \left(\frac{{\theta_1}}{2}\right) \sin \left(\frac{{\theta_2}}{2}\right)\right)+3 {g_s} {N_f}+4 \pi
   \right)\nonumber\\
   && +8 \pi ^2 N\Bigr)\Bigr)^{\frac{1}{4}}\Bigr)/\Bigl(\sqrt{3} \Bigl(3 \left(-9 h_ 5^2 \sin ^2({\theta_1}) \sin
   ^2({\theta_2})+h_ 5 \sin (2 {\theta_1}) \sin (2 {\theta_2})+\sin ^2({\theta_1}) \sin
   ^2({\theta_2})\right)\nonumber\\
   && +2 \sin ^2({\theta_1}) \cos ^2({\theta_2})+2 \cos ^2({\theta_1}) \sin
   ^2({\theta_2})\Bigr)\Bigr)\nonumber\\
& (v) & B^{IIA}_{\theta1y}=
\Bigl(\sqrt[4]{\frac{2}{\pi }}  {f_ 1}({\theta_1}) \cos ({\theta_1}) (3 h_ 5 \cot ({\theta_1})+\cot
   ({\theta_2})) \Bigl({g_s} \Bigl(18 {g_s}^2 {M_{eff}}^2 {N_f} \log ^2(r)+ \nonumber\\
   &&3 {g_s} {M_{eff}}^2
   \log (r) \left({g_s} {N_f} \log \left(\sin \left(\frac{{\theta_1}}{2}\right) \sin
   \left(\frac{{\theta_2}}{2}\right)\right)+3 {g_s} {N_f}+4 \pi \right)+8 \pi ^2 N\Bigr)\Big)^{\frac{1}{4}}\Bigr)/\Bigl(\sqrt{3} \Bigl(-27
   h_ 5^2 \nonumber\\
   && +12 h_ 5 \cot ({\theta_1}) \cot ({\theta_2})+2 \cot ^2({\theta_1})+2 \cot
   ^2({\theta_2})+3\Bigr)\Bigr)\nonumber\\
& (vi) & B^{IIA}_{\theta_2 y}=
\Bigl(\sqrt[4]{\frac{2}{\pi }}  {f_ 1}({\theta_1}) \cos ({\theta_1}) (3 h_ 5 \cot ({\theta_1})+\cot
   ({\theta_2})) \Bigl({g_s} \Bigl(18 {g_s}^2 {M_{eff}}^2 {N_f} \log ^2(r)\nonumber\\
   && +3 {g_s} {M_{eff}}^2
   \log (r) \left({g_s} {N_f} \log \left(\sin \left(\frac{{\theta_1}}{2}\right) \sin
   \left(\frac{{\theta_2}}{2}\right)\right)+3 {g_s} {N_f}+4 \pi \right)+8 \pi ^2 N\Bigr)\Bigr)/\Bigl(\sqrt{3} \Bigl(-27
   h_ 5^2\nonumber\\
   && +12 h_ 5 \cot ({\theta_1}) \cot ({\theta_2})+2 \cot ^2({\theta_1})+2 \cot
   ^2({\theta_2})+3\Bigr)\Bigr)\nonumber\\
& (vii) & B^{IIA}_{\theta_1 \theta_2}=
\frac{1}{4 \pi  r^2}\Bigl({g_s} M \Bigl(-\Bigl(9 \sin ({\theta_1}) \Bigl(\frac{1}{2} {g_s} {N_f} r {f_ 1}({\theta_1}) \log (r)
   \sin ({\theta_1}) \left(\cos \left(\frac{{\theta_2}}{2}\right)+\cos \left(\frac{3 {\theta_2}}{2}\right)\right) \csc
   \left(\frac{{\theta_2}}{2}\right)\nonumber\\
   &&  \left(36 a^2 \log (r)+r\right)+3 h_ 5 {f_ 2}({\theta_2}) \sin
   ({\theta_2}) \Bigl({g_s} {N_f} r \log (r) \cos ({\theta_1}) \cot \left(\frac{{\theta_1}}{2}\right)
   \left(108 a^2 \log (r)+r\right)\nonumber\\
   && -2 \left(3 a^2-r^2\right) \sin ({\theta_1}) \Bigl(2 \log (r) \left({g_s} {N_f}
   \log \left(\sin \left(\frac{{\theta_1}}{2}\right) \sin \left(\frac{{\theta_2}}{2}\right)\right)+2 \pi \right)+9
   {g_s} {N_f} \log ^2(r)\nonumber\\
   && +{g_s} {N_f} \log \left(\sin \left(\frac{{\theta_1}}{2}\right) \sin
   \left(\frac{{\theta_2}}{2}\right)\right)\Bigr)\Bigr)\Bigr)\Bigr)/\Bigl(\cos (2 {\theta_1})-5\Bigr)-\nonumber\\
   && \Bigl({f_ 2}({\theta_2})
   \sin ({\theta_2}) \left(\sin ({\theta_2}) \left(27 h_ 5^2-2 \cot ^2({\theta_1})-3\right)-6 h_ 5 \cot
   ({\theta_1}) \cos ({\theta_2})\right) \Bigl(\frac{9}{2} h_ 5 \sin ({\theta_2}) \nonumber\\
   && \Bigl({g_s} {N_f} r
   \log (r) \cos ({\theta_1}) \csc ^2\left(\frac{{\theta_1}}{2}\right) \left(108 a^2 \log (r)+r\right)-4 \left(3
   a^2-r^2\right) \nonumber\\
   && \left(2 \log (r) \left({g_s} {N_f} \log \left(\sin \left(\frac{{\theta_1}}{2}\right) \sin
   \left(\frac{{\theta_2}}{2}\right)\right)+2 \pi \right)+9 {g_s} {N_f} \log ^2(r)+{g_s} {N_f} \log
   \left(\sin \left(\frac{{\theta_1}}{2}\right) \sin \left(\frac{{\theta_2}}{2}\right)\right)\right)\Bigr)+\cos
   ({\theta_2}) \Bigl(4 \left(3 a^2-r^2\right) \cot ({\theta_1}) \Bigl(2 \log (r)\nonumber\\
   &&  \left({g_s} {N_f} \log
   \left(\sin \left(\frac{{\theta_1}}{2}\right) \sin \left(\frac{{\theta_2}}{2}\right)\right)+2 \pi \right)+9 {g_s}
   {N_f} \log ^2(r)\nonumber\\
   && +{g_s} {N_f} \log \left(\sin \left(\frac{{\theta_1}}{2}\right) \sin
   \left(\frac{{\theta_2}}{2}\right)\right)\Bigr)+3 {g_s} {N_f} r \log (r) \cot
   \left(\frac{{\theta_1}}{2}\right) \left(108 a^2 \log (r)+r\right)\Bigr)\Bigr)\Bigr)/\Bigl(\sin ^2({\theta_2}) \nonumber\\
   && \left(-27
   h_ 5^2+2 \cot ^2({\theta_1})+3\right)+6 h_ 5 \cot ({\theta_1}) \sin (2 {\theta_2})+2 \cos
   ^2({\theta_2})\Bigr)-\Bigl(2 {f_ 1}({\theta_1}) \cos ({\theta_1}) \sin ({\theta_2})\nonumber\\
   &&  (3 h_ 5 \cot
   ({\theta_1})+\cot ({\theta_2})) \Bigl(\left(r^2-3 a^2\right) (\cos (2 {\theta_1})-5) \csc ^2({\theta_1}) \csc
   ({\theta_2}) \left(9 a^4 {g_s}-3 a^2 {g_s} r^2-1\right) \nonumber\\
   && \left(2 \log (r) \left({g_s} {N_f} \log
   \left(\sin \left(\frac{{\theta_1}}{2}\right) \sin \left(\frac{{\theta_2}}{2}\right)\right)+2 \pi \right)+9 {g_s}
   {N_f} \log ^2(r)+{g_s} {N_f} \log \left(\sin \left(\frac{{\theta_1}}{2}\right) \sin
   \left(\frac{{\theta_2}}{2}\right)\right)\right)\nonumber\\
   && - \frac{1}{8} {g_s} {N_f} r \log (r) \cos ({\theta_2}) \csc
   ^3\left(\frac{{\theta_2}}{2}\right) \sec \left(\frac{{\theta_2}}{2}\right) \left(36 a^2 \log (r)+r\right) \Bigl(-18
   h_ 5 \csc ({\theta_1}) \sin ({\theta_2})\nonumber\\
   && +4 \cot ({\theta_1}) \csc ({\theta_1}) \cos ({\theta_2})+\cot
   ^2({\theta_1})-5 \csc ^2({\theta_1})-1\Bigr)\Bigr)\Bigr)/\Bigl(-27 h_ 5^2+12 h_ 5 \cot ({\theta_1}) \cot
   ({\theta_2})\nonumber\\
   && +2 \cot ^2({\theta_1})+2 \cot ^2({\theta_2})+3\Bigr)\Bigr)\Bigr)
\end{eqnarray}
The T-dualized NS-NS components and metric components after the aforementioned triple T-dualities  can be easily evaluated, using which, one obtains the following type $IIA$ metric components.
The exact expressions are extremely long and not particularly illuminating.  The simplified expressions of the same in  the (i) weak($g_s$) coupling - large t'Hooft couplings limit:
$g_s\sim\epsilon^d, M\sim\epsilon^{-\frac{3d}{2}}, N\sim\epsilon^{-19d},\epsilon\leq0.01$, as well
as the (ii) `MQGP limit':$g_s\sim\epsilon^d, M\sim\epsilon^{-\frac{3d}{2}}, N\sim\epsilon^{-39d}, \epsilon\lesssim 1$, are as follows:
{\small
\begin{eqnarray}
\label{metric-mirror}
& (i) & G^{IIA}_{\theta_1\theta_1}\sim \frac{2 \pi \sqrt[]{{g_s} N} \left({f_1}({\theta_1})^2 \sin ^2({\theta_1})+1\right)}{\sqrt[3]{3} } \nonumber\\
   & (ii)& G^{IIA}_{\theta_1\theta_2}\sim -\frac{2 \pi \sqrt[]{{g_s} N} {f_1}({\theta_1}) {f_2}({\theta_2}) \sin ^2({\theta_1})(\cos (2 {\theta_1})-5)^{-1} (\cos (3 {\theta_1})-9 \cos ({\theta_1})) \sin ^2({\theta_2}) \cos
   ({\theta_2})}{\sqrt[3]{3}   \left(3 \left({h_5} \sin (2 {\theta_1}) \sin (2 {\theta_2})+\sin
   ^2({\theta_1}) \sin ^2({\theta_2})\right)+2 \sin ^2({\theta_1}) \cos ^2({\theta_2})+2 \cos ^2({\theta_1}) \sin ^2({\theta_2})\right)}  \nonumber\\
   & (iii) &G^{IIA}_{\theta_2\theta_2}\sim \frac{2 \pi  \sqrt[]{{g_s} N} \cos ({\theta_1}) \cos ({\theta_2}) \left({f_2}({\theta_2})^2 \sin ^2({\theta_1})+1\right)}{9 \sqrt[3]{3}
   \left(\frac{\sin ^2({\theta_1})}{6}+\frac{\cos ^2({\theta_1})}{9}\right)} \nonumber\\
   & (iv) &  G^{IIA}_{x\theta_1}=\frac{{g_s}^{7/4}}{4 \sqrt{2} \pi ^{3/4}\sqrt[4]{N} r}\Bigl(\sqrt[3]{3}   M {N_f} ln (r) \cot \left(\frac{{\theta_1}}{2}\right) \csc ({\theta_1}) \csc
   ({\theta_2}) \left(108 a^2 ln (r)+r\right)  \nonumber\\
   && \left(9 {h_5}+\left(3 \sqrt{6}-2 \cot ({\theta_1})\right) \cot
   ({\theta_2})\right) (2 \cos ({\theta_1}) \cos ({\theta_2})-9 {h_5} \sin ({\theta_1}) \sin ({\theta_2}))\Bigr)\nonumber\\
   & (v) & G^{IIA}_{x\theta_2}=\frac{1}{6 \sqrt{2}
   \pi ^{5/4} r (\cos (2 {\theta_1})-5)^2 \sqrt[4]{{g_s} N}}\Bigl({g_s}^2 M {N_f} ln (r) \sin ({\theta_1}) \cot \left(\frac{{\theta_2}}{2}\right) \csc ({\theta_2}) \left(36 a^2 ln (r)+r\right)\nonumber\\
   && \Bigl(27
   \sqrt{6} \sin ^2({\theta_1}) \cos ({\theta_1}) \sin ({\theta_2})-2 \cos ^3({\theta_1}) \left(\cos ({\theta_2}) \Bigl(6 \sqrt{6} \cot
   ({\theta_2})-4 \cot ({\theta_1}) \cot ({\theta_2})\Bigr)-9 \sqrt{6} \sin ({\theta_2})\right)\nonumber\\
   && +12 \sin ({\theta_1}) \cos ^2({\theta_1}) \cos
   ({\theta_2}) \cot ({\theta_2})\Bigr) \left(-2 \cot ({\theta_1}) \csc ({\theta_1}) \cos ({\theta_2})+2 \cot ^2({\theta_1})+3\right)\Bigr)\nonumber\\
    & (vi) & G^{IIA}_{y\theta_1}= \frac{9\ 3^{5/6} (g_s)^{3/4}  M ln (r) \csc ({\theta_1}) \sin ^4({\theta_2})  \left(36 \pi
    {h_5} \left(3 a^2-r^2\right) \sin({\theta_1})-8 \pi  \left(3 a^2-r^2\right)  \cos ({\theta_1}) \cot ({\theta_2})\right)}{2 N^{1/4}
   \pi ^{3/4} r^2 \left(3 \left({h_5} \sin (2 {\theta_1}) \sin (2 {\theta_2})+\sin ^2({\theta_1}) \sin ^2({\theta_2})\right)+2 \sin ^2({\theta_1})
   \cos ^2({\theta_2})+2 \cos ^2({\theta_1}) \sin ^2({\theta_2})\right)}
    \nonumber\\
   & (vii) & G^{IIA}_{y\theta_2}=\frac{\sqrt{2} \pi ^{3/4}  (g_s N)^{1/4} {f_2}({\theta_2}) (\cos (2 {\theta_1})-5) \sin ^2({\theta_2}) \cos ({\theta_2})
    }{  \left(3 \left({h_5} \sin (2 {\theta_1}) \sin (2 {\theta_2})+\sin ^2({\theta_1}) \sin
   ^2({\theta_2})\right)+2 \sin ^2({\theta_1}) \cos ^2({\theta_2})+2 \cos ^2({\theta_1}) \sin ^2({\theta_2})\right)}\nonumber\\
    & (viii) & G^{IIA}_{z\theta_1 }=\frac{{g_s}^2 M {N_f} ln (r) \cot \left(\frac{{\theta_1}}{2}\right) \csc ^2({\theta_1}) \left(108 a^2 ln (r)+r\right) \left(\sin ^2({\theta_1})
   \left(2 \cot ^2({\theta_2})+3\right)+2 \cos ^2({\theta_1})\right)}{8 \sqrt{2} \pi ^{5/4} r \sqrt[4]{{g_s} N}}\nonumber\\
   & (ix) &  G^{IIA}_{z\theta_2 }=\frac{1}{4 \sqrt{2} \pi ^{5/4} r \sqrt[4]{{g_s} N} (\cos (2 {\theta_1})-5)}\Bigl({g_s}^2 M {N_f} ln (r) \cot \left(\frac{{\theta_2}}{2}\right) \left(36 a^2 ln (r)+r\right) \nonumber\\
   && \left(-2 \cot ({\theta_1}) \csc ({\theta_1})
   \cos ({\theta_2})+2 \cot ^2({\theta_1})+3\right) \left(6 {h_5} \sin (2 {\theta_1}) \cot ({\theta_2})+\sin ^2({\theta_1}) \left(2 \cot
   ^2({\theta_2})+3\right)+2 \cos ^2({\theta_1})\right)\Bigr)\nonumber\\
   & (x) & G^{IIA}_{yy}=\frac{1}{1-9 {h_5}^2} \nonumber\\
   & (xi) & G^{IIA}_{zz}=\frac{1}{27} \left(12 {h_5} \cot ({\theta_1}) \cot ({\theta_2})+2 \cot ^2({\theta_1})+2 \cot ^2({\theta_2})+3\right) \nonumber\\
  & (xii) & G^{IIA}_{xz}= \frac{2 \left(2 \cos ^2({\theta_1}) \left(4 \cot ^2({\theta_1}) \cot ^2({\theta_2})+\cot ({\theta_1}) \left(9 \sqrt{6}-6 \sqrt{6} \cot
   ^2({\theta_2})\right)\right)+27 \sqrt{6} \sin ({\theta_1}) \cos ({\theta_1})\right)}{81 (\cos (2 {\theta_1})-5)}\nonumber\\
  & (xiii) & G^{IIA}_{yz}=-\frac{1}{3} \sqrt{\frac{2}{3}} (3 {h_5} \cot ({\theta_1})+\cot ({\theta_2}))\nonumber\\
   & (xiv) & G^{IIA}_{xy}=-\frac{2 \sin ({\theta_2})}{27
   (\cos (2 {\theta_1})-5) \left(3 \sin ^2({\theta_1}) \sin ^2({\theta_2})+2 \sin ^2({\theta_1}) \cos ^2({\theta_2})+2 \cos ^2({\theta_1}) \sin
   ^2({\theta_2})\right)}\times\nonumber\\
    &&{\hskip -0.2in} \left(8 \sqrt{6} \cos ^4({\theta_1}) \cos ({\theta_2}) \cot ^2({\theta_2})-72 \sin ({\theta_1}) \cos ^3({\theta_1})
   \cos ({\theta_2}) \cot ^2({\theta_2})+12 \sqrt{6} \sin ^2({\theta_1}) \cos ^2({\theta_1}) \cos ({\theta_2}) \cot ^2({\theta_2})\right).\nonumber\\
\end{eqnarray}}

\subsection{Triple T-duals of Type IIB $F_{1,3}$ and Self-Dual $F_5$ to Obtain Type IIA $F_2$}

The components of (\ref{Ftildetildetildeythetai}) are explicitly worked out below (these expressions
not being as long as the mirror type IIA mirror are given in their exact form without simplifying them
in any limit as was done for the mirror type IIA metric):
{\small
\begin{eqnarray}
\label{Ftildetildetildeytheta1}
& (i) & \tilde{\tilde{\tilde{F}}}_{y\theta_1} = \nonumber\\
& & \frac{1}{3 \Bigl(\frac{\cos
   ^2(\theta_2)}{9}+\frac{\sin
   ^2(\theta_2)}{6}\Bigr)}\Bigl[\Bigl(\sqrt{6} \cos (\theta_2) \csc
   (\theta_1)- \nonumber\\
   & &   \frac{1}{6
   \Bigl(\frac{\cos ^2(\theta_1)}{9}+\frac{\sin
   ^2(\theta_1)}{6}\Bigr)}\Bigl(\sin ^2(\theta_1) \Bigl(3 \sqrt{6} \cot
   (\theta_1) \csc ^2(\theta_1) \Bigl(\frac{2}{9} \cos
   (\theta_1) \cos (\theta_2)-{h_5} \sin (\theta_1)
   \sin (\theta_2)\Bigr)- \nonumber\\
   && 9 \csc ^4(\theta_1)
   \Bigl(\frac{2}{9} \cos (\theta_1) \cos (\theta_2)-{h_5}
   \sin (\theta_1) \sin (\theta_2)\Bigr)^2\Bigr)\Bigr)\Bigr)\nonumber\\
   & & \times \Bigl(\frac{1}{\sqrt[4]{h}
   r}\Bigl(2 \sqrt{6} \csc
   (\theta_1) \Bigl(\frac{3 {A_5} {B_5} M {N_f} \cos
   (\theta_1) \cot \Bigl(\frac{\theta_1}{2}\Bigr) ln (r)
   {g_s}^2}{8 \pi }+3 {A_4}{g_s} M \sin
   (\theta_1)  \Bigl(\frac{9 {g_s} {N_f}
   ln ^2(r)}{4 \pi }+ln (r)\nonumber\\
   & & +\frac{{g_s} {N_f} (2 ln (r)+1)
   ln \Bigl(\sin \Bigl(\frac{\theta_1}{2}\Bigr) \sin
   \Bigl(\frac{\theta_2}{2}\Bigr)\Bigr)}{4 \pi }\Bigr) \Bigr) \times\nonumber\\
   & &  \Bigl(-\frac{1}{3 \Bigl(\frac{\cos
   ^2(\theta_1)}{9}+\frac{\sin
   ^2(\theta_1)}{6}\Bigr)}\Bigl(\Bigl(\frac{3 \sqrt{3}
   {N_f} \csc ^3(\theta_1) \Bigl(\frac{\cos
   ^2(\theta_1)}{9}+\frac{\sin
   ^2(\theta_1)}{6}\Bigr)}{\sqrt{2} \sqrt[4]{h} \pi  r}- \nonumber\\
   && \frac{3
   \sqrt{3} {N_f} \csc ^3(\theta_1) \Bigl(\frac{2}{9} \cos
   (\theta_1) \cos (\theta_2)-{h_5} \sin (\theta_1)
   \sin (\theta_2)\Bigr)}{2 \sqrt{2} \sqrt[4]{h} \pi  r}\Bigr)
   \sin ^2(\theta_1)\Bigr) -\frac{\sqrt{3} {N_f} \csc
   (\theta_1)}{2 \sqrt{2} \sqrt[4]{h} \pi  r}\Bigr)\Bigr)- \nonumber\\
   & &\frac{9 A_3 {g_s} M {N_f} \cot
   \Bigl(\frac{\theta_2}{2}\Bigr) \csc (\theta_1) \sin
   (\theta_2)}{2 \sqrt{h} \pi  r^2}\Bigr) \sin
   ^2(\theta_1)\Bigr]\nonumber\\
   & & +\frac{27 \sqrt{\frac{3}{2}}
   \Bigl(\frac{3 {g_s} {N_f} ln (r)}{2 \pi
   }+1\Bigr)}{\sqrt{h} r^2}-\frac{1}{\sqrt[4]{h} r}\Bigl[6 \sqrt{6} \csc (\theta_1)
   \Bigl(\frac{3 {A_5} {B_5} M {N_f} \cos (\theta_1)
   \cot \Bigl(\frac{\theta_1}{2}\Bigr) ln (r) {g_s}^2}{8 \pi
   } \nonumber\\
   & &  +3 {A_4}{g_s} M \sin
   (\theta_1) \Bigl(\frac{9 {g_s} {N_f} ln ^2(r)}{4 \pi
   }+ln (r)+\frac{{g_s} {N_f} (2 ln (r)+1) ln \Bigl(\sin
   \Bigl(\frac{\theta_1}{2}\Bigr) \sin
   \Bigl(\frac{\theta_2}{2}\Bigr)\Bigr)}{4 \pi }\Bigr) \Bigr) \Bigl(\frac{9 {N_f}}{4
   \sqrt[4]{h} \pi  r}\nonumber\\
   & & -\frac{\sin ^2(\theta_1)}{3 \Bigl(\frac{\cos
   ^2(\theta_1)}{9}+\frac{\sin
   ^2(\theta_1)}{6}\Bigr)} \Bigl(-\frac{27
   {N_f} \Bigl(\frac{\cos ^2(\theta_1)}{9}+\frac{\sin
   ^2(\theta_1)}{6}\Bigr) \csc ^2(\theta_1)}{2 \sqrt[4]{h}
   \pi  r}-\frac{3 {N_f} \cot (\theta_1) \csc
   (\theta_1)}{2 \sqrt[4]{h} \pi  r}\Bigr)\Bigr)\Bigr]\nonumber\\
   & & -\frac{\Bigl(\frac{\cos
   ^2(\theta_2)}{9}+\frac{\sin ^2(\theta_2)}{6}\Bigr)}{12 \csc ^2(\theta_1) \Bigl(\frac{\cos
   ^2(\theta_2)}{9}+\frac{\sin ^2(\theta_2)}{6}\Bigr)^2- 3
   \csc ^2(\theta_1) \Bigl(\frac{2}{9} \cos (\theta_1) \cos
   (\theta_2)-{h_5} \sin (\theta_1) \sin
   (\theta_2)\Bigr)^2}\Bigl[2
   \Bigl(\sqrt{6} \cos (\theta_2) \nonumber\\
   & & \csc
   (\theta_1)-\frac{\sqrt{\frac{3}{2}} \cot (\theta_1)
   \Bigl(\frac{2}{9} \cos (\theta_1) \cos (\theta_2)-{h_5}
   \sin (\theta_1) \sin (\theta_2)\Bigr)}{\frac{\cos
   ^2(\theta_1)}{9}+\frac{\sin ^2(\theta_1)}{6}}\Bigr) \times   \Bigl(\frac{9 A_3 {g_s} M {N_f} \cot
   \Bigl(\frac{\theta_2}{2}\Bigr) \csc (\theta_1) \sin
   (\theta_2)}{2 \sqrt{h} \pi  r^2}\nonumber\\
   & & -\frac{1}{\sqrt[4]{h}
   r}\Bigl(2 \sqrt{6} \csc
   (\theta_1) \Bigl(\frac{3 {A_5} {B_5}{g_s}^2 M {N_f} \cos
   (\theta_1) \cot \Bigl(\frac{\theta_1}{2}\Bigr) ln (r)
   }{8 \pi }\nonumber\\
   & & +3 {A_4}  {g_s}M \sin
   (\theta_1) \Bigl(\frac{9 {g_s} {N_f}
   ln ^2(r)}{4 \pi }+ln (r)+\frac{{g_s} {N_f} (2 ln (r)+1)
   ln \Bigl(\sin \Bigl(\frac{\theta_1}{2}\Bigr) \sin
   \Bigl(\frac{\theta_2}{2}\Bigr)\Bigr)}{4 \pi }\Bigr) \Bigr) \nonumber\\
   & & \times\Bigl(\frac{\sin ^2(\theta_1)}{3 \Bigl(\frac{\cos
   ^2(\theta_1)}{9}+\frac{\sin
   ^2(\theta_1)}{6}\Bigr)}\Bigl( \frac{3
   \sqrt{3} {N_f} \csc ^3(\theta_1) \Bigl(\frac{2}{9} \cos
   (\theta_1) \cos (\theta_2)-{h_5} \sin (\theta_1)
   \sin (\theta_2)\Bigr)}{2 \sqrt{2} \sqrt[4]{h} \pi  r}- \nonumber\\
   && \frac{3 \sqrt{3}
   {N_f} \csc ^3(\theta_1) \Bigl(\frac{\cos
   ^2(\theta_1)}{9}+\frac{\sin
   ^2(\theta_1)}{6}\Bigr)}{\sqrt{2} \sqrt[4]{h} \pi  r}
   \Bigr)-\frac{\sqrt{3} {N_f} \csc
   (\theta_1)}{2 \sqrt{2} \sqrt[4]{h} \pi  r}\Bigr)\Bigr)\Bigr)\Bigr]
   \end{eqnarray}

   \begin{eqnarray}
   \label{Ftildetildetildeytheta2}
   & (ii) & \tilde{\tilde{\tilde{F}}}_{y\theta_2} = \nonumber\\
   & &   \frac{1}{2 \sqrt[4]{h} \pi
   r}\Bigl[9 \sqrt{\frac{3}{2}} {A_5}{g_s}^2 M {N_f} \cos (\theta_2)
   \cot \Bigl(\frac{\theta_2}{2}\Bigr) \csc (\theta_1) ln
   (r) \Bigl(\frac{9 {N_f}}{4 \sqrt[4]{h} \pi  r}-\frac{\sin
   ^2(\theta_1)}{3
   \Bigl(\frac{\cos ^2(\theta_1)}{9}+\frac{\sin
   ^2(\theta_1)}{6}\Bigr)}\nonumber\\
   &&  \Bigl(-\frac{27 {N_f} \Bigl(\frac{\cos
   ^2(\theta_1)}{9}+\frac{\sin ^2(\theta_1)}{6}\Bigr) \csc
   ^2(\theta_1)}{2 \sqrt[4]{h} \pi  r}-\frac{3 {N_f} \cot
   (\theta_1) \csc (\theta_1)}{2 \sqrt[4]{h} \pi  r}\Bigr)\Bigr) \Bigr]\nonumber\\
   & & +\frac{1}{3
   \Bigl(\frac{\cos ^2(\theta_2)}{9}+\frac{\sin
   ^2(\theta_2)}{6}\Bigr)}\Bigl[\sin ^2(\theta_1) \Bigl(\sqrt{6} \cos (\theta_2)
   \csc (\theta_1)- \nonumber\\
   & &   \frac{1}{6
   \Bigl(\frac{\cos ^2(\theta_1)}{9}+\frac{\sin
   ^2(\theta_1)}{6}\Bigr)} \Bigl(3 \sqrt{6}
   \cot (\theta_1)  \Bigl(\frac{2}{9} \cos
   (\theta_1) \cos (\theta_2)-{h_5} \sin (\theta_1)
   \sin (\theta_2)\Bigr)-  9\csc ^2(\theta_1) \nonumber\\
   & &
   \Bigl(\frac{2}{9} \cos (\theta_1) \cos (\theta_2)-{h_5}
   \sin (\theta_1) \sin (\theta_2)\Bigr)^2 \Bigr)\Bigr) \Bigl(-\frac{1}{2
   \sqrt[4]{h} \pi  r}\Bigl(3
   \sqrt{\frac{3}{2}} {A_5} {g_s}^2 M {N_f} \cos (\theta_2) \cot
   \Bigl(\frac{\theta_2}{2}\Bigr) \csc (\theta_1) ln (r)\nonumber\\
   & & \times
   \Bigl(\frac{\sin ^2(\theta_1)}{3 \Bigl(\frac{\cos
   ^2(\theta_1)}{9}+\frac{\sin
   ^2(\theta_1)}{6}\Bigr)}  \Bigl(\frac{3
   \sqrt{3} {N_f} \csc ^3(\theta_1) \Bigl(\frac{2}{9} \cos
   (\theta_1) \cos (\theta_2)-{h_5} \sin (\theta_1)
   \sin (\theta_2)\Bigr)}{2 \sqrt{2} \sqrt[4]{h} \pi  r}- \nonumber\\
   && \frac{3 \sqrt{3} {N_f} \csc ^3(\theta_1)
   \Bigl(\frac{\cos ^2(\theta_1)}{9}+\frac{\sin
   ^2(\theta_1)}{6}\Bigr)}{\sqrt{2} \sqrt[4]{h} \pi  r}
  \Bigr)  -\frac{\sqrt{3} {N_f} \csc
   (\theta_1)}{2 \sqrt{2} \sqrt[4]{h} \pi  r}\Bigr) \Bigr) -\frac{9 A_3 {B_3}{g_s} M {N_f} \cot
   \Bigl(\frac{\theta_1}{2}\Bigr) \csc (\theta_1) \sin
   (\theta_2) }{2 \sqrt{h} \pi  r^2}\Bigr)\Bigr]\nonumber\\
   & & -\frac{{2
   \Bigl(\frac{\cos ^2(\theta_1)}{9}+\frac{\sin
   ^2(\theta_1)}{6}\Bigr)}}{12 \csc
   ^2(\theta_1) \Bigl(\frac{\cos ^2(\theta_2)}{9}+\frac{\sin
   ^2(\theta_2)}{6}\Bigr)^2-\Bigl(3 \csc ^2(\theta_1)
   \Bigl(\frac{2}{9} \cos (\theta_1) \cos (\theta_2)-{h_5}
   \sin (\theta_1) \sin (\theta_2)\Bigr)^2\Bigr)}\times\nonumber\\
   & & \Bigl[2 \Bigl(\sqrt{6} \cos
   (\theta_2) \csc (\theta_1)-\frac{\sqrt{\frac{3}{2}} \cot
   (\theta_1) \Bigl(\frac{2}{9} \cos (\theta_1) \cos
   (\theta_2)-{h_5} \sin (\theta_1) \sin
   (\theta_2)\Bigr)}{\frac{\cos ^2(\theta_1)}{9}+\frac{\sin
   ^2(\theta_1)}{6}}\Bigr)\nonumber\\
   & &  \times \Bigl(\frac{1}{2
   \sqrt[4]{h} \pi  r}\Bigl(3 \sqrt{\frac{3}{2}}
   {A_5}{g_s}^2 M {N_f} \cos (\theta_2) \cot
   \Bigl(\frac{\theta_2}{2}\Bigr) \csc (\theta_1) ln (r)   \Bigl(\frac{\sin ^2(\theta_1)}{3 \Bigl(\frac{\cos
   ^2(\theta_1)}{9}+\frac{\sin
   ^2(\theta_1)}{6}\Bigr)}\times\nonumber\\
& & \Bigl(\frac{3
   \sqrt{3} {N_f} \csc ^3(\theta_1) \Bigl(\frac{2}{9} \cos
   (\theta_1) \cos (\theta_2)-{h_5} \sin (\theta_1)
   \sin (\theta_2)\Bigr)}{2 \sqrt{2} \sqrt[4]{h} \pi  r}-  \frac{3 \sqrt{3} {N_f} \csc ^3(\theta_1)
   \Bigl(\frac{\cos ^2(\theta_1)}{9}+\frac{\sin
   ^2(\theta_1)}{6}\Bigr)}{\sqrt{2} \sqrt[4]{h} \pi  r}\Bigr)
    -\nonumber\\
    && \frac{\sqrt{3} {N_f} \csc
   (\theta_1)}{2 \sqrt{2} \sqrt[4]{h} \pi  r}\Bigr) \Bigr) +\frac{9 A_3 {B_3} M {N_f} \cot
   \Bigl(\frac{\theta_1}{2}\Bigr) \csc (\theta_1) \sin
   (\theta_2) {g_s}}{2 \sqrt{h} \pi  r^2}\Bigr)\Bigr]
   \end{eqnarray}

   \begin{eqnarray}
   \label{Ftildetildetildeyr}
   &  & {\hskip -4.0in} (iii) \tilde{\tilde{\tilde{F}}}_{yr} = -\frac{81 \sqrt{\frac{3}{2}} {A2} {B_2} {g_s} M {N_f}
   \cot \Bigl(\frac{\theta_1}{2}\Bigr)}{2 \sqrt{h} \pi  r^3}.
   \end{eqnarray}}

The components of (\ref{Ftildetildetildezthetair}) are given as under:
{\small
 \begin{eqnarray}
   \label{Ftildetildetildeztheta1}
   & (iv) & \tilde{\tilde{\tilde{F}}}_{z\theta_1} =\nonumber\\
   & & \frac{1}{\sqrt[4]{h}
   r}\Bigl[2 \sqrt{6} \csc (\theta_1) \Bigl(\frac{3 {A_5} {B_5}
  {g_s}^2 M {N_f} \cos (\theta_1) \cot
   \Bigl(\frac{\theta_1}{2}\Bigr) ln (r) }{8 \pi }\nonumber\\
   & & +3
   {A_4} M  {g_s} \sin
   (\theta_1) \Bigl(\frac{9 {g_s} {N_f} ln ^2(r)}{4 \pi
   }+ln (r)+\frac{{g_s} {N_f} (2 ln (r)+1) ln \Bigl(\sin
   \Bigl(\frac{\theta_1}{2}\Bigr) \sin
   \Bigl(\frac{\theta_2}{2}\Bigr)\Bigr)}{4 \pi }\Bigr) \Bigr)\nonumber\\
   & & \times \Bigl(\frac{ \sin ^2(\theta_1)}{3 \Bigl(\frac{\cos
   ^2(\theta_1)}{9}+\frac{\sin
   ^2(\theta_1)}{6}\Bigr)}\Bigl(\Bigl(\frac{3
   \sqrt{3} {N_f} \csc ^3(\theta_1) \Bigl(\frac{2}{9} \cos
   (\theta_1) \cos (\theta_2)-{h_5} \sin (\theta_1)
   \sin (\theta_2)\Bigr)}{2 \sqrt{2} \sqrt[4]{h} \pi  r}- \nonumber\\
   && \frac{3 \sqrt{3}
   {N_f} \csc ^3(\theta_1) \Bigl(\frac{\cos
   ^2(\theta_1)}{9}+\frac{\sin
   ^2(\theta_1)}{6}\Bigr)}{\sqrt{2} \sqrt[4]{h} \pi  r}\Bigr)
  \Bigr) -\frac{\sqrt{3} {N_f} \csc
   (\theta_1)}{2 \sqrt{2} \sqrt[4]{h} \pi  r}\Bigr)\Bigr] -\frac{9 A_3 {g_s} M {N_f} \cot
   \Bigl(\frac{\theta_2}{2}\Bigr) \csc (\theta_1) \sin
   (\theta_2)}{2 \sqrt{h} \pi  r^2} \nonumber\\
   \end{eqnarray}

   \begin{eqnarray}
   \label{Ftildetildetildeztheta2}
   & (v) & \tilde{\tilde{\tilde{F}}}_{z\theta_2} = -\frac{1}{2
   \sqrt[4]{h} \pi  r}\Bigl[3 \sqrt{\frac{3}{2}} {A_5} {g_s}^2 M {N_f} \cos (\theta_2)
   \cot \Bigl(\frac{\theta_2}{2}\Bigr) \csc (\theta_1) ln
   (r) \Bigl(\frac{\sin ^2(\theta_1)}{3 \Bigl(\frac{\cos
   ^2(\theta_1)}{9}+\frac{\sin
   ^2(\theta_1)}{6}\Bigr)}\times
  \nonumber\\
  & &\Bigl(\Bigl(\frac{3
   \sqrt{3} {N_f} \csc ^3(\theta_1) \Bigl(\frac{2}{9} \cos
   (\theta_1) \cos (\theta_2)-{h_5} \sin (\theta_1)
   \sin (\theta_2)\Bigr)}{2 \sqrt{2} \sqrt[4]{h} \pi  r}- \frac{3 \sqrt{3} {N_f} \csc
   ^3(\theta_1) \Bigl(\frac{\cos ^2(\theta_1)}{9}+\frac{\sin
   ^2(\theta_1)}{6}\Bigr)}{\sqrt{2} \sqrt[4]{h} \pi  r}\Bigr)
\Bigr)\nonumber\\
   && -\frac{\sqrt{3} {N_f} \csc
   (\theta_1)}{2 \sqrt{2} \sqrt[4]{h} \pi  r}\Bigr) \Bigr] -\frac{9 A_3 {B_3}{g_s} M {N_f} \cot
   \Bigl(\frac{\theta_1}{2}\Bigr) \csc (\theta_1) \sin
   (\theta_2) }{2 \sqrt{h} \pi  r^2}
   \end{eqnarray}

\begin{eqnarray}
\label{Ftildetildetildezr}
&  & {\hskip -5.8in} (vi) \tilde{\tilde{\tilde{F}}}_{zr}=0.
\end{eqnarray}}

Equation (\ref{eq:Ftildetildetildexthetair}) yields:
{\small
\begin{eqnarray}
\label{Ftildetildetildextheta1}
& (vii) & \tilde{\tilde{\tilde{F}}}_{x\theta_1}= \nonumber\\
& &  \frac{\sin
   ^2(\theta_1)}{2 \Bigl(\frac{\cos
   ^2(\theta_1)}{9}+\frac{\sin
   ^2(\theta_1)}{6}\Bigr)}\Bigl[\Bigl(\frac{27 \sqrt{\frac{3}{2}} \csc ^2(\theta_1)
   \Bigl(\frac{3 {g_s} {N_f} ln (r)}{2 \pi }+1\Bigr)
   \Bigl(\frac{2}{9} \cos (\theta_1) \cos (\theta_2)-{h_5}
   \sin (\theta_1) \sin (\theta_2)\Bigr)}{\sqrt{h}
   r^2}-\nonumber\\
   && \frac{9 \sqrt{\frac{3}{2}} A_3 {g_s} M {N_f} \cot
   (\theta_1) \cot \Bigl(\frac{\theta_2}{2}\Bigr) \csc
   (\theta_1) \sin (\theta_2)}{\sqrt{h} \pi  r^2}\Bigr) \Bigr]\nonumber\\
   & & +\frac{1}{2 \Bigl(\frac{\cos
   ^2(\theta_1)}{9}+\frac{\sin
   ^2(\theta_1)}{6}\Bigr)}\Bigl[3 \Bigl(\frac{1}{\sqrt[4]{h} r}\Bigl(2 \sqrt{6} \csc
   (\theta_1)\Bigl(\frac{3 {A_5} {B_5}{g_s}^2 M {N_f}  \cos
   (\theta_1) \cot \Bigl(\frac{\theta_1}{2}\Bigr) ln (r)
   }{8 \pi }\nonumber\\
   & & +3 {A_4}{g_s} M \sin
   (\theta_1)\Bigl(\frac{9 {g_s} {N_f}
   ln ^2(r)}{4 \pi }+ln (r)+\frac{{g_s} {N_f} (2 ln (r)+1)
   ln \Bigl(\sin \Bigl(\frac{\theta_1}{2}\Bigr) \sin
   \Bigl(\frac{\theta_2}{2}\Bigr)\Bigr)}{4 \pi }\Bigr) \Bigr) \Bigl(\frac{9 {N_f}}{4
   \sqrt[4]{h} \pi  r}\nonumber\\
   & & -\frac{\sin ^2(\theta_1)}{3 \Bigl(\frac{\cos
   ^2(\theta_1)}{9}+\frac{\sin
   ^2(\theta_1)}{6}\Bigr)}  \Bigl(-\frac{27
   {N_f} \Bigl(\frac{\cos ^2(\theta_1)}{9}+\frac{\sin
   ^2(\theta_1)}{6}\Bigr) \csc ^2(\theta_1)}{2 \sqrt[4]{h}
   \pi  r}-\frac{3 {N_f} \cot (\theta_1) \csc
   (\theta_1)}{2 \sqrt[4]{h} \pi  r} \Bigr)\Bigr)\Bigr) \nonumber\\
   && -\frac{9
   \sqrt{\frac{3}{2}} \Bigl(\frac{3 {g_s} {N_f} ln (r)}{2 \pi
   }+1\Bigr)}{\sqrt{h} r^2}\Bigr) \Bigl(\frac{2}{9} \cos
   (\theta_1) \cos (\theta_2)-{h_5} \sin (\theta_1)
   \sin (\theta_2)\Bigr)\Bigr]-\nonumber\\
   & & \frac{2 \Bigl(\frac{\cos
   ^2(\theta_1)}{9}+\frac{\sin ^2(\theta_1)}{6}\Bigr)}{12 \csc ^2(\theta_1) \Bigl(\frac{\cos
   ^2(\theta_2)}{9}+\frac{\sin ^2(\theta_2)}{6}\Bigr)^2- {3
   \csc ^2(\theta_1) \Bigl(\frac{2}{9} \cos (\theta_1) \cos
   (\theta_2)-{h_5} \sin (\theta_1) \sin
   (\theta_2)\Bigr)^2}} \times \nonumber\\
   &&   \Bigl[2 \Bigl(\Bigl(\frac{9
   \sqrt{\frac{3}{2}} \Bigl(\frac{3 {g_s} {N_f} ln (r)}{2 \pi
   }+1\Bigr)}{\sqrt{h} r^2}  + 2 \Bigl(\frac{9 {N_f}}{4 \sqrt[4]{h}
   \pi  r}-\frac{\sin ^2(\theta_1)}{3 \Bigl(\frac{\cos
   ^2(\theta_1)}{9}+\frac{\sin
   ^2(\theta_1)}{6}\Bigr)}\Bigl(-\frac{27 {N_f}
   \Bigl(\frac{\cos ^2(\theta_1)}{9}+\frac{\sin
   ^2(\theta_1)}{6}\Bigr) \csc ^2(\theta_1)}{2 \sqrt[4]{h}
   \pi  r}- \nonumber\\
   &&  \frac{3 {N_f} \cot (\theta_1) \csc
   (\theta_1)}{2 \sqrt[4]{h} \pi  r}\Bigr)\Bigr)\Bigr) \Bigl(\frac{3 \csc ^2(\theta_1)
   \Bigl(\frac{2}{9} \cos (\theta_1) \cos (\theta_2)-   {h_5}
   \sin (\theta_1) \sin (\theta_2)\Bigr)^2}{2
   \Bigl(\frac{\cos ^2(\theta_1)}{9}+\frac{\sin
   ^2(\theta_1)}{6}\Bigr)}- \nonumber\\
   & & 6 \csc
   ^2(\theta_1) \Bigl(\frac{\cos ^2(\theta_2)}{9}+\frac{\sin
   ^2(\theta_2)}{6}\Bigr)\Bigr)-\frac{\sqrt{\frac{3}{2}} \cot
   (\theta_1) \Bigl(\frac{2}{9} \cos (\theta_1) \cos
   (\theta_2)-{h_5} \sin (\theta_1) \sin
   (\theta_2)\Bigr)}{\frac{\cos ^2(\theta_1)}{9}+\frac{\sin
   ^2(\theta_1)}{6}}\Bigr)- \nonumber\\
   && \Bigl(\sqrt{6} \cos
   (\theta_2) \csc (\theta_1)  \times \Bigl(-\frac{9 A_3 {g_s} M
   {N_f} \cot \Bigl(\frac{\theta_2}{2}\Bigr) \csc
   (\theta_1) \sin (\theta_2)}{2 \sqrt{h} \pi  r^2}- \nonumber\\
   && \frac{1}{\sqrt[4]{h}
   r}\Bigl(2
   \sqrt{6} \csc (\theta_1) \Bigl(\frac{3 {A_5} {B_5}{g_s}^2 M
   {N_f} \cos (\theta_1) \cot
   \Bigl(\frac{\theta_1}{2}\Bigr) ln (r) }{8 \pi } +3
   {A_4}{g_s}M \sin
   (\theta_1) \Bigl(\frac{9 {g_s} {N_f} ln ^2(r)}{4 \pi
   }+ ln (r)+ \nonumber\\
   & & \frac{{g_s} {N_f} (2 ln (r)+1) ln \Bigl(\sin
   \Bigl(\frac{\theta_1}{2}\Bigr) \sin
   \Bigl(\frac{\theta_2}{2}\Bigr)\Bigr)}{4 \pi }\Bigr) \Bigr) \Bigl(\frac{\sqrt{3} {N_f} \csc
   (\theta_1) \Bigl(\frac{2}{9} \cos (\theta_1) \cos
   (\theta_2)-{h_5} \sin (\theta_1) \sin
   (\theta_2)\Bigr)}{4 \sqrt{2} \sqrt[4]{h} \pi  r
   \Bigl(\frac{\cos ^2(\theta_1)}{9}+\frac{\sin
   ^2(\theta_1)}{6}\Bigr)}- \nonumber\\
   & &  \frac{\sqrt{3} {N_f} \csc
   (\theta_1)}{2 \sqrt{2} \sqrt[4]{h} \pi  r}\Bigr)\Bigr)\Bigr)\Bigr)\Bigr]
\end{eqnarray}

\begin{eqnarray}
\label{Ftildetildetildextheta2}
& (viii) & \tilde{\tilde{\tilde{F}}}_{x\theta_2}= -\frac{1}{4 \sqrt[4]{h} \pi  r
   \Bigl(\frac{\cos ^2(\theta_1)}{9}+\frac{\sin
   ^2(\theta_1)}{6}\Bigr)}\Bigl[9 \sqrt{\frac{3}{2}} {A_5}{g_s}^2 M {N_f} \cos (\theta_2)
   \cot \Bigl(\frac{\theta_2}{2}\Bigr) \csc (\theta_1) ln
   (r) \Bigl(\frac{9 {N_f}}{4 \sqrt[4]{h} \pi  r}\nonumber\\
   & &  -\frac{\sin
   ^2(\theta_1)}{3
   \Bigl(\frac{\cos ^2(\theta_1)}{9}+\frac{\sin
   ^2(\theta_1)}{6}\Bigr)}  \Bigl(-\frac{27 {N_f} \Bigl(\frac{\cos
   ^2(\theta_1)}{9}+\frac{\sin ^2(\theta_1)}{6}\Bigr) \csc
   ^2(\theta_1)}{2 \sqrt[4]{h} \pi  r}-\frac{3 {N_f} \cot
   (\theta_1) \csc (\theta_1)}{2 \sqrt[4]{h} \pi  r} \Bigr)\Bigr) \nonumber\\
   & & \Bigl(\frac{2}{9} \cos
   (\theta_1) \cos (\theta_2)-{h_5} \sin (\theta_1)
   \sin (\theta_2)\Bigr) \Bigr] -\frac{9 \sqrt{\frac{3}{2}} A_3
   {B_3}{g_s} M {N_f} \cos (\theta_1) \cot
   \Bigl(\frac{\theta_1}{2}\Bigr) \sin (\theta_2)
   }{2 \sqrt{h} \pi  r^2 \Bigl(\frac{\cos
   ^2(\theta_1)}{9}+\frac{\sin
   ^2(\theta_1)}{6}\Bigr)}\nonumber\\
   & & -\frac{2 \Bigl(\frac{\cos ^2(\theta_2)}{9}+\frac{\sin
   ^2(\theta_2)}{6}\Bigr)}{12 \csc ^2(\theta_1)
   \Bigl(\frac{\cos ^2(\theta_2)}{9}+\frac{\sin
   ^2(\theta_2)}{6}\Bigr)^2- 3 \csc ^2(\theta_1)
   \Bigl(\frac{2}{9} \cos (\theta_1) \cos (\theta_2)-{h_5}
   \sin (\theta_1) \sin (\theta_2)\Bigr)^2 } \times \nonumber\\
   && \Bigl[2 \Bigl(\Bigl(\sqrt{6} \cos
   (\theta_2) \csc (\theta_1) -\frac{\sqrt{\frac{3}{2}} \cot
   (\theta_1) \Bigl(\frac{2}{9} \cos (\theta_1) \cos
   (\theta_2)-{h_5} \sin (\theta_1) \sin
   (\theta_2)\Bigr)}{\frac{\cos ^2(\theta_1)}{9}+\frac{\sin
   ^2(\theta_1)}{6}}\Bigr) \times\nonumber\\
   & &   \Bigl(\frac{-1}{2 \sqrt[4]{h}
   \pi  r}\bigl(3 \sqrt{\frac{3}{2}}
   {A_5} {g_s}^2 M {N_f} \cos (\theta_2) \cot
   \Bigl(\frac{\theta_2}{2}\Bigr) \csc (\theta_1) ln (r)
   \Bigl(\frac{\sqrt{3} {N_f} \csc
   (\theta_1)}{2 \sqrt{2} \sqrt[4]{h} \pi  r}-\nonumber\\
   && \frac{\sqrt{3} {N_f} \csc (\theta_1) \Bigl(\frac{2}{9}
   \cos (\theta_1) \cos (\theta_2)-{h_5} \sin
   (\theta_1) \sin (\theta_2)\Bigr)}{4 \sqrt{2} \sqrt[4]{h}
   \pi  r \Bigl(\frac{\cos ^2(\theta_1)}{9}+\frac{\sin
   ^2(\theta_1)}{6}\Bigr)}\Bigr)\Bigr)-\frac{9 {A_3} {B_3} {g_s} M {N_f} \cot
   \Bigl(\frac{\theta_1}{2}\Bigr) \csc (\theta_1) \sin
   (\theta_2)}{2 \sqrt{h} \pi  r^2}\Bigr)- \nonumber\\
   && 2 \Bigl(\frac{9
   {N_f}}{4 \sqrt[4]{h} \pi  r}-\frac{\sin ^2(\theta_1)}{3
   \Bigl(\frac{\cos ^2(\theta_1)}{9}+\frac{\sin
   ^2(\theta_1)}{6}\Bigr)}
   \Bigl(-\frac{27 {N_f} \Bigl(\frac{\cos
   ^2(\theta_1)}{9}+\frac{\sin ^2(\theta_1)}{6}\Bigr) \csc
   ^2(\theta_1)}{2 \sqrt[4]{h} \pi  r}-\frac{3 {N_f} \cot
   (\theta_1) \csc (\theta_1)}{2 \sqrt[4]{h} \pi  r}\Bigr)\Bigr)\nonumber\\
   & & \times \Bigl(6 \csc ^2(\theta_1)
   \Bigl(\frac{\cos ^2(\theta_2)}{9}+\frac{\sin
   ^2(\theta_2)}{6}\Bigr)-\frac{3 \csc ^2(\theta_1)
   \Bigl(\frac{2}{9} \cos (\theta_1) \cos (\theta_2)-{h_5}
   \sin (\theta_1) \sin (\theta_2)\Bigr)^2}{2
   \Bigl(\frac{\cos ^2(\theta_1)}{9}+\frac{\sin
   ^2(\theta_1)}{6}\Bigr)}\Bigr)\Bigr)\Bigr]
   \end{eqnarray}

   \begin{eqnarray}
   \label{Ftildetildetildexr}
   & (ix) & \tilde{\tilde{\tilde{F}}}_{xr} =  -2 \Bigl[\frac{27 \sqrt{\frac{3}{2}} {A_2} {B_2} {g_s} M
   {N_f} \cot \Bigl(\frac{\theta_1}{2}\Bigr)}{2 \sqrt{h} \pi
   r^3}-2 \Bigl(\frac{9 {N_f}}{4 \sqrt[4]{h} \pi  r}+ \frac{\sin
   ^2(\theta_1)} {3
   \Bigl(\frac{\cos ^2(\theta_1)}{9}+\frac{\sin
   ^2(\theta_1)}{6}\Bigr)} \Bigl(\frac{3 {N_f} \cot
   (\theta_1) \csc (\theta_1)}{2 \sqrt[4]{h} \pi  r}\nonumber\\
   && + \frac{27 {N_f} \Bigl(\frac{\cos
   ^2(\theta_1)}{9}+\frac{\sin ^2(\theta_1)}{6}\Bigr) \csc
   ^2(\theta_1)}{2 \sqrt[4]{h} \pi  r} \Bigr)\Bigr)\Bigr]
   \end{eqnarray}}

Further, (\ref{eq:Fitldetildetildebeta1beta2}) yields:
{\small
\begin{eqnarray}
\label{Ftildetildetildeyx}
& (x) & \tilde{\tilde{\tilde{F_1}}}_{yx} = \frac{1}{\Lambda_3}\times[9 {N_f} \Bigl(\frac{8}{81} \Bigl(\frac{2 \cos
   \theta_2}{9}-\frac{\sin ^2\theta_2}{2}\Bigr) \cos
   ^4\theta_1-\frac{4}{81}\cos ^3\theta_1 \Bigl(\frac{2 \cos
   ^2\theta_2}{9}+6 {h_5} \sin \theta_1 \sin
   \theta_2 \cos \theta_2 \nonumber\\
   && + \frac{\sin
   ^2\theta_2}{2}\Bigr) -\frac{2}{9} \sin
   \theta_1 \Bigl(\frac{2}{9} \sin \theta_1 \cos
   ^2\theta_2+\frac{4}{9} \Bigl({h_5} \sin
   \theta_2-\frac{2 \sin \theta_1}{3}\Bigr) \cos
   \theta_2+3 \Bigl(\frac{2}{9}-{h_5}^2\Bigr) \sin
   \theta_1 \sin ^2\theta_2\Bigr) \cos
   ^2\theta_1\nonumber\\
   & & -\frac{1}{9} \sin ^2\theta_1 \Bigl(\frac{10
   \cos ^2\theta_2}{27}+4 {h_5} \sin \theta_1 \sin
   \theta_2 \cos \theta_2+3
   \Bigl(\frac{1}{9}-{h_5}^2\Bigr) \sin ^2\theta_2\Bigr)
   \cos \theta_1+\frac{1}{3} \sin ^3\theta_1
   \Bigl(-\frac{2}{9} \sin \theta_1 \cos
   ^2\theta_2\nonumber\\
   & & +\frac{4}{27} \sin \theta_1 \cos
   \theta_2-3 \Bigl(\frac{1}{9}-{h_5}^2\Bigr) \sin
   \theta_1 \sin ^2\theta_2+\frac{1}{9} {h_5} \sin (2
   \theta_2)\Bigr)\Bigr)\Bigr],
   \end{eqnarray}}
   where
   {\small
   \begin{eqnarray}
   & & \Lambda_3\equiv 4 \sqrt[4]h\left(r,\theta_1,\theta_2\right) \pi  r \Bigl(\frac{\cos
   ^2\theta_1}{9}+\frac{\sin ^2\theta_1}{6}\Bigr)
   \Bigl(\frac{2}{27} \cos ^2\theta_2 \sin
   ^2\theta_1-{h_5}^2 \sin ^2\theta_2 \sin
   ^2\theta_1+\frac{1}{9} \sin ^2\theta_2 \sin
   ^2\theta_1\nonumber\\
   & & +\frac{2}{27} \cos ^2\theta_1 \sin
   ^2\theta_2+\frac{1}{9} {h_5} \sin (2 \theta_1) \sin
   (2 \theta_2)\Bigr);
   \end{eqnarray}}
   and
   {\small
   \begin{eqnarray}
   \label{Ftildetildetildezx}
   & & \tilde{\tilde{\tilde{F_1}}}_{zx}=\frac{\sqrt{3} {N_f} \Bigl(\frac{4}{9} \cos \theta_1 \cot
   \theta_1-\frac{2}{9} \cos \theta_2 \cot
   \theta_1+\frac{2 \sin \theta_1}{3}+{h_5} \sin
   \theta_2\Bigr)}{2 \sqrt{2} \sqrt[4]h\left(r,\theta_1,\theta_2\right) \pi  r
   \Bigl(\frac{\cos ^2\theta_1}{9}+\frac{\sin
   ^2\theta_1}{6}\Bigr)}.
   \end{eqnarray}}

Now, (\ref{eq:HodgedualFtildetildetildebeta1beta2}) yields:
{\small
\begin{eqnarray}
   \label{HodgeFtildetildetildetheta1theta2}
   & (xi) & \tilde{\tilde{\tilde{F_5}}}_{\theta_1\theta_2}=   \Bigl[ h^{3/2} \Bigl(\frac{9 {M_{\rm eff}}^2 {N_f} ln (r) {g_s}^3}{\pi  r}+4 {g_s} N \pi  \left(h r^4-4 {g_s} N \pi \right) -4 h r^3\Bigr) \times \nonumber\\
   &&
   \bigl(-\Bigr(r^4 \sin ^2({\theta_2}) \Bigr(4 \left({h_5}^2-4 {h_2} {h_4}\right) \nonumber\\
   &&  \csc ({\theta_1}) {f_1}({\theta_1})
   {f_2}({\theta_2}) \sin ({\theta_2}) {h_5}^2-4 \left({h_5}^4-5 {h_2} {h_4} {h_5}^2+4 {h_2}^2 {h_4}^2\right) \csc
   ^2({\theta_1})+\left(4 {h_2} {h_4}-{h_5}^2\right)
   {f_2}({\theta_2})^2  \nonumber\\
   && \left({h_5}^2 {f_1}({\theta_1})^2 \sin ^2({\theta_2})+4
   {h_2} {h_4} \left(\csc ^2({\theta_1}) \sin ^2({\theta_2})-1\right)\right)\Bigr)\Bigr)/\bigl(\sqrt{h} {h_1} {h_2} {h_4}\bigr)\bigr])^{\frac{1}{2}} \Bigr]/480 {g_s}
   r^4 + \nonumber\\
   && 4 \Bigr[\Bigr(3 {A_5} \sqrt{{h_2}}{g_s}^2 M {N_f} \cos ({\theta_2}) \cot \left(\frac{{\theta_2}}{2}\right) ln (r) \sin ({\theta_1}) \nonumber\\
    &&
   \Bigr(\frac{{A_1} M \csc ^2({\theta_1}) \left(\frac{3 {g_s} {N_f} ln (r)}{2 \pi }+1\right)  (2 {h_1} \cos ({\theta_1}) \cos
   ({\theta_2})-{h_5} \sin ({\theta_1}) \sin ({\theta_2}))}{2 \sqrt{h} {h_1} {h_2} \sqrt{{h_4}} r^2}- \nonumber\\
   && \frac{3 {A_3} {g_s} M
   {N_f} \cot ({\theta_1}) \cot \left(\frac{{\theta_2}}{2}\right) \csc ({\theta_1})  \sin ({\theta_2})}{4 \sqrt{h} {h_2} \sqrt{{h_4}} \pi
   r^2}\Bigr) \Bigr)/\Bigr(4 \sqrt[4]{h} \pi  r \left({h_1} \cos ^2({\theta_1})+{h_2} \sin ^2({\theta_1})\right)\Bigr)+ \nonumber\\
   &&
   \Bigr(\csc ({\theta_1})
   \Bigr(\frac{3 {A_5} {B_5} {g_s}^2 M {N_f} \cos ({\theta_1})  \cot \left(\frac{{\theta_1}}{2}\right) ln (r) }{8 \pi} +   3 {A_4 {g_s} M }\sin ({\theta_1})
   \Bigr(\frac{ {g_s} {N_f} ln ^2(r)}{4 \pi } +ln (r)+ \nonumber\\
    && \frac{{g_s} {N_f} (2 ln (r)+1) ln \left(\sin \left(\frac{{\theta_1}}{2}\right)  \sin
   \left(\frac{{\theta_2}}{2}\right)\right)}{4 \pi }\Bigr)   \Bigr) \nonumber\\
   && \Bigr(-\Bigr(2 {h_2} \Bigl(\frac{2 {A_1} {B_1} M
   \csc ^3({\theta_1}) \left(\frac{3 {g_s} {N_f} ln (r)}{2 \pi }+1\right) \left({h_1} \cos ^2({\theta_1})+{h_2} \sin
   ^2({\theta_1})\right) \sin ({\theta_2})}{\sqrt{h} {h_1} {h_2} \sqrt{{h_4}} r^2}- \nonumber\\
   && \frac{3 {A_3} {B_3} {g_s} M {N_f} \cot
   \left(\frac{{\theta_1}}{2}\right) \cot ({\theta_1}) \csc ({\theta_1}) \sin ({\theta_2})}{4 \sqrt{h} {h_2} \sqrt{{h_4}} \pi  r^2}\Bigr) \sin
   ^2({\theta_1})\Bigr)/\bigl({h_1} \cos ^2({\theta_1}) {h_2} \sin ^2({\theta_1})\bigr)-  \nonumber\\
   && \frac{2 {A_1} {B_1} M \csc ({\theta_1}) \left(\frac{3 {g_s}
   {N_f} ln (r)}{2 \pi }+1\right) \sin ({\theta_2})}{\sqrt{h} {h_1} \sqrt{{h_4}} r^2}\Bigr)\Bigr)/\bigl(\sqrt[4]{h} \sqrt{{h_2}} r\bigr)\Bigr] + \nonumber\\
   && 3
    \Bigr[\Bigl(\csc ({\theta_1}) \Bigl(\frac{3 {A_5} {B_5} {g_s}^2 M {N_f} \cos ({\theta_1}) \cot \left(\frac{{\theta_1}}{2}\right) ln (r)
  }{8 \pi }+  3 {A_4} {g_s}M \sin ({\theta_1}) \Bigl(\frac{9 {g_s} {N_f} ln ^2(r)}{4 \pi }+ln (r)+ \nonumber\\
   && \frac{{g_s} {N_f} (2 ln (r)+1) ln \left(\sin
   \left(\frac{{\theta_1}}{2}\right) \sin \left(\frac{{\theta_2}}{2}\right)\right)}{4 \pi }\Bigr) \Bigr)  (2 {h_1} \cos
   ({\theta_1}) \cos ({\theta_2})-{h_5} \sin ({\theta_1}) \sin ({\theta_2}))\Bigr)/ \nonumber\\
   && \Bigl(2 \sqrt[4]{h} \sqrt{{h_4}} r \left({h_1} \cos
   ^2({\theta_1})+{h_2} \sin ^2({\theta_1})\right)\Bigr)-\frac{3 {A_5} {B_5} {g_s}^2 M {N_f} \cos ({\theta_2}) \cot
   \left(\frac{{\theta_1}}{2}\right) \csc ({\theta_1}) ln (r)}{8 \sqrt[4]{h} \sqrt{{h_4}} \pi  r}\Bigr] \nonumber\\
   && \Bigr[\Bigr(9 {A_5} M {N_f} \cos
   ({\theta_2}) \cot \left(\frac{{\theta_2}}{2}\right) \csc ({\theta_1}) ln (r) \nonumber\\
   && \Bigr(\frac{{N_f}}{4 \sqrt[4]{h} {h_1} \pi  r}-\frac{{h_2}
   \sin ^2({\theta_1})}{{h_1} \cos
   ^2({\theta_1})+{h_2} \sin ^2({\theta_1})}  \Bigl(-\frac{{N_f} \left({h_1} \cos ^2({\theta_1})+{h_2} \sin ^2({\theta_1})\right) \csc ^2({\theta_1})}{4
   \sqrt[4]{h} {h_1} {h_2} \pi  r}- \nonumber\\
   && \frac{{N_f} \cot ({\theta_1}) \csc ({\theta_1})}{4 \sqrt[4]{h} {h_2} \pi  r}\Bigr) \Bigr) {g_s}^2\Bigr)/\bigl(4 \sqrt[4]{h} \sqrt{{h_2}} \pi  r\bigr)+ \nonumber\\
   &&
   \frac{\sqrt{h_4} {h_5} {N_f} \cos ({\theta_2})
   {f_2}({\theta_2}) \sin ({\theta_2})}{4  \pi
   \left(4 {\csc ^2({\theta_1}) \left({h_1} \cos ^2({\theta_2})+{h_4} \sin ^2({\theta_2})^2\right)}- {\csc ^2({\theta_1}) (2
   {h_1} \cos ({\theta_1}) \cos ({\theta_2})-{h_5} \sin ({\theta_1}) \sin ({\theta_2}))^2}\right)}\Bigr]\nonumber\\
   && - \Bigr(\frac{3 {A_5} {g_s}^2 M {N_f} \cos ({\theta_2}) \cot
   \left(\frac{{\theta_2}}{2}\right) \csc ({\theta_1}) ln (r) ({h_5} \sin ({\theta_1}) \sin
   ({\theta_2}))-2 {h_1} \cos ({\theta_1}) \cos ({\theta_2}) }{16 \sqrt[4]{h} \sqrt{{h_4}} \pi  r \left({h_1} \cos ^2({\theta_1})+{h_2} \sin ^2({\theta_1})\right)}- \nonumber\\
   &&  \frac{\csc
   ({\theta_1})}{\sqrt[4]{h}
   \sqrt{{h_4}} r} \Bigl(\frac{-3 {g_s}^2 {A_5} M {N_f} \cos ({\theta_2}) \cot \left(\frac{{\theta_2}}{2}\right) ln (r) }{8 \pi }-3 {A_4}
   {B_4} {g_s}M \sin ({\theta_2})  \bigl(\frac{9 {g_s} {N_f} ln ^2(r)}{4 \pi }+ln (r)+ \nonumber\\
   && \frac{{g_s} {N_f} (2 ln (r)+1) ln \left(\sin
   \left(\frac{{\theta_1}}{2}\right) \sin \left(\frac{{\theta_2}}{2}\right)\right)}{4 \pi }\Bigr) \Bigr)\Bigr) \times \nonumber\\
   && \Bigr(\frac{3 {A_1} M \left(\frac{3 {g_s} {N_f} ln (r)}{2 \pi }+1\right)}{\sqrt{h} {h_1} \sqrt{{h_2}}
   r^2}- \frac{6 \csc ({\theta_1})}{\sqrt[4]{h}   \sqrt{{h_2}} r} \Bigl(\frac{3 {A_5} {B_5}{g_s}^2 M {N_f} \cos ({\theta_1}) \cot \left(\frac{{\theta_1}}{2}\right) ln (r)
   }{8 \pi }+ \nonumber\\
   && 3 {A_4}{g_s} M \sin ({\theta_1})  \Bigl(\frac{9 {g_s} {N_f} ln ^2(r)}{4 \pi }+ln (r) \frac{{g_s} {N_f} (2 ln (r)+1) ln \left(\sin
   \left(\frac{{\theta_1}}{2}\right) \sin \left(\frac{{\theta_2}}{2}\right)\right)}{4 \pi }\Bigr) \Bigr) \times \nonumber\\
  & & \Bigl(\frac{{N_f}}{4 \sqrt[4]{h} {h_1} \pi  r}-\frac{{h_2} \sin ^2({\theta_1})}{{h_1} \cos ^2({\theta_1})+{h_2} \sin ^2({\theta_1})} \Bigl(-\frac{{N_f} \left({h_1} \cos
   ^2({\theta_1})+{h_2} \sin ^2({\theta_1})\right) \csc ^2({\theta_1})}{4 \sqrt[4]{h} {h_1} {h_2} \pi  r}-\frac{{N_f} \cot ({\theta_1})
   \csc ({\theta_1})}{4 \sqrt[4]{h} {h_2} \pi  r}\Bigr)\Bigr)
   \nonumber\\
   &&   + \frac{\sqrt{{h_4}} {h_5} {N_f} \cos ({\theta_2}) {f_2}({\theta_2}) \sin ({\theta_2})}{4  \pi  \left( 4{\csc ^2({\theta_1}) \left({h_1} \cos ^2({\theta_2})+{h_4} \sin
   ^2({\theta_2})\right)^2} - {\csc ^2({\theta_1}) (2 {h_1} \cos ({\theta_1}) \cos ({\theta_2})-{h_5} \sin ({\theta_1}) \sin
   ({\theta_2}))^2}\right)}\Bigr)\Bigr)\Bigr)\nonumber\\
   & &
   \end{eqnarray}
\begin{eqnarray}
\label{HodgeFtildetildetildertheta2}
 & (xii) &  \tilde{\tilde{\tilde{F_5}}}_{r \theta_2}=\nonumber\\
   && -\frac{9 {A_2} {A_5} {B_2} {g_s}^3 M^2 {N_f}^2 \cos ({\theta_2}) \cot \left(\frac{{\theta_1}}{2}\right) \cot \left(\frac{{\theta_2}}{2}\right) \csc
   ({\theta_1}) ln (r) (2 {h_1} \cos ({\theta_1}) \cos ({\theta_2})-{h_5} \sin ({\theta_1}) \sin ({\theta_2})) }{8 h^{3/4}
   {h_1} \sqrt{{h_4}} \sqrt{{h_4}} \pi ^2 r^4 \left({h_1} \cos ^2({\theta_1})+{h_4} \sin ^2({\theta_1})\right)} + \nonumber\\
   &&  \bigg[\Bigr({h_4}
   {M_{\rm eff}}^2 {N_f} \cot \left(\frac{{\theta_1}}{2}\right) \csc ^2({\theta_1}) \left((\cos (2 {\theta_1})-\cos (2 {\theta_2}))
   {f_2}({\theta_2})^2-2\right) ln (r)
   {g_s}^2\Bigr)/\Bigr(320 \sqrt{h} \pi  r^2 \left(r^4-{r_h}^4\right) \nonumber\\
   && \Bigr(-4 \csc ({\theta_1}) {f_1}({\theta_1}) {f_2}({\theta_2}) \sin ({\theta_2}) \
   {h_5}^2+4 \left({h_5}^2-{h_4} {h_4}\right) \nonumber\\
    && \csc ^2({\theta_1})+{f_2}({\theta_2})^2 \left({h_5}^2 {f_1}({\theta_1})^2 \sin
   ^2({\theta_2})+4 {h_4} {h_4} \left(\csc ^2({\theta_1}) \sin ^2({\theta_2})-1\right)\right)\Bigr)\Bigr)\bigg] \times \nonumber\\
  &&   \bigg[-\Bigr(r^4 \sin ^2({\theta_2}) \Bigr(4 \left({h_5}^2-4 {h_4} {h_4}\right) \csc ({\theta_1})
   {f_1}({\theta_1}) {f_2}({\theta_2}) \sin ({\theta_2}) {h_5}^2-4 \left({h_5}^4-5 {h_4} {h_4} {h_5}^2+4 {h_4}^2
   {h_4}^2\right) \csc ^2({\theta_1}) \nonumber\\
   && +\left(4 {h_4} {h_4}-{h_5}^2\right) {f_2}({\theta_2})^2 \left({h_5}^2 {f_1}({\theta_1})^2
   \sin ^2({\theta_2})+4 {h_4} {h_4} \left(\csc ^2({\theta_1}) \sin ^2({\theta_2})-1\right)\right)\Bigr)\Bigr)/\sqrt{h} {h_1} {h_4} {h_4}\bigg]^{\frac{1}{2}} \times \nonumber\\
   && -3 \Bigr(\frac{3 {A_5} {g_s}^2 M {N_f} \cos
   ({\theta_2}) \cot \left(\frac{{\theta_2}}{2}\right) \csc ({\theta_1}) ln (r) ({h_5} \sin
   ({\theta_1}) \sin ({\theta_2})-2 {h_1} \cos ({\theta_1}) \cos ({\theta_2})) }{16 \sqrt[4]{h} \sqrt{{h_4}} \pi  r \left({h_1} \cos ^2({\theta_1})+{h_4} \sin
   ^2({\theta_1})\right)}- \nonumber\\
   &&     \frac{\csc ({\theta_1})}{\sqrt[4]{h} \sqrt{{h_4}} r} \Bigl(-\frac{3 {A_5} {g_s}^2 M {N_f} \cos ({\theta_2}) \cot \left(\frac{{\theta_2}}{2}\right) ln (r)
   }{8 \pi }-3 {A_4} {B_4}
   {g_s} M \sin ({\theta_2}) \Bigl(\frac{9 {g_s} {N_f} ln ^2(r)}{4 \pi }+ln (r)+ \nonumber\\
   &&\frac{{g_s} {N_f} (2 ln (r)+1) ln
   \left(\sin \left(\frac{{\theta_1}}{2}\right) \sin \left(\frac{{\theta_2}}{2}\right)\right)}{4 \pi }\Bigr) \Bigr)\Bigr) \times \nonumber\\
    &&  \Bigr(\frac{ \sqrt{{h_4}} {h_5} {N_f} \cos ({\theta_2}) {f_2}({\theta_2}) \sin ({\theta_2})}{4
    \pi  \left( {4 \csc ^2({\theta_1}) \left({h_1} \cos
   ^2({\theta_2})+{h_4} \sin ^2({\theta_2})\right)^2} - {\csc ^2({\theta_1}) (2 {h_1} \cos ({\theta_1}) \cos
   ({\theta_2})-{h_5} \sin ({\theta_1}) \sin ({\theta_2}))^2}\right)} \nonumber\\
  &&  -\frac{9 {A_2} {B_2} {g_s} M {N_f} \cot \left(\frac{{\theta_1}}{2}\right)}{4 \sqrt{h} {h_1}
   \sqrt{{h_4}} \pi  r^3}\Bigr)
   \end{eqnarray}

 \begin{eqnarray}
 \label{HodgeFtildetildetildertheta1}
 & (xiii) &  \tilde{\tilde{\tilde{F_5}}}_{r \theta_1}=
  \nonumber\\
  && -\Bigr( {h_2} {g_s}^2 {M_{\rm eff}}^2 {N_f} \cot \left(\frac{{\theta_2}}{2}\right) ln (r)  \csc ^2({\theta_1})\Bigr)/\Bigr(160 h^{3/2} \pi  r^2 \left(r^4-{r_h}^4\right) \Bigr(-4 \csc ({\theta_1}) {f_1}({\theta_1}) {f_2}({\theta_2}) \sin ({\theta_2}) {h_5}^2\nonumber\\
   && +4 \left({h_5}^2-{h_2} {h_4}\right) \csc
   ^2({\theta_1})+{f_2}({\theta_2})^2   \left({h_5}^2 {f_1}({\theta_1})^2 \sin ^2({\theta_2})+4 {h_2} {h_4} \left(\csc
   ^2({\theta_1}) \sin ^2({\theta_2})-1\right)\right)\Bigr)\Bigr) \times \nonumber\\
   && \Bigl[-\Bigl(r^4 \sin ^2({\theta_2}) \Bigr(4
   \left({h_5}^2-4 {h_2} {h_4}\right) \csc ({\theta_1}) {f_1}({\theta_1}) {f_2}({\theta_2}) \sin ({\theta_2}) {h_5}^2-4
   \left({h_5}^4-5 {h_2} {h_4} {h_5}^2+4 {h_2}^2 {h_4}^2\right) \nonumber\\
    && \csc ^2({\theta_1})+\left(4 {h_2} {h_4}-{h_5}^2\right)
   {f_2}({\theta_2})^2 \left({h_5}^2 {f_1}({\theta_1})^2 \sin ^2({\theta_2})+4 {h_2} {h_4} \left(\csc ^2({\theta_1}) \sin
   ^2({\theta_2})-1\right)\right)\Bigr)\bigr)/\sqrt{h} {h_1} {h_2} {h_4}\Bigr]^{\frac{1}{2}}+ \nonumber\\
   &&
   \Bigr(\frac{3 {A_2} {B_2} {g_s} M {N_f} \cot \left(\frac{{\theta_1}}{2}\right)
   }{h^{3/4} {h_1} \sqrt{{h_2}} \sqrt{{h_4}} \pi  r^4
   \left({h_1} \cos ^2({\theta_1})+{h_2} \sin ^2({\theta_1})\right)} \Bigr(\frac{3 {A_5} {B_5} M {N_f} \cos ({\theta_1}) \cot \left(\frac{{\theta_1}}{2}\right) ln (r) {g_s}^2}{8 \pi } \nonumber\\
   && +3 {A_4} M {g_s}
   \Bigr(\frac{9 {g_s} {N_f} ln ^2(r)}{4 \pi }+  ln (r)+\frac{{g_s} {N_f} (2 ln (r)+1) ln \left(\sin \left(\frac{{\theta_1}}{2}\right) \sin
   \left(\frac{{\theta_2}}{2}\right)\right)}{4 \pi }\Bigr)\Bigr) \sin ({\theta_1})\Bigr)/ \nonumber\\
    && (2 {h_1} \cos ({\theta_1}) \cos
   ({\theta_2})-{h_5} \sin ({\theta_1}) \sin ({\theta_2})) \csc ({\theta_1}))-3 \bigg[\Bigr(\csc ({\theta_1}) \Bigr(\frac{3 {A_5} {B_5} M {N_f}
   \cos ({\theta_1}) \cot \left(\frac{{\theta_1}}{2}\right) ln (r) {g_s}^2}{8 \pi }+ \nonumber\\
   && 3 {A_4} M \Bigr(\frac{9 {g_s} {N_f} ln ^2(r)}{4 \pi
   }+ln (r)+\frac{{g_s} {N_f} (2 ln (r)+1) ln \left(\sin \left(\frac{{\theta_1}}{2}\right) \sin \left(\frac{{\theta_2}}{2}\right)\right)}{4 \pi
   }\Bigr) \sin ({\theta_1}) {g_s}\Bigr) \nonumber\\
   && (2 {h_1} \cos ({\theta_1}) \cos ({\theta_2})-{h_5} \sin ({\theta_1}) \sin ({\theta_2}))\Bigr)/\Bigl(2
   \sqrt[4]{h} \sqrt{{h_4}} r \left({h_1} \cos ^2({\theta_1})+{h_2} \sin ^2({\theta_1})\right)\Bigr)- \nonumber\\
   && \frac{3 {A_5} {B_5} {g_s}^2 M
   {N_f} \cos ({\theta_2}) \cot \left(\frac{{\theta_1}}{2}\right) \csc ({\theta_1}) ln (r)}{8 \sqrt[4]{h} \sqrt{{h_4}} \pi  r}\bigg] \times \nonumber\\
   &&
   \Bigr(-\frac{9 {A_2} {B_2} {g_s} M {N_f} \cot \left(\frac{{\theta_1}}{2}\right)}{4 \sqrt{h} {h_1} \sqrt{{h_2}} \pi
   r^3}- \nonumber\\
   &&  \frac{{h_2}\sqrt{{h_4}} {N_f} \cos ({\theta_2}) {f_1}({\theta_1}) \sin ({\theta_1})}{2  \pi   \left( {4 \csc ^2({\theta_1}) \left({h_1} \cos ^2({\theta_2})+{h_4} \sin
   ^2({\theta_2})\right)^2} - {\csc ^2({\theta_1}) (2 {h_1} \cos ({\theta_1}) \cos ({\theta_2})-{h_5} \sin ({\theta_1}) \sin
   ({\theta_2}))^2} \right)}\Bigr)
   \end{eqnarray}
}

\end{document}